\documentclass[fleqn,usenatbib]{mnras}

\usepackage{newtxtext,newtxmath}

\usepackage[T1]{fontenc}

\DeclareRobustCommand{\VAN}[3]{#2}
\let\VANthebibliography\thebibliography
\def\thebibliography{\DeclareRobustCommand{\VAN}[3]{##3}\VANthebibliography}

\usepackage{graphicx}	
\usepackage{amsmath}	
\usepackage{gensymb}
\usepackage{color}
\usepackage{threeparttable}

\definecolor{burntorange}{rgb}{0.8, 0.33, 0.0}

\newcommand\lsim{\mathrel{\rlap{\lower4pt\hbox{\hskip1pt$\sim$}}
    \raise1pt\hbox{$<$}}}
\newcommand\gsim{\mathrel{\rlap{\lower4pt\hbox{\hskip1pt$\sim$}}
    \raise1pt\hbox{$>$}}}

\newcommand{\jhat}{\hat{\textbf{\j}}}

\makeatletter
\newlength{\abovecaptionskip}%
\makeatother

\title[Star-disc alignmment]{Capture of stars into gaseous discs around massive black holes: Alignment, circularization and growth}

\author[Generozov \& Perets]{
A. Generozov,$^{1}$\thanks{E-mail: al.generozov@campus.technion.ac.il}
H. B. Perets$^{1}$  
\\
$^{1}$Technion - Israel Institute of Technology, Haifa, Israel, 3200003\\
}

\date{Accepted XXX. Received YYY; in original form ZZZ}

\pubyear{2022}

\begin{document}
\label{firstpage}
\pagerange{\pageref{firstpage}--\pageref{lastpage}}
\maketitle

\begin{abstract}
The majority of massive black holes (MBHs) likely hosted gas discs during their lifetimes.
These could either be long-lived active galactic nuclei (AGN) discs, or shorter-lived discs formed following singular gas infall events, as was likely the case in our own Galactic Center. Stars and compact objects in such environments are therefore expected to interact with the gaseous disc as they go through it, and potentially become aligned and fully embedded within it. The interactions of embedded stars with the gas could give rise to a plethora of physical processes affecting the stars, including growth through accretion of gas, migration in the disc, stellar captures, and mergers with other stars. The impact of such processes strongly depends on the population of stars that eventually align with the disc and become embedded in it. 
Here we make use of analytic tools to analyze the alignment process, accounting for both geometric drag and gas dynamical friction. We find that up to $\sim 50\%$ of main sequence stars and stellar mass black holes in the central 0.1 pc can align with AGN discs in the Galactic Center and similar galactic nuclei. The orbits of aligned stars are typically circularized and are prograde with respect to the AGN disc. Furthermore, alignment and accretion are intimately linked, and the capture of stars by an AGN disc can potentially explain the origin of the young stellar disc in the Galactic Center with a top-heavy mass function, even without the need for a star-formation event.
\end{abstract}




\section{Introduction}
Stellar dynamics in a galactic nucleus can be significantly perturbed by a gaseous nuclear disc (for which we use the term AGN disc hereafter). Many aspects of such stars-disc interactions have been explored over the last decades, with significant renewed interest over the last few years, due to their potential role in producing gravitational-waves (GW) and other electromagnetic transient events \citep[e.g.][and references therein]{syer+1991,Artymowicz1993,vokrouhlicky&karas1998,subr&karas1999,karas&subr2001,miralde-escude&kollmeier2005,McKernan2012,just+2012,kennedy+2016,bartos+2017,Stone2017,panamarev+2018,fabj+2020}.

Considering their deep potential wells, galactic nuclei can retain compact remnants, even following natal kicks. Together with mass-segregation processes, this could give rise to a large population of compact objects near MBHs \citep{bahcall&wolf1977,miralda-escude&gould2000,alexander&hopman2009,preto&amaro-seoane2010, aharon&perets2015, antonini&rasio2016, alexander_tal2017}.
Thereby both stars and compact objects are likely to interact with an AGN disc.

Once embedded in an AGN disc, binary interactions with the gas may catalyze stellar \citep{BaruteauCuadraLin2011}, and compact object mergers \citep{McKernan2012,li+2021,li+2022}. Gas-assisted formation of binaries \citep{Tagawa2020, rozner+2022,deLaurentiis+2022,rowan+2022}, migration in the AGN disc, and accumulation at migration traps \citep{McKernan2012, McKernan2018} are also possible.  
In addition, migration and secular perturbations \citep{von-Zeipel1910,lidov1962, kozai1962} by the disc could drive stars and compact objects into close interactions with the central MBH, potentially increasing the rates of tidal disruption events and extreme mass ratio GW inspirals \citep{karas&subr2007, kocsis+2011_emri, mckernan+2021_starfall,pan&yang2021,secunda+2021}

Finally, the capture of background stars by an AGN disc may account for the young clockwise stellar disc in the Galactic Center. In particular, this disc is a few Myr old and has a top-heavy mass function \citep{levin&beloborodov2003,paumard+2006,bartko+2010,lu+2013}.\footnote{Additional young, disc-like features have been identified at large radii, but they are consistent with a standard mass function \citep{vonFellenberg+2022}.} As discussed in \S~\ref{sec:accretion} alignment of stars with an AGN disc is likely associated with significant accretion. Thus, the stars in the observed clockwise disc may be captured old stars that were rejuvenated after accreting gas (see also \citealt{panamarev+2018,davies&lin2020}).

In this paper, we provide analytic tools for analyzing inclined and eccentric stellar orbits that are perturbed by an AGN disc, accounting for both gas dynamical friction (GDF) and geometric gas drag. These are also complemented by and compared with few-body simulations. We provide recipes for the timescale for orbits to align, and for the semimajor axis of orbits after alignment that are generally accurate to within a factor of a few (as described in \S~\ref{sec:ecc} we expect orbits to evolve towards a prograde, circular configuration). {Some of our results follow up on previous studies by \citep{Artymowicz1993} and \citep{subr&karas1999}, but also provide a more comprehensive analysis of the disc alignment process.}

The remainder of this paper is organized as follows. In \S~\ref{sec:alignAnalytic} we provide analytic estimates for the alignment timescale of stellar mass objects with an AGN disc. In \S~\ref{sec:accretion} we discuss the connection between the alignment process and accretion. In \S~\ref{sec:ecc} we discuss eccentricity evolution during the alignment process. In \S~\ref{sec:num} we validate our analytic results with few-body simulations and apply our machinery to model galactic nuclei. We summarize our results in \S~\ref{sec:summary}.

\section{Alignment timescale}
\label{sec:alignAnalytic}
We introduce analytic estimates for the alignment timescale in different regimes below. Taking the minimum of these estimates gives an approximation that is within a factor of a few of the alignment time measured in our numerical few-body simulations. This provides a convenient and computationally efficient way to determine which orbits will align with an AGN disc.

Let us consider the case where the orbital energy of a star on an inclined orbit with respect to the disc does not significantly change during its alignment. In this case, the characteristic alignment timescale is the time to dissipate the vertical velocity through repeating, dissipative encounters with the disc, viz.

\begin{align}
t_{\rm align}=\frac{v_z}{f_{\rm drag,z}} \frac{P_{\rm orb}}{t_{\rm cross}},
\label{eq:talign1a}
\end{align}
where $v_{z}$ is the star's velocity perpendicular to the disc as it crosses through it, $f_{\rm drag}$ is the (specific) drag force, and $t_{\rm cross}$ is the timescale for the star to cross through the disc, viz.

\begin{align}
    t_{\rm cross}=\frac{2 (h/r) r}{v_z},
\end{align}
where $h/r$ is the aspect ratio of the disc and $r$ is the radius of the orbital node. In general, equation~\eqref{eq:talign1a} gives different results at the ascending and descending node of the orbit. 

One source of drag is the dissipative force due to gas dynamical friction (GDF). We use the prescription in \citet{ostriker1999} to model this force, viz.
\begin{align}
    \mathbf{f_{\rm drag}}=\mathbf{f_{\rm gdf}}=-\frac{4 \pi G^2 m_* \rho_g}{v_{\rm rel}^3} I({v_{\rm rel}/c_s}) \mathbf{v_{\rm rel}},
    \label{eq:fgdf}
\end{align}
where $\rho_g$ is the gas density, $m_*$ is the stellar mass, and $v_{\rm rel}$ is the relative velocity between the star and the disc. The function $I$ is 

\begin{align}
I(\mathcal M)=
    \begin{cases}
        1/2 \log(1-1/\mathcal M^2)+\ln \Lambda, & \mathcal M> 1\\
        1/2 \log\left(\frac{1+\mathcal M}{1-\mathcal M}\right)-\mathcal M, & \mathcal M<1\\
    \end{cases}
    \label{eq:I}
\end{align}
For $\mathcal M\gg 1$, $I$ is nearly independent of the Mach number. In order to avoid the divergence at $M=1$, we use 
\begin{align}
    I(\mathcal M)=
    \begin{cases}
        \ln(\Lambda) & \mathcal M\geq 1\\
        \min\left[\ln \Lambda, \frac{1}{2}\log\left(\frac{1+\mathcal M}{1-\mathcal M}\right)-\mathcal M\right] & \mathcal M<1.
    \end{cases}
    \label{eq:Ib}
\end{align}
Above $\ln \Lambda$ is the Coulomb logarithm. Following \citet{Tagawa2020} we assume $\ln \Lambda=3.1$. The sound speed in the disc is approximately $h/r$ times the local Keplerian velocity. Thus, stars outside the disc (with inclinations greater than $h/r$) will be supersonic. Thus, we use $I=\ln \Lambda$ for our analytic estimates. 

For GDF the characteristic alignment time is then

\begin{align}
    &t_{\rm align}=k_o \frac{v_{\rm c}(a)^3}{4 \pi G^2 \rho_g(a) m_* \ln \Lambda} 
    \left(\frac{v_{\rm rel}}{v_{\rm c}}\right)^{3} \left(\frac{P_{\rm orb}}{t_{\rm cross}}\right) \left(\frac{r}{a}\right)^{\gamma}\nonumber\\
    &\resizebox{0.96\columnwidth}{!}{
    $\frac{v_{\rm rel}}{v_{\rm c}}=\sqrt{\frac{e^2 \pm e \cos (\omega ) \left(2 \cos (i) \sqrt{1\pm e \cos (\omega )}-3\right)-2 \cos (i) \sqrt{1\pm e \cos (\omega )}+2}{1-e^2}}$
    }\nonumber\\
    &\frac{P}{t_{\rm cross}}=\frac{\pi (1 \pm e \cos(\omega))^2  \sin(i) }{(1-e^2)^{3/2} (h/r)}\nonumber\\
    &\frac{r}{a}=\frac{1-e^2}{1 \pm e \cos(\omega)}
    \label{eq:talign1}
\end{align}
Here, $e$, $a$, $i$, and $\omega$ are the eccentricity, semimajor axis, inclination, and argument of pericenter of the stellar orbit, and $-\gamma$ is the power-law index of the gas density profile; $v_c(a)=\sqrt{\frac{G M}{a}}$. We assume that the aspect ratio of the disc, $h/r$, is constant. In general, the relative velocities differ at the ascending and descending node of the orbit. Therefore, some terms can be either positive or negative. For an $r^{-3/2}$ gas density profile, the alignment timescale in the GDF regime is independent of the semimajor axis. To improve the accuracy of equation~\eqref{eq:talign1} we include the constant factor $k_o$, which is chosen to match the numerical results in \S~\ref{sec:num}. 

For a zero eccentricity orbit equation~\eqref{eq:talign1} simplifies to

\begin{align}
    t_{\rm align}=k_o \frac{v_{\rm c}(a)^3 \sin^3 (i/2)  \sin(i)}{G^2 \rho_g m_{\star} (h/r) \ln \Lambda},
    \label{eq:talignCirc}
\end{align}
For low inclinations $t_{\rm align} \propto i^4$. Note that equation~\eqref{eq:talignCirc} includes the contributions of both the ascending and the descending node.

At small radii, where the relative velocity exceeds the stellar escape speed, geometric drag will dominate GDF. In this case the drag force is 
\begin{equation}
    f_{\rm drag}=f_{\rm geo}= \left(\frac{v_{\rm rel}}{v_{\rm esc}}\right)^4 \frac{Q}{I} f_{\rm gdf},
    \label{eq:fgeo}
\end{equation}
where $v_{\rm esc}$ is the escape speed from the surface of the star. The corresponding alignment time is 
\begin{align}
    \resizebox{0.96\columnwidth}{!}{$t_{\rm align,geo}=k_1 \frac{v_{\rm c}(a)^3}{4 \pi G^2 \rho_g(a) m_\star \ln \Lambda} \left(\frac{\ln \Lambda}{Q_d}\right)  \left(\frac{v_{\rm c}}{v_{\rm rel}}\right) \left(\frac{P_{\rm orb}}{t_{\rm cross}}\right) \left(\frac{r}{a}\right)^{\gamma} \left(\frac{v_{\rm esc}}{v_{\rm c}}\right)^4$}
\end{align}
For circular orbits, the alignment time is 
\begin{align}
   \resizebox{0.96\columnwidth}{!}{$t_{\rm align,geo}=k_1 \frac{v_{\rm c}(a)^3}{4 G^2 \rho_g(a) m_* \ln \Lambda (h/r)} 
    \left(\frac{\ln \Lambda}{Q_d}\right) \left(\frac{\sin(i)}{\sqrt{2-2 \cos(i)}}\right) \left(\frac{v_{\rm esc}}{v_{c}(a)}\right)^4.$}
\end{align}
Interestingly, the alignment time is nearly independent of the inclination in the low inclination limit.

Above, we used the timescale to change the z-component of a star's momentum as a proxy for the alignment time. This is an accurate estimate at low inclinations. However, it breaks down at high inclinations, since orbits inspiral significantly in semimajor axis before they align. For high inclinations, we find the timescale to change the energy of an orbit is a more accurate proxy for the alignment timescale, viz.

\begin{align}
    t_{\rm E}=\frac{E}{\mathbf{f_{\rm drag} \cdot v}} \frac{P_{\rm orb}}{t_{\rm cross}}
\end{align}
In the GDF regime this is
\begin{align}
   t_{\rm E, GDF} = k_2 \frac{v_{\rm c}(a)^3}{4 \pi G^2 \rho_g(a) m_\star \ln (\Lambda)} \left(\frac{v_{\rm rel}}{v_{\rm c}(a)}\right) \left(\frac{P}{t_{\rm cross}}\right) \left(\frac{r}{a}\right)^{\gamma}
\end{align}
In the geometric regime 
\begin{align}
    t_{\rm E,geo}\approx k_3 \frac{v_{\rm c}(a)^3}{4 \pi G^2 \rho_g(a) m_\star Q_d} \left(\frac{v_{\rm c}(a)}{v_{\rm rel}}\right)^3 \left(\frac{P}{t_{\rm cross}}\right) \left(\frac{r}{a}\right)^{\gamma} \left(\frac{v_{\rm esc}}{v_{\rm c}(a)}\right)^4,
    \label{eq:tegeo}
\end{align}
where we use the approximation $\mathbf{v_{\rm rel} \cdot v_{\rm node}} \approx 1/2 v_{\rm rel}^2$.\footnote{This is exact for a circular orbit.} In practice, the alignment timescale can be estimated (within a factor of a few) as 
\begin{align}
    t_{\rm align, all}={\rm min} \left[t_{\rm align}, t_{\rm align, geo}, t_{\rm E, GDF}, t_{\rm E,geo}\right],
    \label{eq:talignGen}
\end{align}
with $k_o=0.18$, $k_1=3$, $k_2=1$, and $k_3=1$. These values are chosen to reproduce the results of the numerical simulations in \S~\ref{sec:num} and Appendix~\ref{app:numValid2}. For each term in equation~\eqref{eq:talignGen}, we use the harmonic mean of the timescales at the ascending and the descending node (we assume both nodes are within the disc). Note that without the last two terms in brackets, this would be equivalent to the alignment timescales estimated by \citet{Artymowicz1993}, modulo constant and logarithmic terms (see their equations 6 and 7). In practice, these terms are critical for reproducing the results of our few-body simulations.

Alternatively, \citet{subr&karas1999} provide an exact expression for the alignment timescale for orbits rotated over their minor axis (with argument of pericenter, $\omega=\pi/2$) under the influence of geometric drag. Their expression can be readily generalized for the GDF case. However, we find that equation~\eqref{eq:talignGen} is more reliable in this regime since the alignment timescale changes at the order of magnitude level as a function of argument of pericenter. Thus, the $\omega=\pi/2$ solution cannot be applied to arbitrary orientations.

\citet{subr&karas1999} also present an approximate analytic relation between the orbital angular momentum and inclination which is exact for circular orbits, and eccentric ones rotated over their minor axes (see also \citealt{rauch1995}).
In Appendix~\ref{app:coupleSol}, we present a simple derivation of this result for completeness. 
This relation can be used to estimate the final semimajor axis of orbits that become aligned with the disc. We also find a more accurate expression for the final angular momentum in the prograde, GDF for orbits of arbitrary orientation (see equation~\ref{eq:jfGenCorr}).

We have neglected other physical effects that can lead to alignment between stars and an AGN disc, such as resonant dynamical friction between the disc and stars \citep{ginat+2022}. In that case the alignment time is proportional to $1/\sin(i)$. Naturally, this effect is not present in our study, since we neglect the gravity of the AGN disc.

\section{Mass accretion}
\label{sec:accretion}
So far we have neglected the effect of accretion on the alignment timescale. Below, we show that accretion will change the alignment time by at most a factor of order unity. Also, we show that the accreted mass formally diverges as stars approach alignment with the disc, though this runaway growth is likely to be stopped by feedback. These conclusions are also supported by numerical experiments in \S~\ref{sec:num} (see Figure~\ref{fig:accretion example} and the surrounding discussion).

As stars pass through the disc, they will also accrete material. This accretion will also modify the stars' momentum. The alignment timescale from accretion is (see e.g. \citealt{bartos+2017})

\begin{align}
    t_{\rm align,\dot{m}}=\frac{P_{\rm orb}v_z}{\Delta v_z},
\end{align}
where $\Delta v_z$ is the change in the perpendicular velocity at an orbital node. From momentum conservation

\begin{align}
    \Delta v_z=v_z\frac{\Delta m}{m_{\star}}
\end{align}
Thus,

\begin{equation}
    t_{\rm align, \dot{m}}=\frac{P_{\rm orb} m_{\star}}{\Delta m},
\end{equation}
where $\Delta m$ is the mass accreted in a single crossing. For Bondi-Hoyle-Littleton accretion in the supersonic limit, 
\begin{align}
    \Delta m&=\dot{m} t_{\rm cross}\nonumber\\
    &\approx \frac{4 \pi \rho_g G^2 m_\star^2}{v_{\rm rel}^3} t_{\rm cross}\nonumber\\
    &=m_{\star} \frac{f_{\rm drag, z}}{v_z \ln \Lambda} t_{\rm cross}.
\end{align}
Thus,
\begin{align}
    t_{\rm align, \dot{m}}=\frac{v_z P_{\rm orb}}{f_{\rm drag,z} t_{\rm cross}} \ln \Lambda
    \label{eq:talignMdot}
\end{align}
This exceeds the alignment time from GDF by a factor of $\ln \Lambda$. Thus, we expect accretion to change the alignment timescale by a factor of order unity. 
For moderate inclination orbits ($i\lsim 40^\degree$) evolving under GDF the mass and the inclination are initially related via 
\begin{equation}
    \left(\frac{m_\star}{m_{\star,o}}\right)=\left(\frac{\sin(i)}{\sin(i_o)}\right)^{-1/(\ln \Lambda+1)}.
\end{equation}
Here the subscript `o' indicates the initial condition. This condition no longer applies as the orbit approaches alignment with the disc, and the relative velocity becomes subsonic. Mathematically, the star's mass diverges as it approaches alignment. However, the growth of the star may be stopped by feedback from stellar winds or radiation, particularly if the accretion rate becomes super-Eddington. For a star on a circular orbit accreting at the Bondi rate, the Eddington ratio is
\begin{align}
    \frac{\dot{m}_{\rm Bondi}}{\dot{m}_{\rm Edd}}=5.9\times 10^{-4} &\left(\frac{m_\star}{1 M_{\odot}}\right) \left(\frac{\rho_{\rm g, 1 pc}}{10^6 {\rm M_{\odot} pc^{-3}}}\right) \left(\frac{M}{4\times 10^6 M_{\odot}}\right)^{-3/2}\nonumber\\
    &\left(\frac{r}{{\rm pc}}\right)^{-3/2-\gamma} \sin^{-3}(i/2).
\end{align}
For a $1 M_{\odot}$ star this exceeds unity for $i\lesssim 10^{\circ}$ near $\sim 1$ pc. In the geometric regime,
\begin{equation}
t_{\rm align, m}=t_{\rm align, geo} Q_d
\end{equation}
For moderately inclined orbits, the mass and inclination are related by 
\begin{equation}
    \left(\frac{m}{m_o}\right)=\left(\frac{\sin(i)}{\sin(i_o)}\right)^{-1/(Q_d+1)}.
\end{equation}
Here the mass also diverges as the inclination approaches zero. However, the geometric drag force will go to zero and GDF will dominate in this limit.

The Eddington ratio in the geometric limit is 
\begin{align}
    \frac{\dot{m}_{\rm Geo}}{\dot{m}_{\rm Edd}}=1.9\times 10^{-5} &\left(\frac{R_\star}{1 R_{\odot}}\right)^{2} \left(\frac{m_\star}{1 M_{\odot}}\right)^{-1} \left(\frac{\rho_{\rm g, 1 pc}}{10^6 {\rm M_{\odot} pc^{-3}}}\right) \nonumber\\
    &\left(\frac{M}{4\times 10^6 M_{\odot}}\right)^{1/2} \left(\frac{r}{{\rm pc}}\right)^{-1/2-\gamma} \sin (i/2).
\end{align}
Thus, geometric accretion is sub-Eddington in the outer disc.

\section{Eccentricity evolution}
\label{sec:ecc}
The orbital eccentricity is always damped in the geometric regime. In the GDF regime, the eccentricity is always damped for prograde orbits. 

For retrograde orbits in the GDF regime, the eccentricity evolution depends on the orbital orientation and the slope of the gas density profile. For a given profile and orbit, it is possible to evaluate the direction of eccentricity evolution using the procedure described in Appendix~\ref{app:ecc}. For a density profile steeper than $r^{-2.8}$, the orbit is always circularized, as orbits predominantly evolve due to energy rather than angular momentum dissipation. For shallower density profiles, the orbital orientation determines the direction of eccentricity evolution. For example, for an $r^{-1.5}$ gas density profile, the eccentricity is damped as long as the argument of pericenter satisfies
\begin{align}
    \omega_c<\omega<\pi-\omega_c
\end{align}
or
\begin{align}
    \pi+\omega_c<\omega<2 \pi-\omega_c,
\end{align}
where $\omega_c\approx  (1-e^2)^{0.3}$. Geometrically, orbits that are rotated over their minor axes experience eccentricity damping. Orbits that are rotated over their major axis (i.e. those that have their major axis aligned with the disc plane), experience eccentricity excitation. Numerically, we find retrograde orbits evolve towards the latter configuration under the influence of GDF, and experience eccentricity growth. (See also \citealt{rozner&perets2022}).

However, retrograde orbits evolve towards a prograde orientation. Once they flip they will circularize. Thus, eventually, all orbits become prograde and circular as they align with the disc. Nevertheless, retrograde stars and compact remnants may be consumed by the central SMBH before they flip in inclination. In contrast to \citet{nasim+2022} we do not find orbits retrograde alignment between stellar orbits and an AGN disc.

Also, as pointed out in \citet{vokrouhlicky&karas1998}, the potential of the disc may qualitatively change the direction of eccentricity evolution. For example, with the disc potential eccentricity excitation is possible even in the geometric regime. This effect is outside of the scope of our study.

\section{Numerical validation and application}
\label{sec:num}
We perform few-body simulations with star-disc interactions to validate the analytic estimates for the alignment timescale in \S~\ref{sec:alignAnalytic}. 

In our simulations the disc gas density is 

\begin{align}
    \rho_g(r, z)=  \rho_o
    \begin{cases}
        \left(\frac{r}{r_o}\right)^{-\gamma} \exp\left(-z^2/(2 h^2)\right) & z < 10 h\\
        0                                                                  & z \geq 10 h
    \end{cases}
    \label{eq:discDens}
\end{align}
where $r$ is the cylindrical radius. In this section, $\rho_g(0.1 {\rm pc})=6\times 10^8 {\rm M_{\odot} pc^{-3}}$ and $h/r=0.01$. These parameters are similar to the model discs in \citet{BaruteauCuadraLin2011}. The power law index of the density profile, $-\gamma$, is between -1 and -3.
 The goal is to validate the analytic expressions in \S~\ref{sec:alignAnalytic}, which can later be used to model alignment with arbitrary discs (see \S~\ref{sec:res}).

In general, the gas disc's velocity will not be exactly Keplerian due to pressure effects. Although small, this correction strongly affects the relative velocity (and hence the drag force) of stars that are nearly aligned with the disc. We account for this correction, assuming the azimuthal velocity of the gas is

\begin{align}
    v_{\phi}(r)=\left(1-\left(\frac{h}{r}\right)^2 \right) v_{\rm c}(r).
\end{align}
(see \citealt{Armitage2010}; we neglect a factor of order unity in front of the second term in parentheses.) We also assume the gas velocity is precisely azimuthal, neglecting the vertical and radial components.

The sound speed in the disc is
\begin{equation}
c_s(r) = \left(\frac{h}{r}\right) \sqrt{\frac{G M}{r}}
\end{equation}
We model the interaction of individual stars with this disc using \texttt{REBOUND} N-body framework \citep{rein.liu2012}. 
We include GDF (equation~\ref{eq:fgdf}) and geometric drag force (equation~\ref{eq:fgeo}) using \texttt{REBOUNDX} \citep{tamayo+2019}\footnote{Our implementation is publicly available as the ``gas\_dynamical\_friction'' effect in \texttt{REBOUNDX} (see \url{https://github.com/dtamayo/reboundx}).}, accounting for the subsonic regime which becomes important once stars are embedded in the disc (see equation~\ref{eq:Ib}). Accretion is neglected in these simulations. Later, we show numerically that accretion changes the alignment timescale by a factor $\lsim$2 (see Figure~\ref{fig:accretion example} and the surrounding discussion).

We perform simulations with $10 M_{\odot}$ stars on randomized orbits, with the built-in Burlisch--Stohr integrator in \texttt{REBOUND}. In all cases, the stars orbit a $4\times 10^6 M_{\odot}$ SMBH and are distributed isotropically with a thermal eccentricity distribution and an $a^{-1.5}$ semimajor axis distribution between 0.01 and 1 pc. The potential is purely Keplerian: The gravitational effects of the extended stellar mass and of the AGN disc are neglected. We consider particles to be aligned within our simulations, once their inclination is less than $h/r$ and their eccentricity is less than $2 h/r$. We exclude orbits where the analytic alignment time is less than the initial orbital period from our analysis.

\begin{figure}
    \includegraphics[width=\columnwidth]{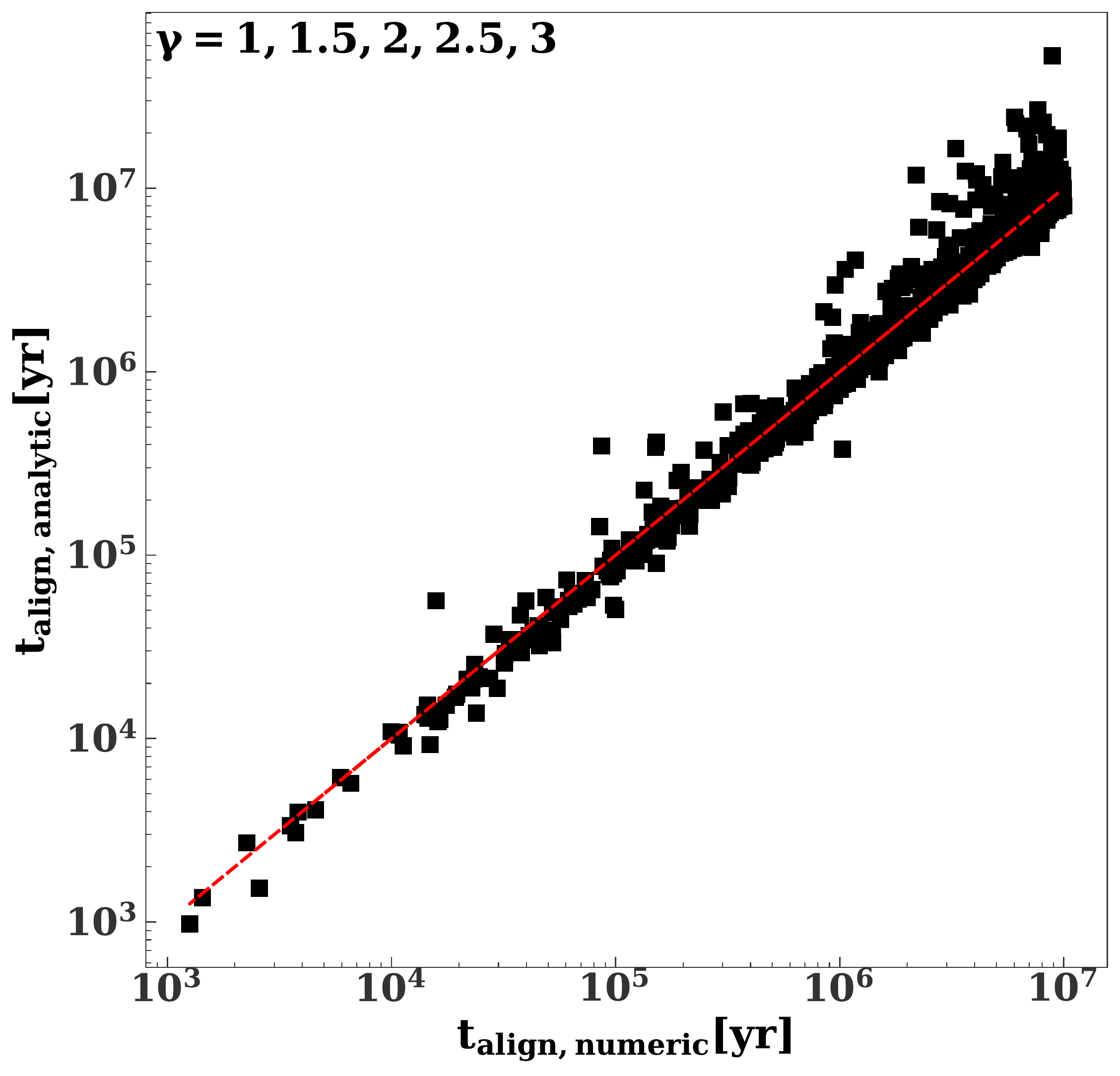}
    \includegraphics[width=\columnwidth]{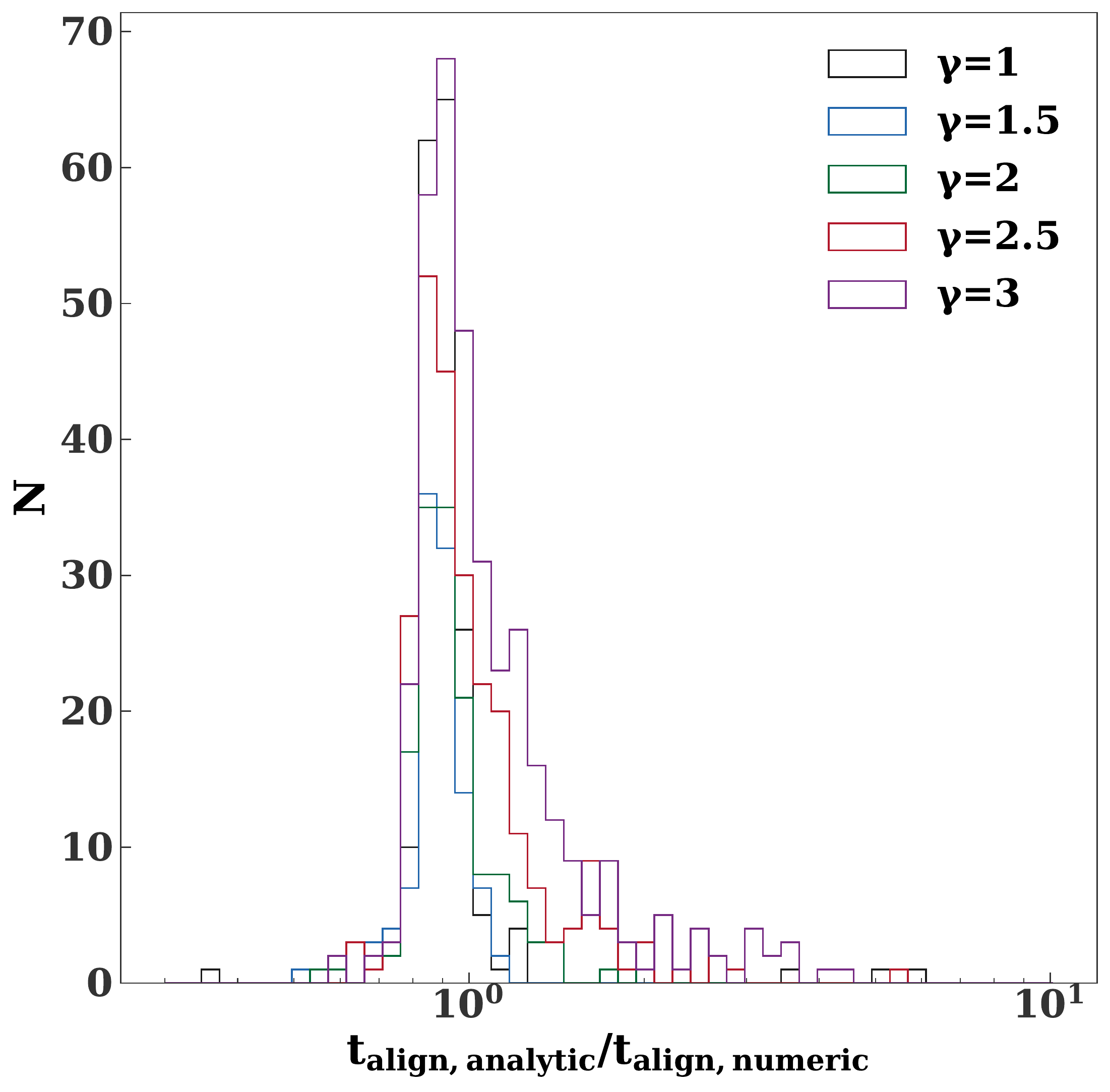}
    \caption{Top panel: The analytic alignment time (eq~\ref{eq:talignGen}) versus the alignment time from the few-body simulations. We stack the results from simulations with different density slopes ($\gamma$; see equation~\ref{eq:discDens}). The bottom panel shows a histogram of the ratio of the analytic alignment time to the alignment time from few-body simulations for each $\gamma$.}
    \label{fig:align1b}
\end{figure}

The top panel of Figure~\ref{fig:align1b} shows a comparison of the analytic and numerical alignment times for stars that align with the disc in simulations with different density slopes. The analytic alignment times fall within a factor of a few of the numerical alignment times. Furthermore, the final semimajor axis of aligned stars in the simulations fall within a factor of 3 of the analytic estimates in Appendix~\ref{app:coupleSol}, except in 1\% of cases, where the error in the final semimajor axis can be up to several orders of magnitude. The large errors occur for very flat ($\gamma=1$) or very steep ($\gamma=2.5$ or $\gamma=3$). The discrepant orbits are all highly retrograde ($i \gsim 170^\circ$) and highly eccentric ($e \gsim 0.97$). 

The above experiments only constrain $t_{\rm align}$ (eq~\ref{eq:talign1}) and $t_{\rm E,geo}$ (eq~\ref{eq:tegeo}). In order to constrain $t_{\rm align, geo}$ and $t_{\rm E,GDF}$, and to validate the timescales in regions of the parameter space, where the alignment timescale is longer, we perform additional numerical experiments, as described in Appendix~\ref{app:numValid2}. To summarize we construct a grid of orbits in eccentricity-inclination space for two different density profiles ($r^{-3}$ and $r^{-1.5}$) and two different arguments of pericenter (0 and 90$^\circ$). Then we simulate each orbit until it aligns with the disc. Generally, the alignment time in these simulations falls within a factor of a few of our analytic prescription. Except for isolated points, the worst discrepancy occurs for retrograde orbits with $\omega \approx 90^\circ$, where our analytic alignment time differs by up to a factor of $\sim$6 from the numerical results. However, this inaccuracy will have a small effect on the fraction of orbits that align with the disc, considering it occurs in a small region of parameter space.

Also, we cannot accurately predict the final semimajor axis of retrograde orbits in the GDF regime. Equation~\ref{eq:jfGenCorr} predicts final semimajor axis monotonically decreases with inclination, and that the final inclination goes to 0 as the inclination goes to $180^{\circ}$. Numerically, this is not always the case  (see Figure~\ref{fig:numValidSma1}). This could be relevant for black holes aligning with an AGN disc, where a non-negligible fraction of retrograde orbits can align via GDF.

\subsection{Consequences for galactic nuclei}
\label{sec:res}
The fraction of stars that align with the AGN disc depends on its density profile.  The density of a marginally Toomre-stable disc around a $4\times 10^6 M_{\odot}$ SMBH is $\rho_g\approx10^9 \left(\frac{r}{{\rm 0.1 pc}}\right)^{-3} M_{\odot}$ pc$^{-3}$ on large scales \citep{thompson+2005,Tagawa2020}. 
In contrast, the gas density in \citet{BaruteauCuadraLin2011} is $\rho_g\approx 10^8-10^9 \left(\frac{r}{{\rm 0.1 pc}}\right)^{-1.5} M_{\odot}$ pc$^{-3}$. Motivated by these models we vary the gas density at $0.1$ pc between $10^8 M_{\odot}$ pc$^{-3}$ and $10^9 M_{\odot}$ pc$^{-3}$, and vary the power law index of the gas density profile from $-1.5$ to $-3$. For simplicity we fix the outer disc radius to 0.1 pc (at larger radii some of these models become Toomre unstable).

We generate mock stellar populations with a realistic mass function and determine the fraction that aligns with these different AGN disc models using equation~\eqref{eq:talignGen}. First, we model $10^5$ stars with stellar masses randomly drawn from a Kroupa mass function distribution between 0.1 and 1 $M_{\odot}$. Truncating the mass function at $1 M_{\odot}$ is a reasonable approximation for the present-day mass function, whether star formation occurred in the distant past or continuously. The stars are isotropically distributed with a thermal eccentricity distribution, and a power law semimajor axis distribution between $10^{-4}$ and 0.1 pc with index, $-\alpha$, between $-1$ and $-2$. The stars have ZAMS radii (interpolated from the MIST stellar evolution tracks; \citealt{dotter+2016,choi+2016,paxton+2011,paxton+2013,paxton+2015,paxton+2018}).

\begin{align}
    &t_{\rm align,all}<t_{\rm agn} 
\end{align}
where $t_{\rm align, all}$ is the (analytic) alignment timescale of the star (cf equation~\ref{eq:talignGen}) and $t_{\rm agn}$ is the AGN lifetime ($\sim 10^7-10^9$ yr; see \citealt{martini&weinberg2001, marconi+2004}). Essentially, there are two populations that align with the disc: low inclination, low eccentricity stars that align due to gas dynamical friction, and high eccentricity stars that align with the disc due to geometric gas drag. For the steepest gas density profiles, all of the stars on small radial scales align with the disc. For example, for $\rho_\star\propto r^{-2}$, $\rho_g \propto r^{-3}$, all stars within $\sim$0.005  (0.01) pc align with the disc after $10^7$ ($10^8$) yr. 

\begin{figure*}
    \includegraphics[width=\columnwidth]{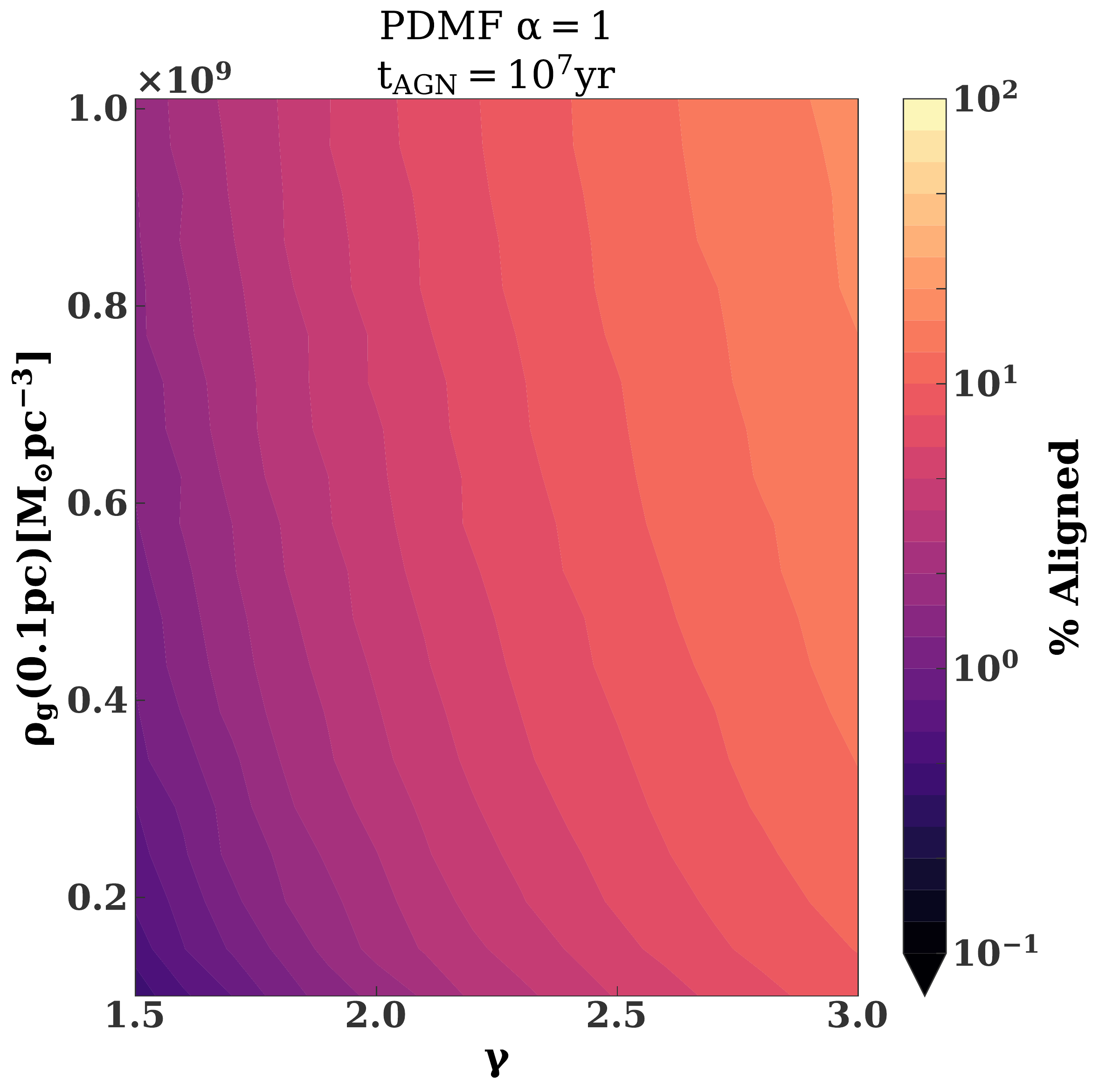}
    \includegraphics[width=\columnwidth]{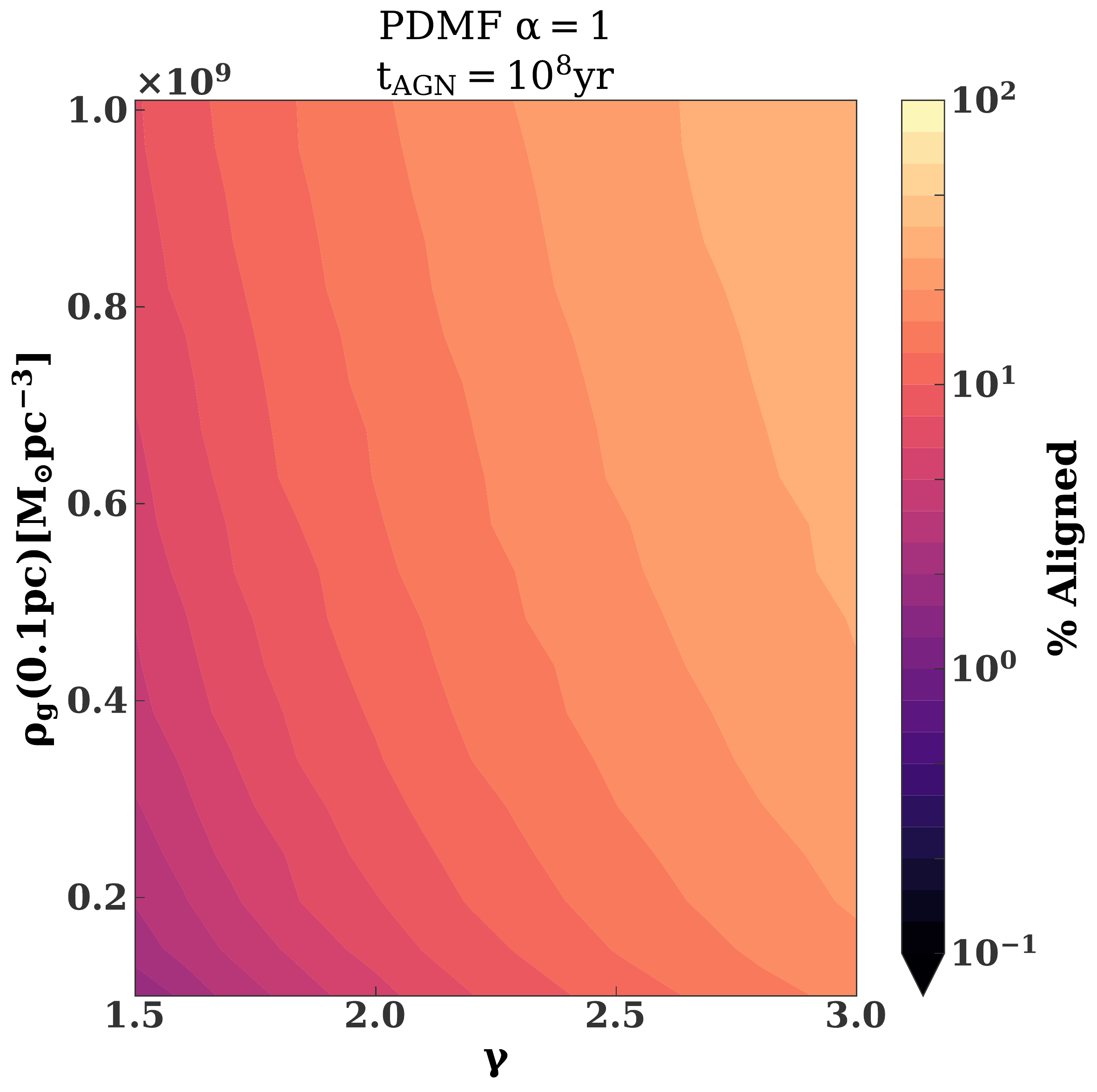}
    \includegraphics[width=\columnwidth]{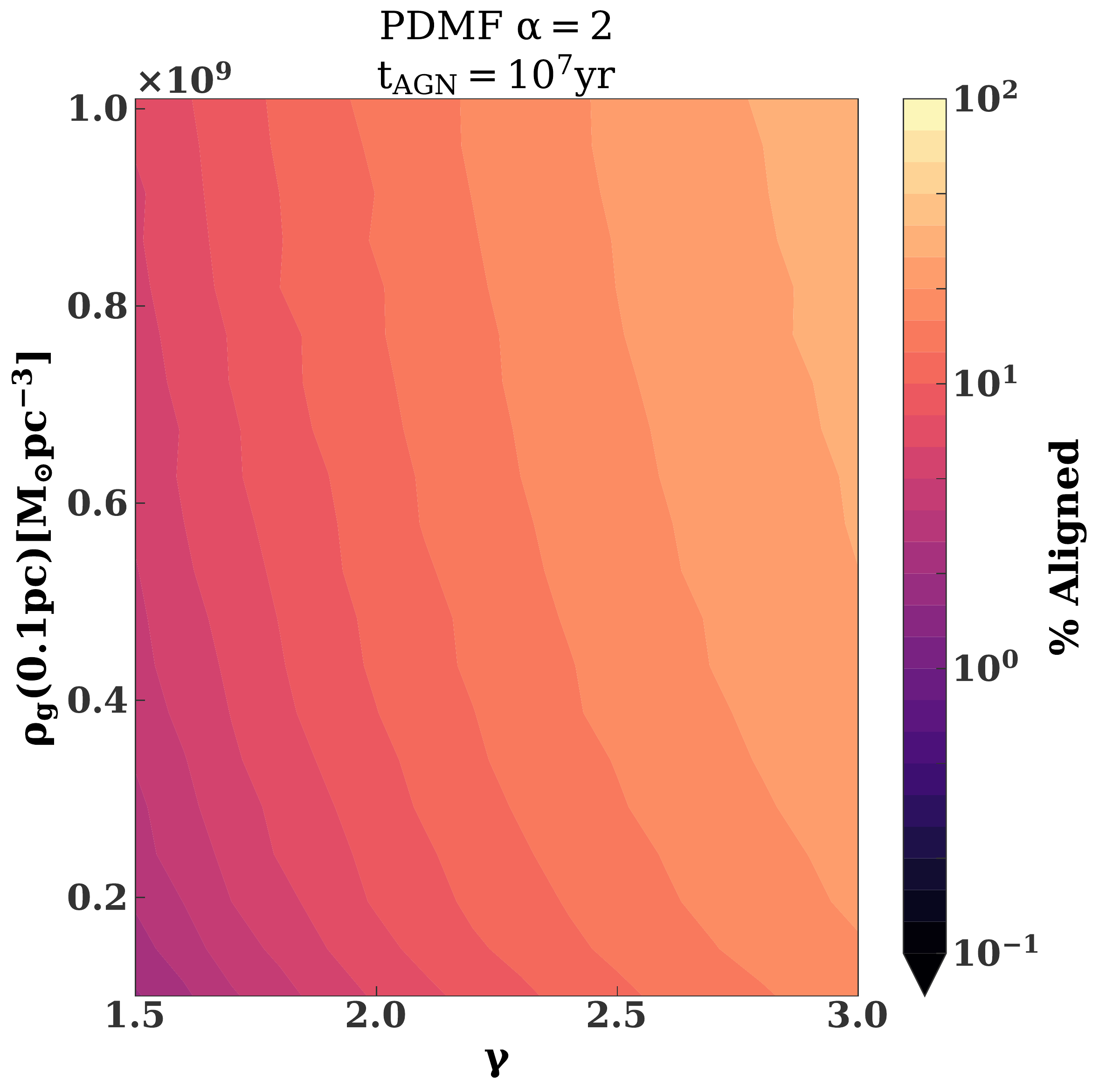}
    \includegraphics[width=\columnwidth]{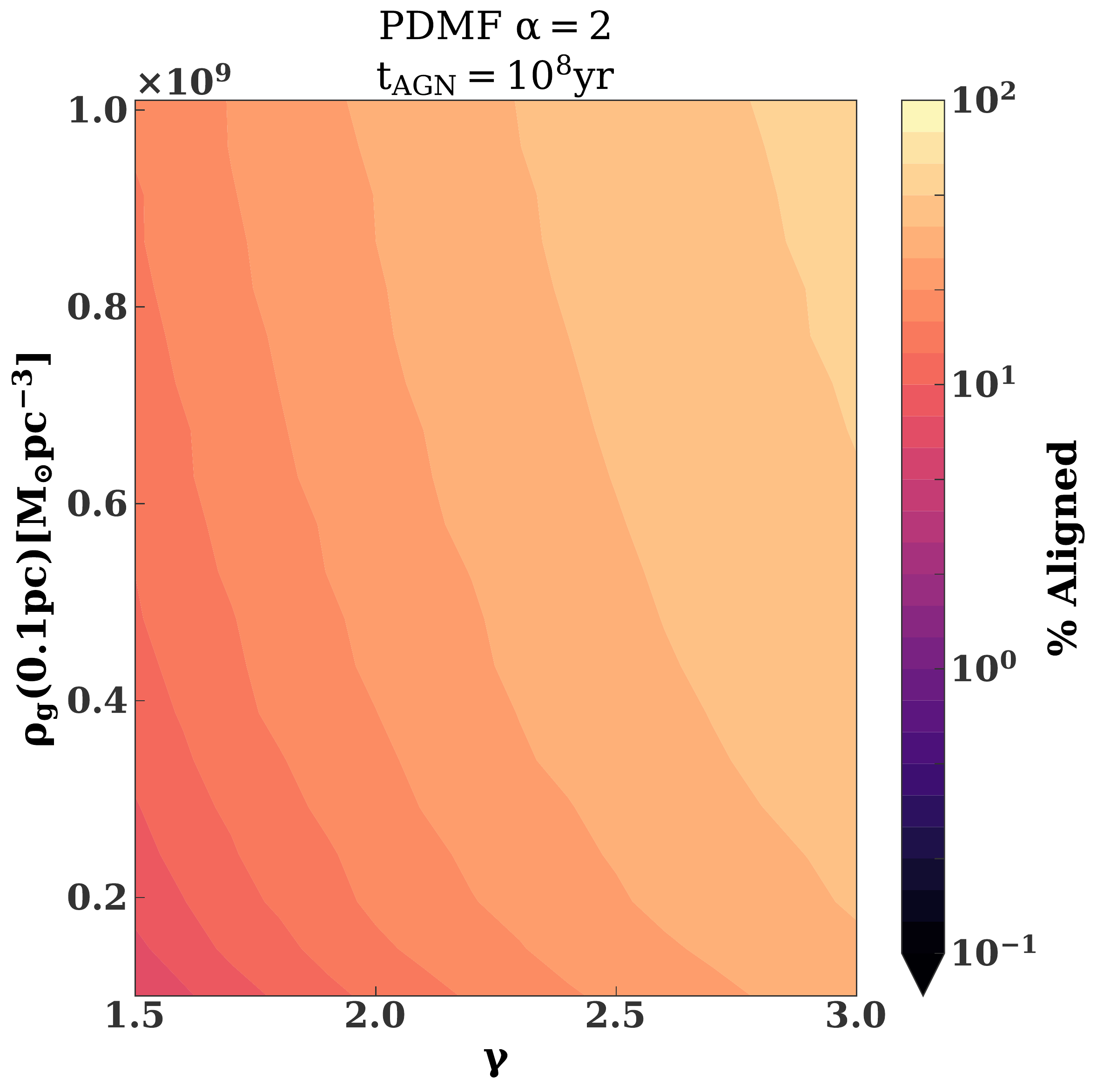}
    \caption{\label{fig:frac} Percent of main sequence stars that align with a AGN disc as a function power-law index of the gas density profile ($-\gamma$), and the density at 0.1 pc. The left (right) column corresponds to an AGN disc lifetime of $10^7$ ($10^8$ yr). The top (bottom) row corresponds to a stellar density profile of $r^{-1}$ ($r^{-2}$). The stellar density profile  The star have masses drawn from a Kroupa mass function extending from 0.1 to 1 $M_{\odot}$ and ZAMS stellar radii.}
\end{figure*}

\begin{figure*}
    \includegraphics[width=\columnwidth]{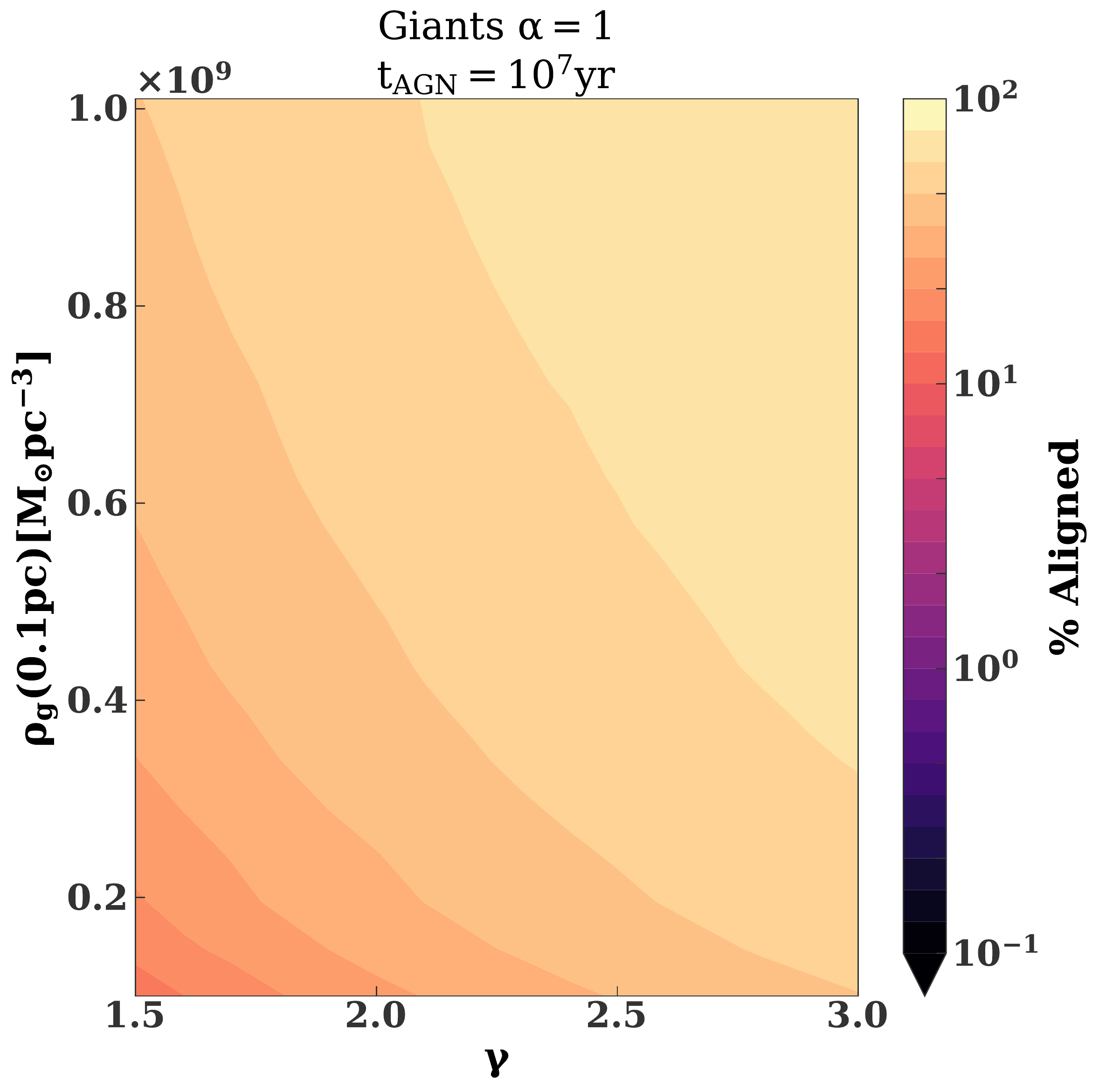}
    \includegraphics[width=\columnwidth]{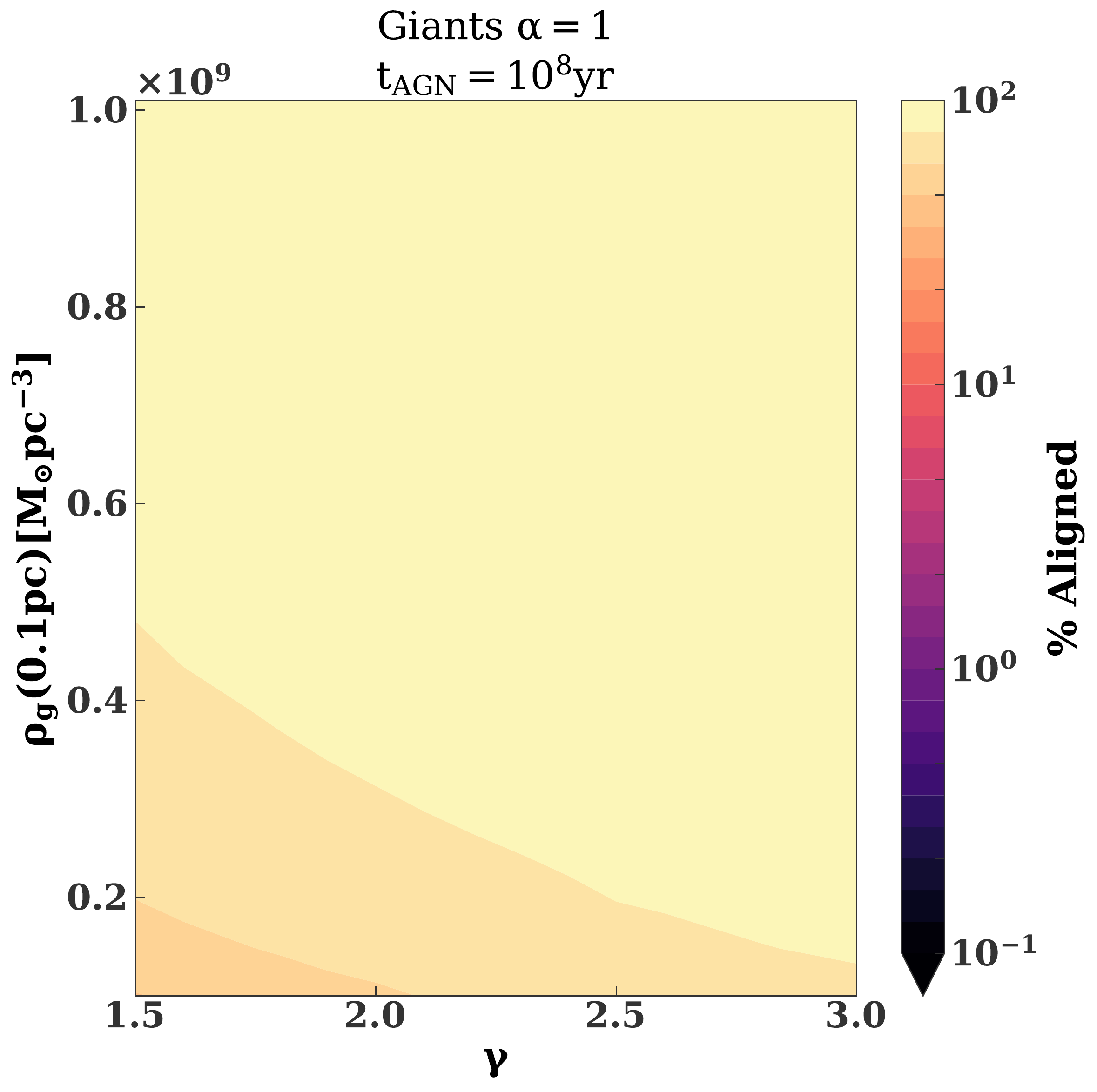}
    \includegraphics[width=\columnwidth]{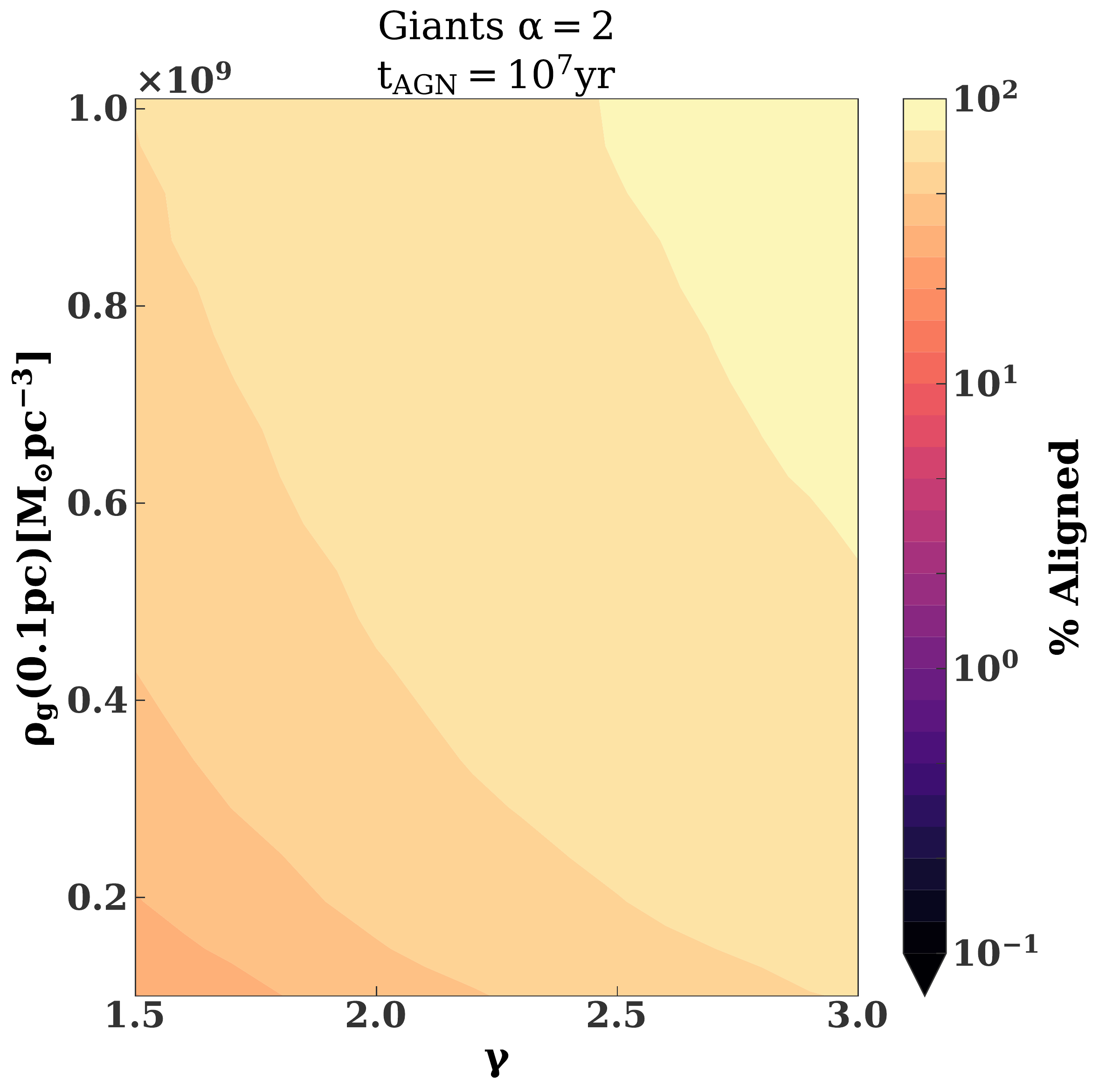}
    \includegraphics[width=\columnwidth]{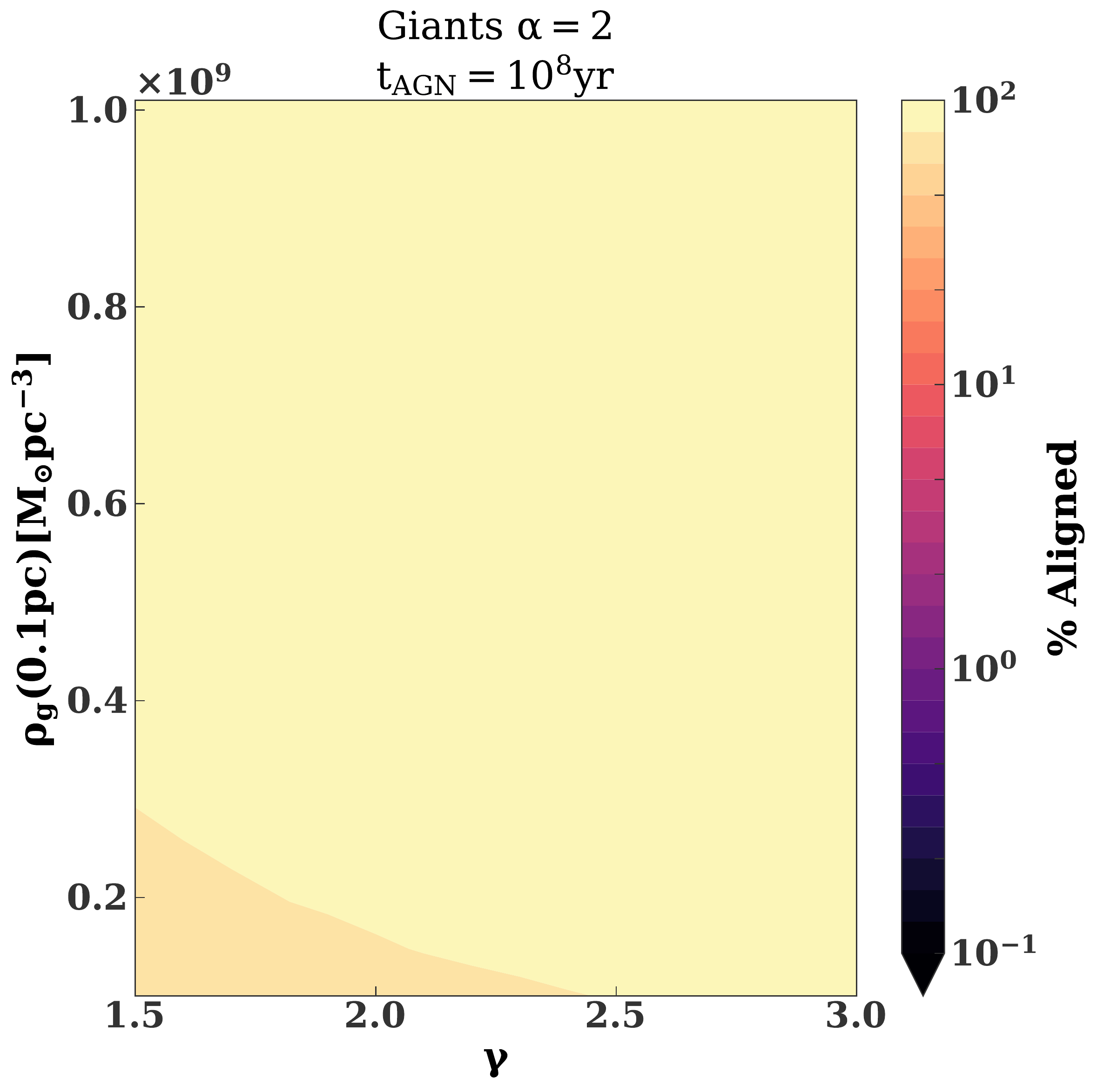}
    \caption{\label{fig:fracGiant} Same as Figure~\ref{fig:frac} except for giant stars. All stars have mass $1 M_{\odot}$ and radius $10 R_{\odot}$. The giant lifetime is $\approx 250$ Myr, which is greater than than the disc lifetime.}
\end{figure*}

\begin{figure*}
    \includegraphics[width=\columnwidth]{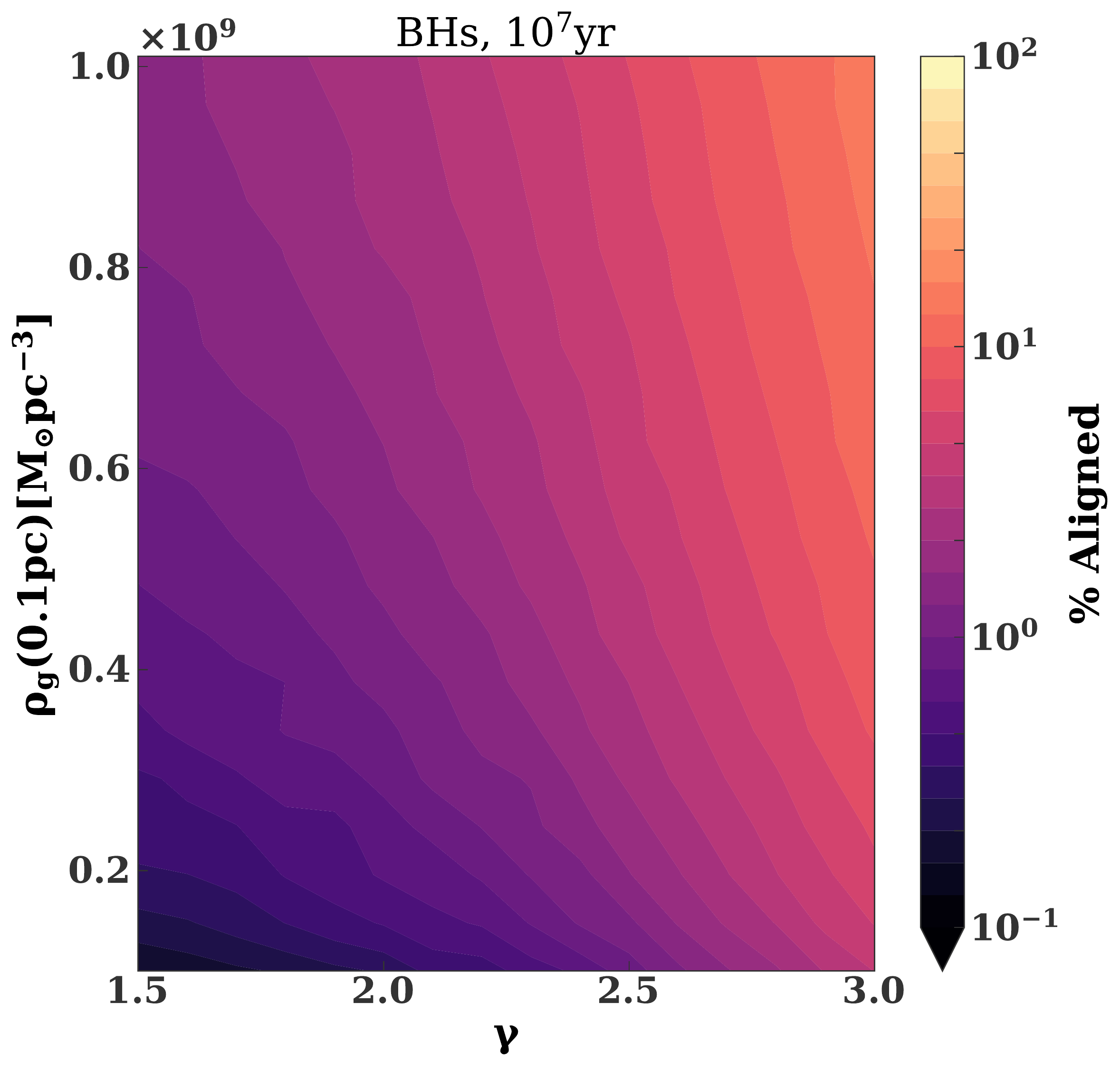}
    \includegraphics[width=\columnwidth]{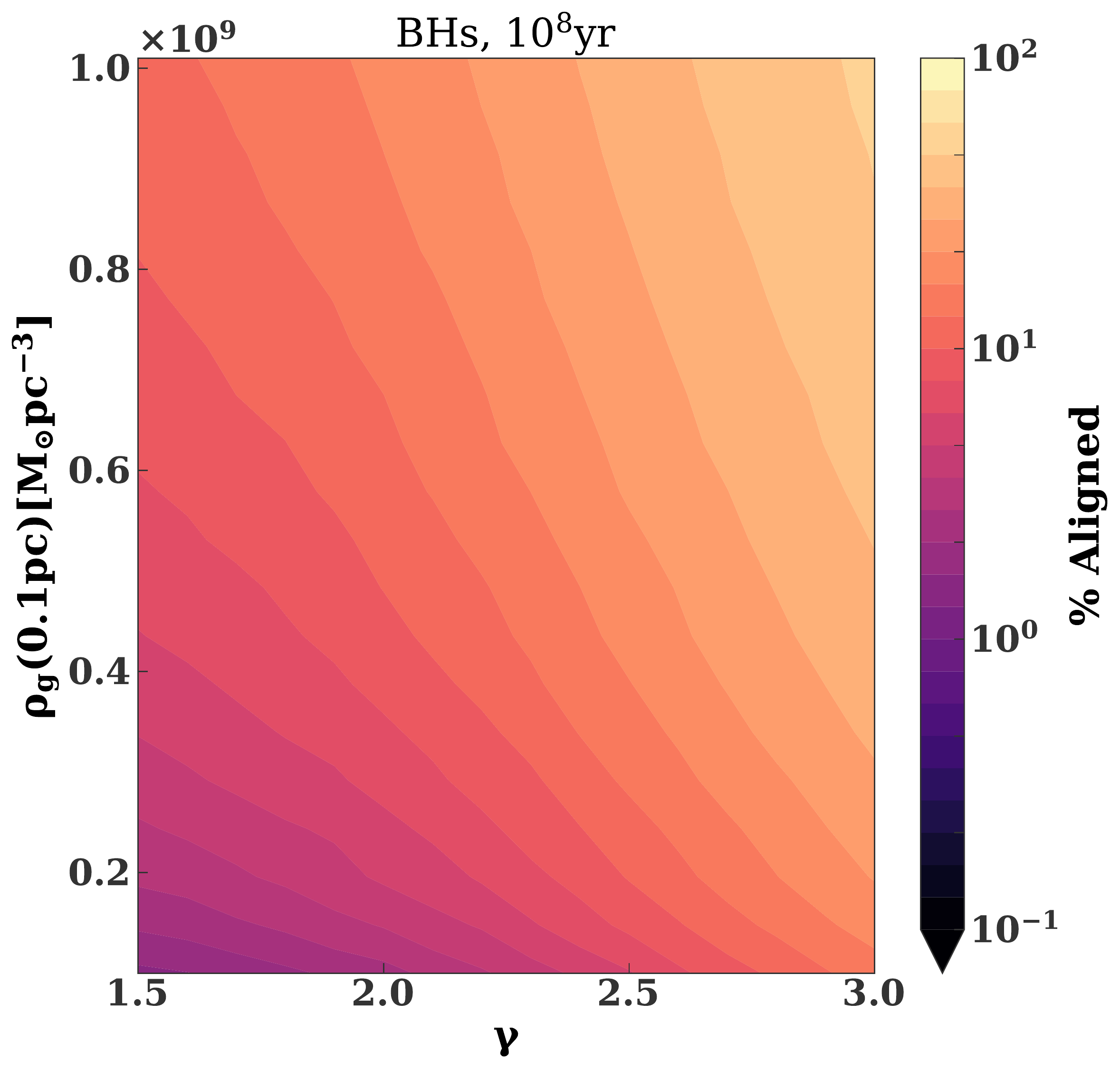}
    \caption{\label{fig:fracBH} Same as Figure~\ref{fig:frac} except for stellar mass black holes with $10 M_{\odot}$. Unlike stars, black holes do not experience geometric drag (only GDF). They have an $r^{-2}$ density profile.}
\end{figure*}

Figure~\ref{fig:frac} shows the fraction of main sequence stars that align with the disc for different density profiles and AGN disc lifetimes. Up to 50\% of stars in the central 0.1 pc align.
The fraction of stars that align with the disc is a weak function of the stellar mass for main sequence stars, because the alignment timescale is more sensitive to the initial orbital orientation than the stellar mass. 
We also estimate the fraction of giant stars and stellar mass black holes that align with the disc (see Figures~\ref{fig:fracGiant} and~\ref{fig:fracBH}). For simplicity, we consider only a handful of discrete population bins. All black holes are 10 $M_{\odot}$ and the density profile is $r^{-2}$, motivated by studies of relaxation in the Galactic Center \citep{freitag+2006,hopman&alexander2006,alexander&hopman2009,preto&amaro-seoane2010, aharon&perets2015,vasiliev2017}. 

Our analytic alignment timescales are only accurate to within a factor of a few. We now quantify the uncertainty from this in the fraction of aligned PDMF stars and black holes using Figures~\ref{fig:numValid2a} and~\ref{fig:numValid2b}. These figures show the ratio between the analytic and few-body alignment times as a function of eccentricity and inclination for $\gamma=1.5$ and $\gamma=3$ discs (for geometric drag and GDF and for two orientations).  For each orbit generated for these profiles, we choose the closest configuration, divide by the corresponding ratio, and recompute the fraction of aligned stars. The aligned fraction changes by at most $\sim 20\%$.

\begin{figure*}
    \includegraphics[width=0.9\columnwidth]{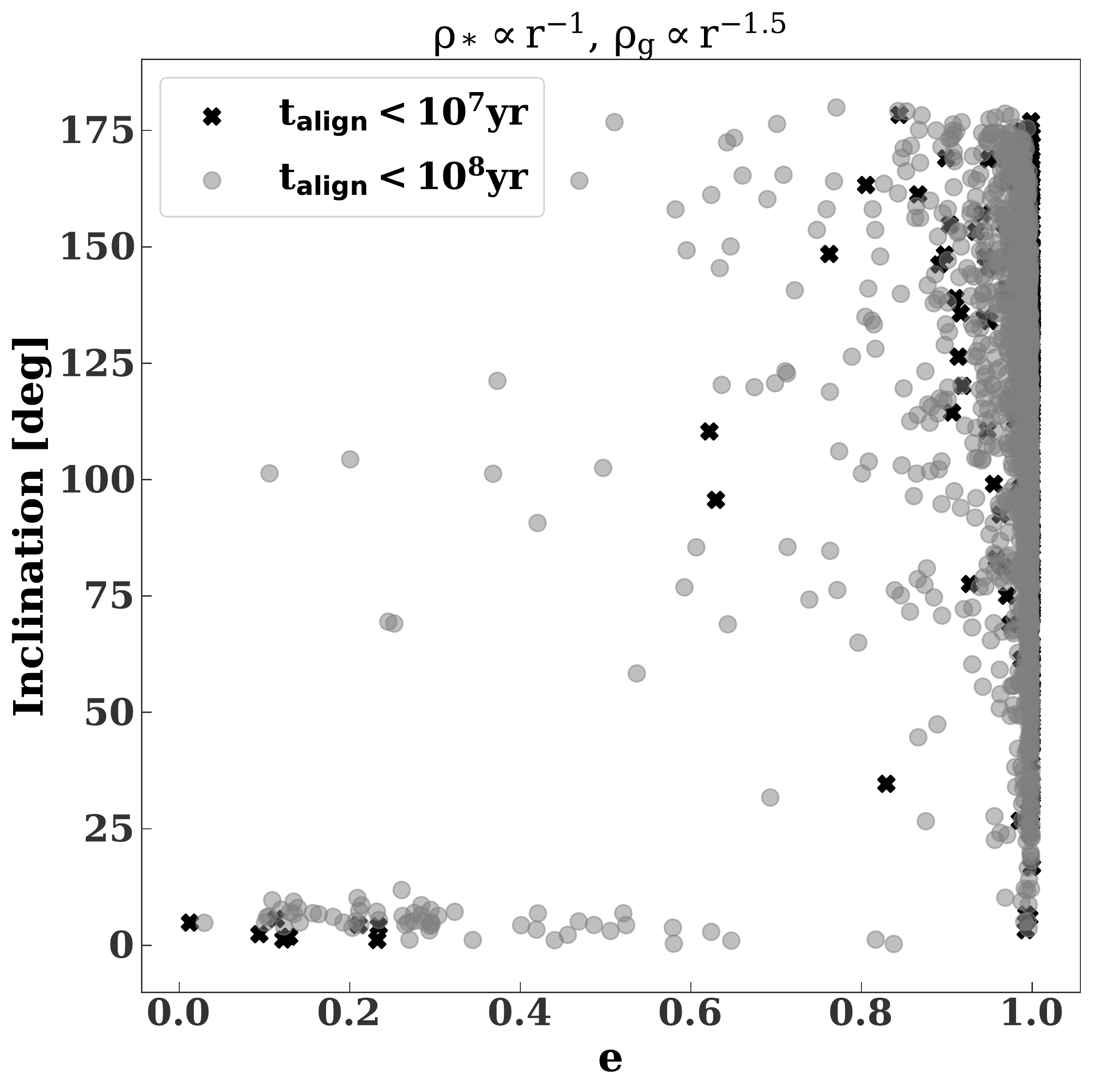}
    \includegraphics[width=0.9\columnwidth]{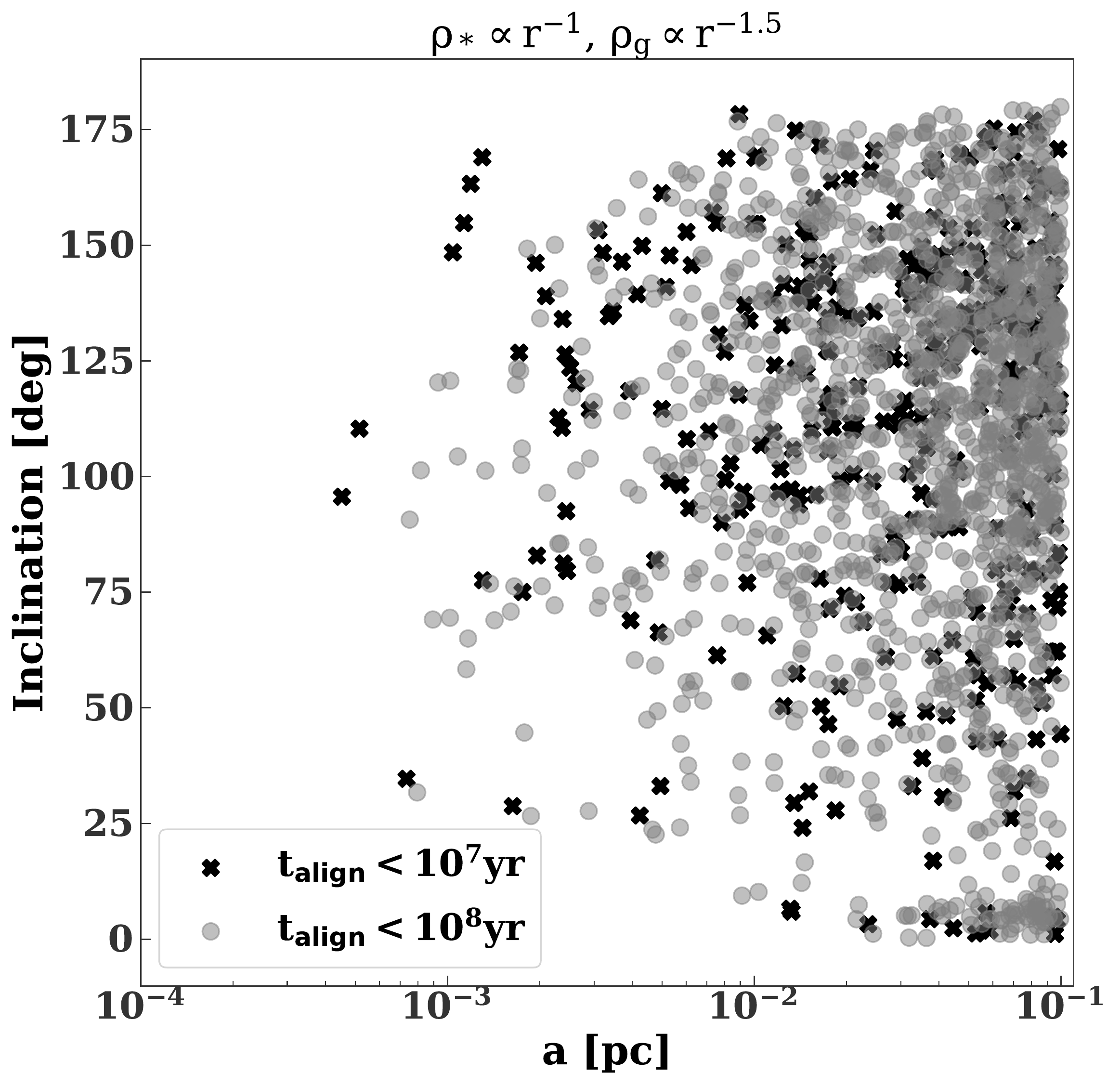}
    \includegraphics[width=0.9\columnwidth]{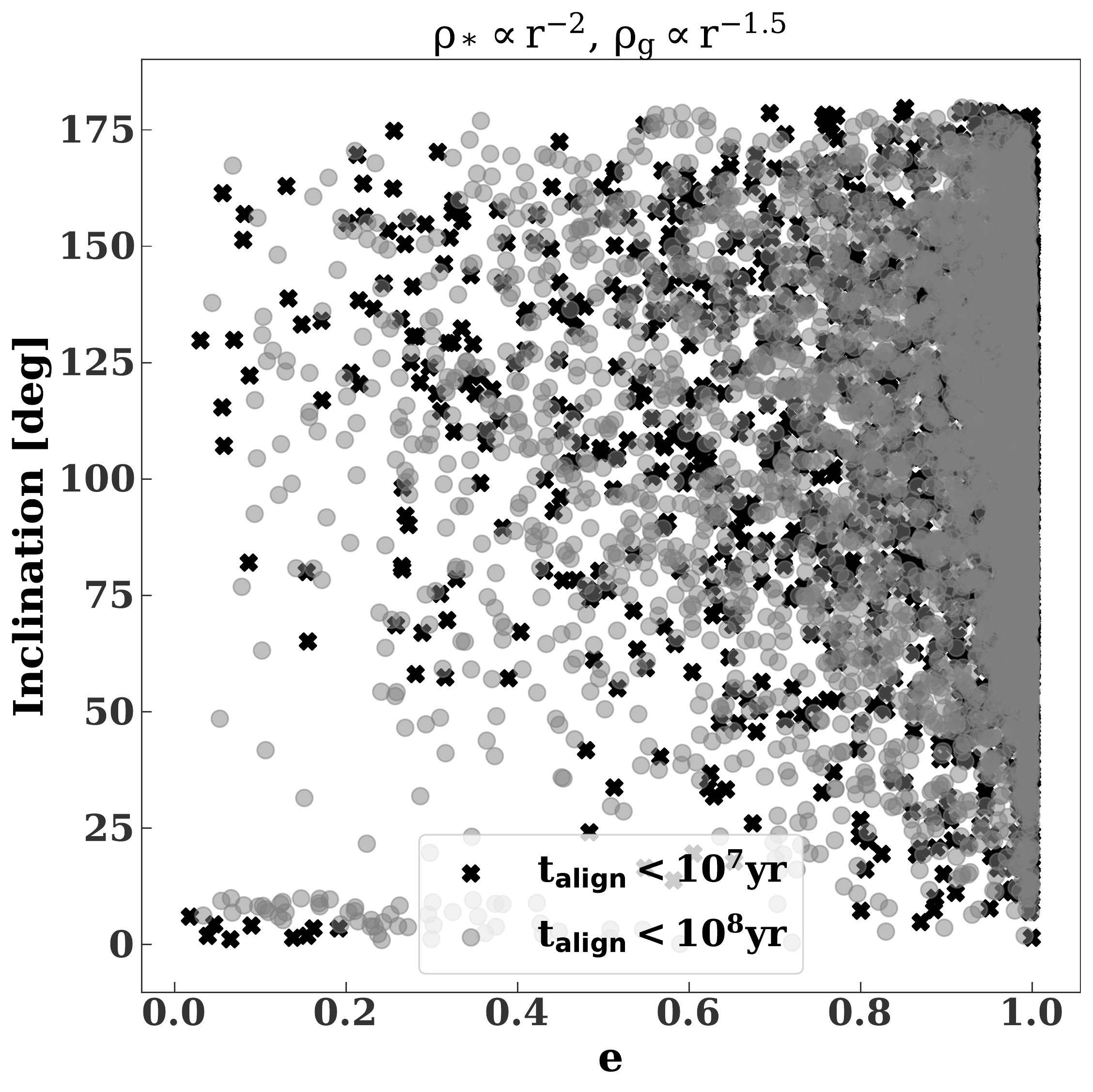}
    \includegraphics[width=0.9\columnwidth]{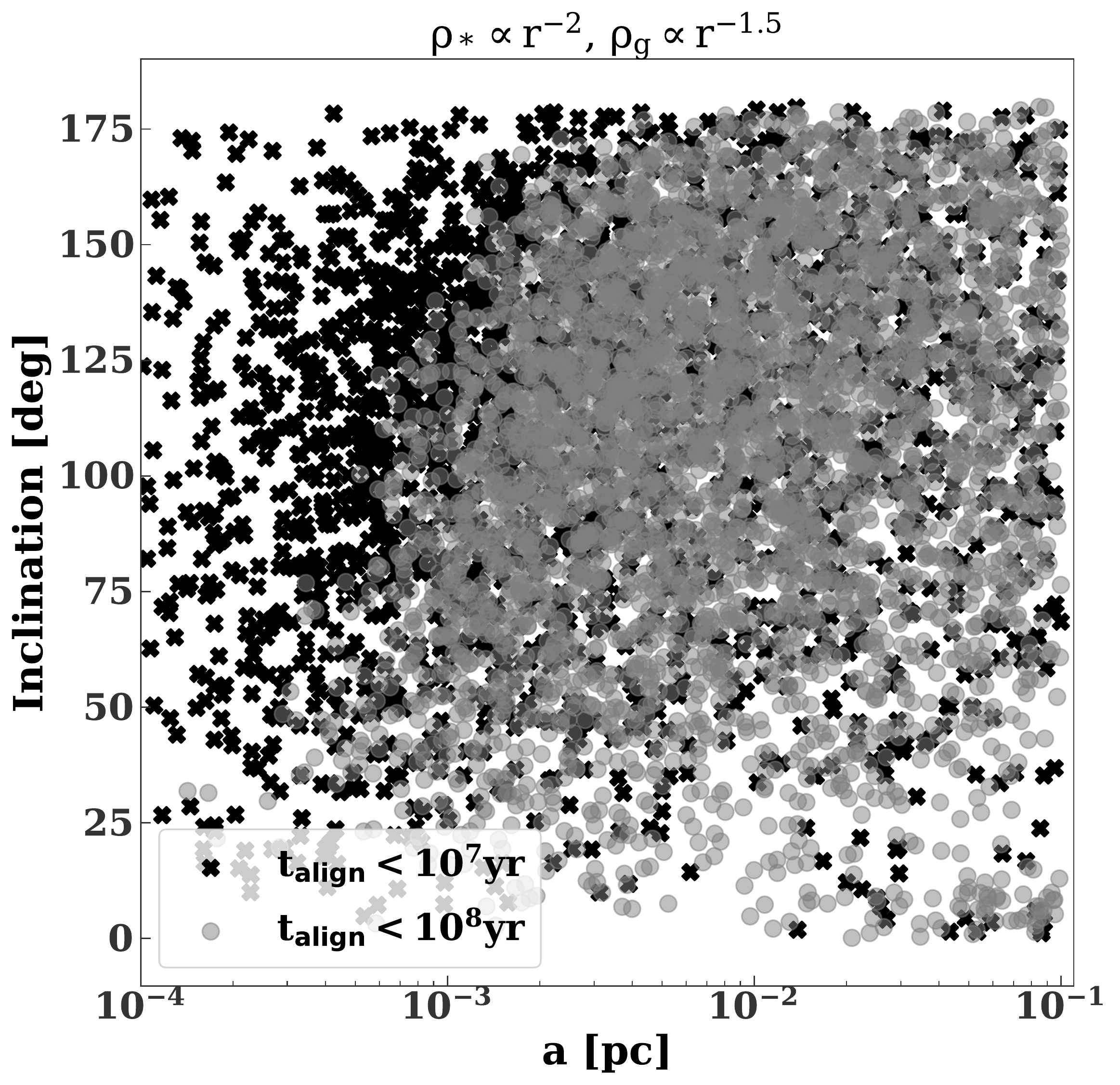}
    \includegraphics[width=0.9\columnwidth]{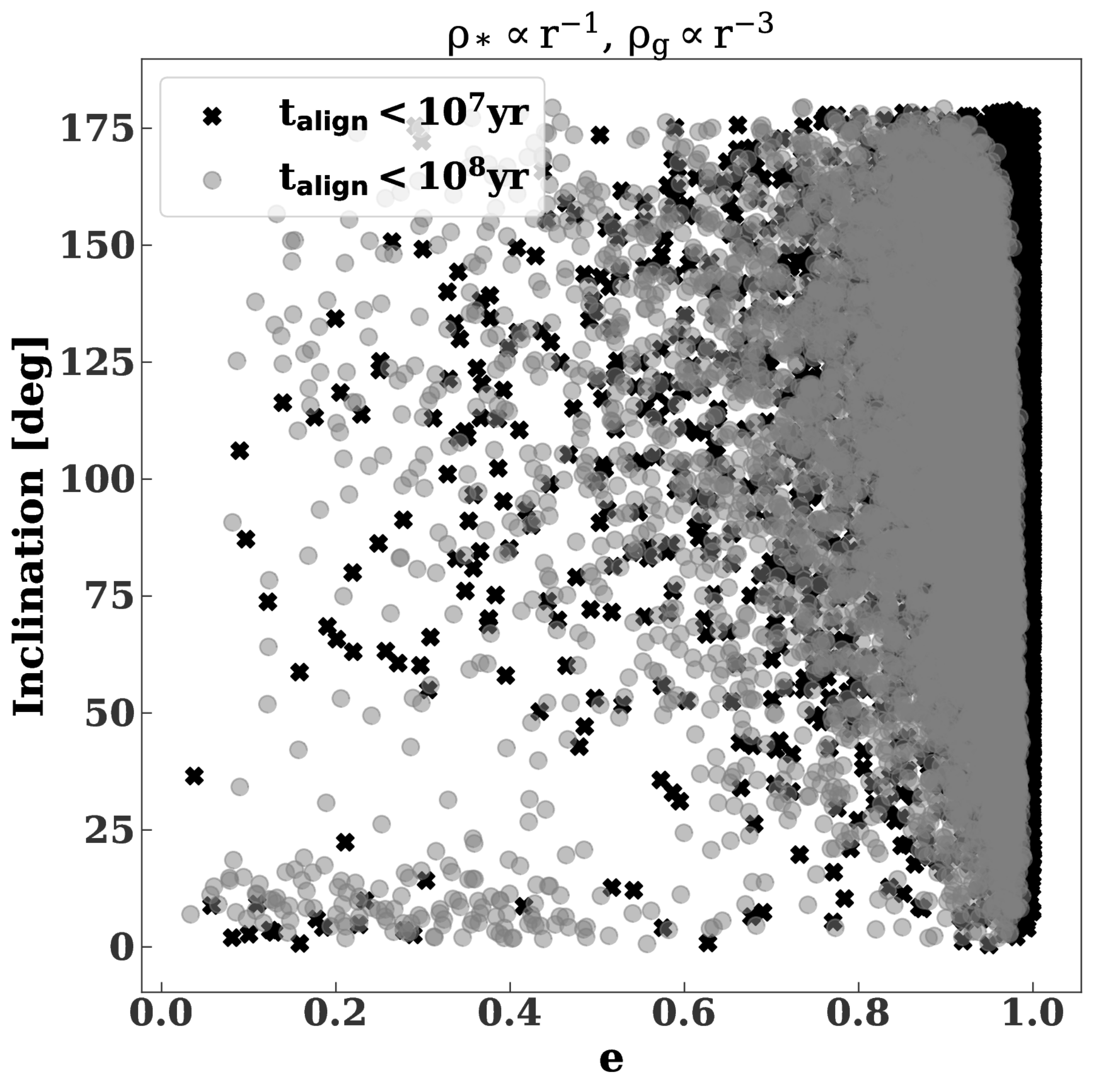}
    \includegraphics[width=0.9\columnwidth]{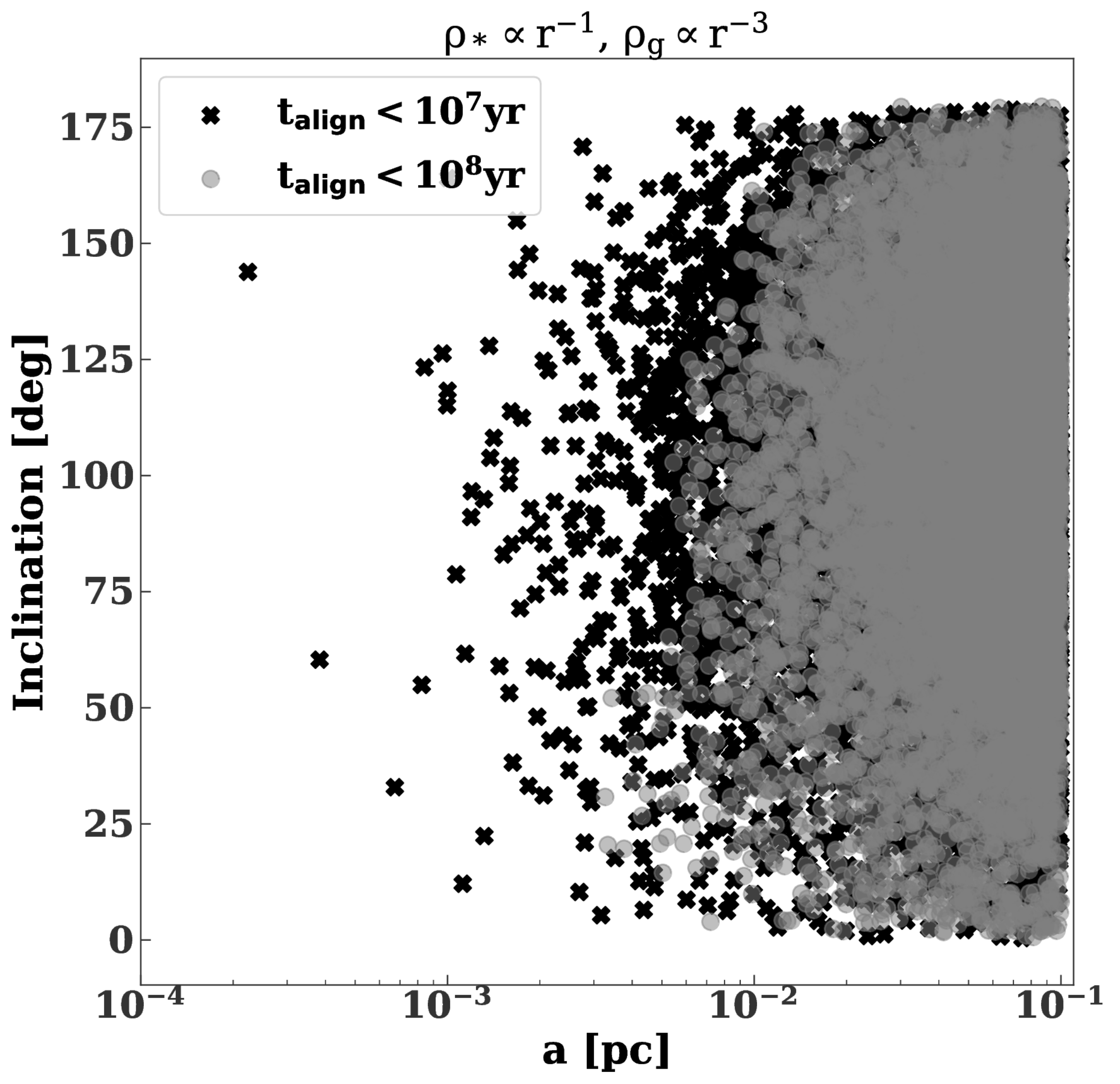}
    \caption{Initial orbital elements of stars with alignment times  $<10^7$ yr (black crosses) and between $10^7$ and $10^8$ yr (gray circles) for different gas density and stellar density profiles. The gas density at 0.1 pc is fixed to $10^8 M_{\odot}$ pc$^{-3}$, and the total number of stars (including those that do not align is $10^5$).}
    \label{fig:align2}
\end{figure*}

\begin{figure*}
    \includegraphics[width=0.9\columnwidth]{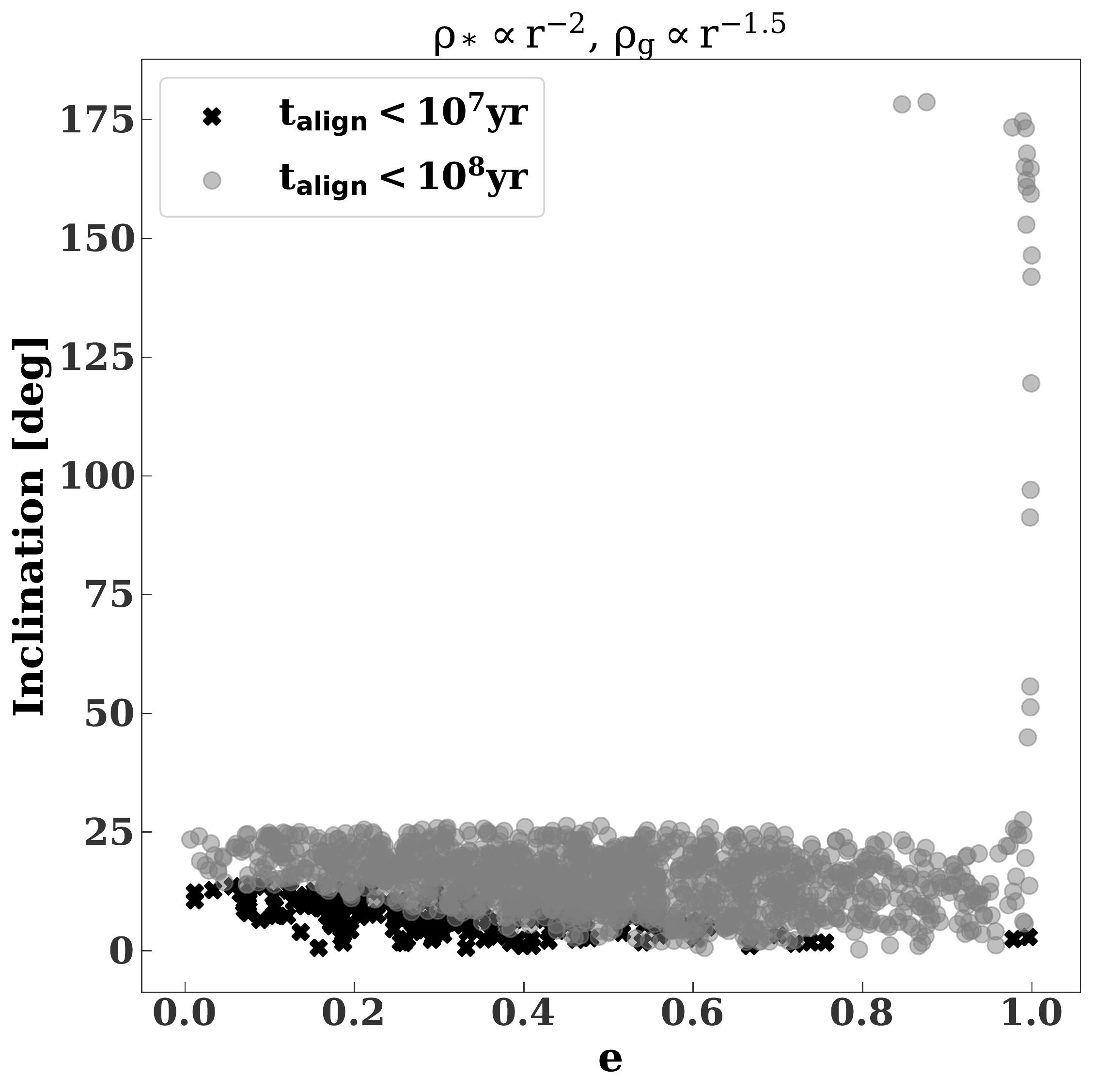}
    \includegraphics[width=0.9\columnwidth]{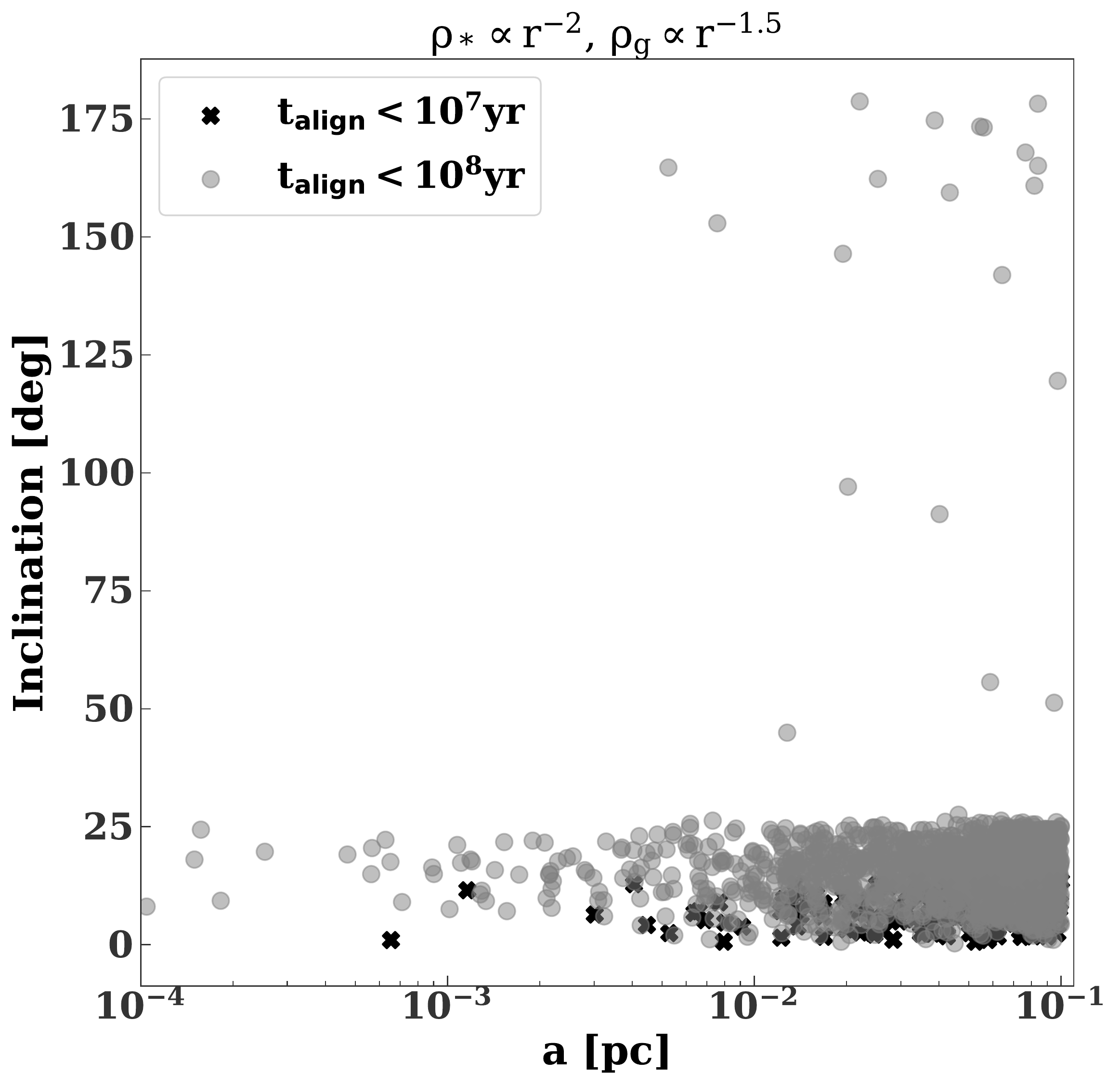}
    \caption{Same as Figure~\ref{fig:align2}, except for $10 M_{\odot}$ black holes.}
    \label{fig:alignBH}
\end{figure*}

Figure~\ref{fig:align2} shows the initial orbital parameters of stars that align with the disc a handful of model discs.
We also show the initial eccentricities and inclinations of black holes that align with an example disc in Figure~\ref{fig:alignBH}. In this case, only GDF is active, and virtually all the black holes that align start at low inclinations.

\begin{figure}
    \includegraphics[width=\columnwidth]{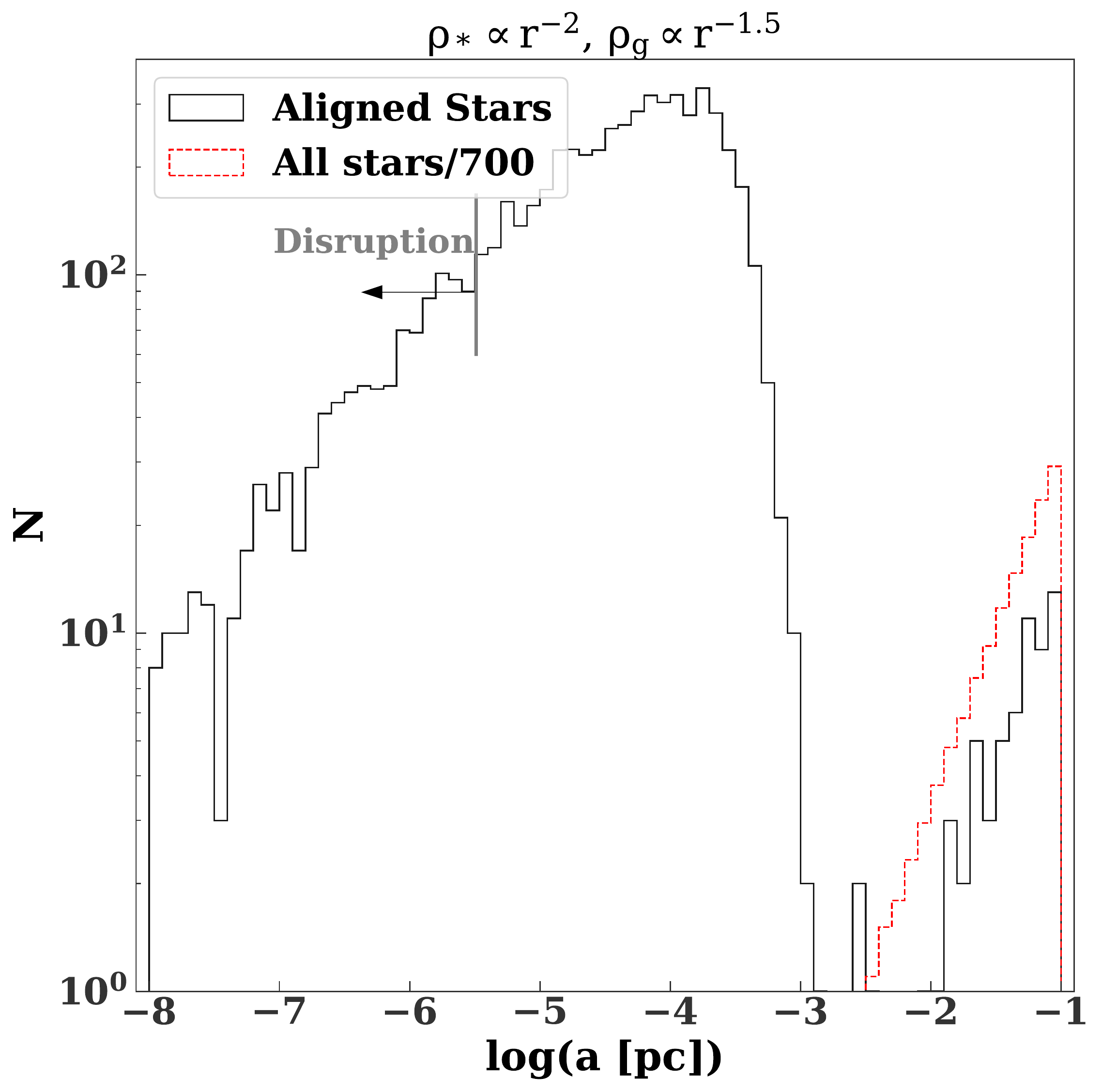}
    \includegraphics[width=\columnwidth]{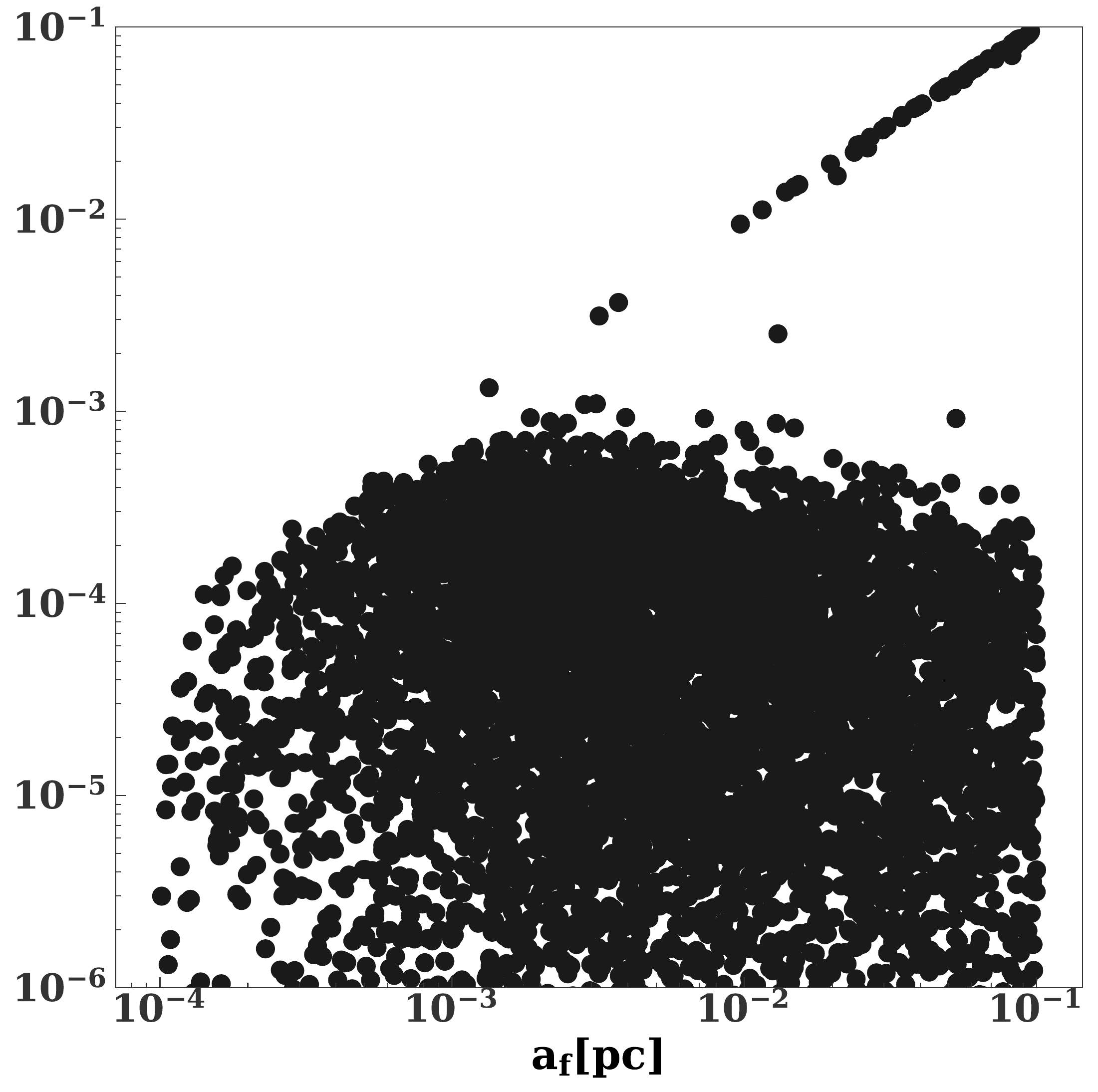}
    \caption{Top panel: The solid, black line shows an example of the semimajor axis distribution of main sequence stars following alignment with an AGN disc around the Galactic Center SMBH. The dashed, red line shows overall semimajor axis distribution of the stellar population. The small peak in the black distribution at large semimajor axes corresponds to stars that align via GDF starting from small inclinations. These stars remain close to their initial semimajor axis. The peak at small semimajor axis corresponds to stars that that align via geometric drag starting from large inclinations. These stars inspiral significantly. To the left of the vertical, gray line stars will be tidally disrupted before they align. 
    Bottom panel: Final versus initial semimajor axis for aligning stars after $10^8$ yr. Note the region inside of $\sim 10^{-6}$ pc is not included.}
    \label{fig:smaF}
\end{figure}

As discussed in \S~\ref{sec:ecc} we expect aligned orbits to become circular and prograde. From Appendix~\ref{app:coupleSol}, for low eccentricity, low inclination orbits the change in the orbit's semimajor axis will be modest prior to alignment with the disc. For orbits that are initially at high inclinations the change in semimajor axis will be significant (see equation~\ref{eq:jfGen}). Figure~\ref{fig:smaF} shows the final semimajor axis distribution of aligned stars from equations~\eqref{eq:jfGen} and~\eqref{eq:jfGenCorr}, for an example density profile.

Once orbits are embedded in the disc they can continue to inspiral until they are disrupted or possibly caught in a migration trap
\citep{McKernan2012,McKernan2018}. Roughly, the inspiral timescale will be the type I migration timescale inside the disc, viz.
\begin{align}
    t_I&\approx \left(\frac{M}{m_\star}\right) \left(\frac{M}{\Sigma_g r^2}\right) \left(\frac{h}{r}\right)^2 \Omega^{-1}.\nonumber\\
    &\approx 1.2 \times 10^9 {\rm yr} \left(\frac{1}{2.7+1.1 \beta}\right)  \left(\frac{h/r}{0.01}\right)^2 \left(\frac{M}{4\times 10^6 M_{\odot}}\right)^{1.5}
    \left(\frac{r}{1 {\rm pc}}\right)^{\beta-1/2} \nonumber\\
    &\left(\frac{m_\star}{1 M_{\odot}}\right)^{-1} \left(\frac{\Sigma_1}{4\, {\rm g \,cm^{-2}}}\right),
\end{align}
where $\Omega$ is the orbital frequency, $\beta$ is the power-law index of the surface density profile, and $\Sigma_1$ is the surface density of the disc at 1 pc (see e.g. \citealt{fogg&nelson2007} and the references therein). 

\begin{figure}
    \includegraphics[width=\columnwidth]{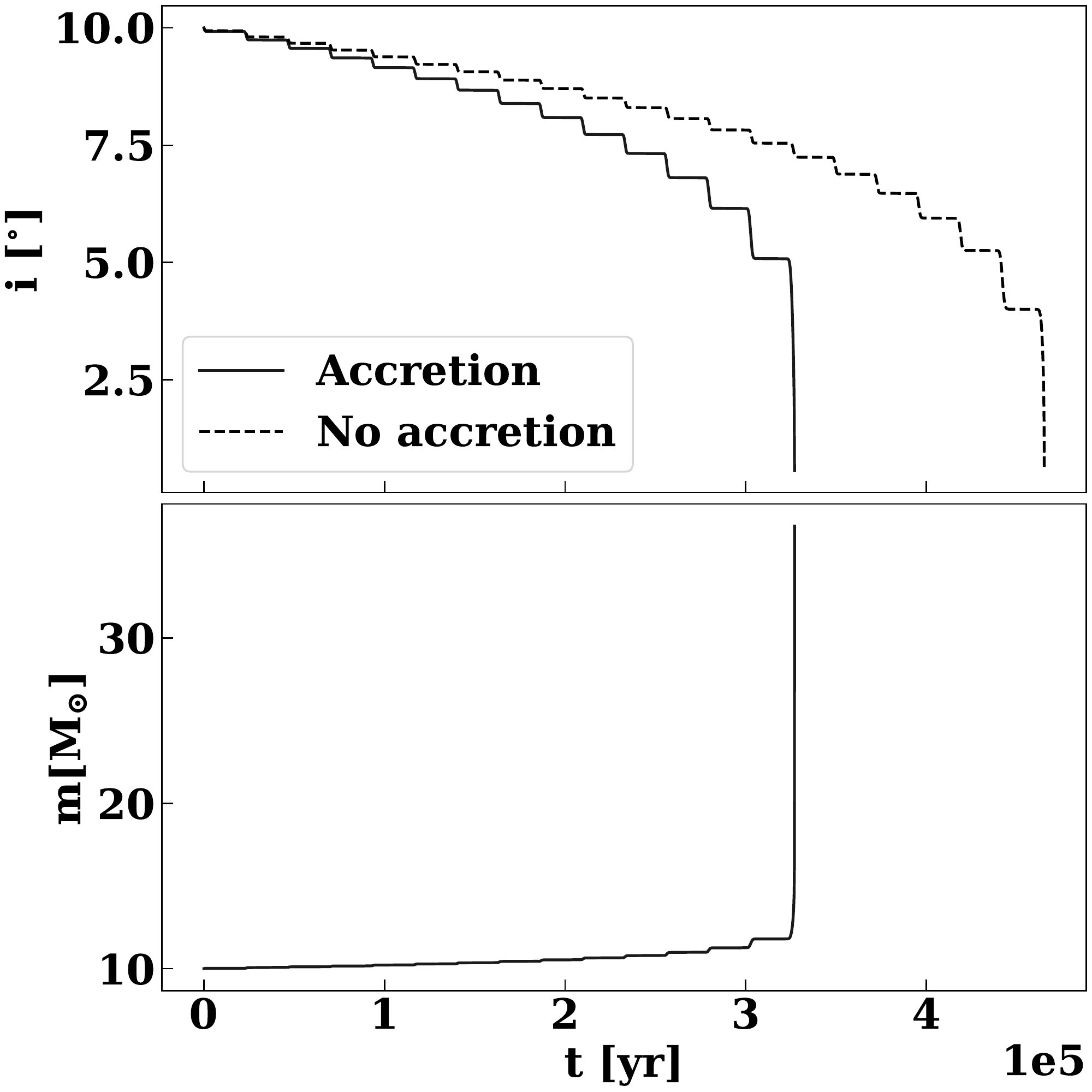}
    \caption{Top panel: inclination evolution of an object with (solid line) and without (dashed line) gas accretion. The particle's initial mass is $10 M_{\odot}$ and its radius is 0, so only GDF and Bondi-Hoyle accretion contribute to the orbital alignment. Bottom panel: Mass of object as a function of time, when accretion is included. The mass diverges, as it comes into alignment with the disc.}
    \label{fig:accretion example}
\end{figure}

As discussed in \S~\ref{sec:alignAnalytic}, the timescale for stars to come into alignment with the disc is comparable to their accretion timescale. Thus, stars that align will likely accrete significantly. This is illustrated in Figure~\ref{fig:accretion example}. The dashed, black line in the top panel shows the inclination of a $10 M_{\odot}$ compact remnant as a function of time due to GDF from an AGN disc (in this example, we set the radius of the object to 0, so there is no geometric drag). The solid line shows the evolution, including Bondi-Hoyle accretion. There is a modest acceleration of the alignment (by a factor of order unity). Also, the mass of the object diverges as it comes into alignment with the disc, as shown in the bottom panel of this figure. 

The true mass accreted may differ significantly from the Bondi--Hoyle (or geometric) estimate due to feedback effects. We leave detailed, quantitative estimates of the mass accreted to future work.

As noted in \S~\ref{sec:ecc}, retrograde orbits in the GDF regime generally increase their eccentricities as they interact with the disc. Objects on such orbits may be consumed or disrupted by the central SMBH before coming into alignment with the disc. However, for the parameters explored here, the alignment of retrograde orbits is dominated by geometric drag, which always circularizes orbits. Therefore, we do not expect disruptions from eccentricity excitation. (Though there is a caveat: including the potential of the disc can potentially lead to secular eccentricity excitation even in the geometric case). In any case, some stars and compact objects will likely be consumed by the central SMBH by inspiraling in semimajor axis. In particular, stars to the left of the vertical, gray line in Figure~\ref{fig:smaF} will be tidally disrupted before they can align.

Overall, we expect gas drag to produce a disc of stars and compact objects on prograde, circular orbits aligned with the AGN disc. Objects that align with the disc may experience significant accretion, resulting in a top-heavy mass function for the stellar disc. Up to $10\%$ of stars may become aligned with the disc over its lifetime, though the stellar disc may subsequently be depleted by radial migration and tidal disruption, as discussed in \citet{panamarev+2018}. As pointed out there, a steady state may develop between capture and radial migration (see also \citealt{karas&subr2001}).

\section{Summary}
\label{sec:summary}
AGN discs will significantly affect stellar dynamics within galactic nuclei, capturing stars and compact objects \citep{syer+1991,Artymowicz1993,vokrouhlicky&karas1998,subr&karas1999,karas&subr2001,miralde-escude&kollmeier2005,just+2012,kennedy+2016,bartos+2017,panamarev+2018,fabj+2020}. Such captures can lead to the formation of binary black holes \citep{McKernan2012,Tagawa2020,rozner+2022}, EMRIs \citep{kocsis+2011_emri,pan&yang2021,secunda+2021}, and unusual TDEs \citep{kathirgamara+2017,chankrolik+2019, chankrolik+2021,mckernan+2021_starfall}.

In this paper, we presented analytic tools for modeling the capture of stellar mass objects by an AGN disc, building on earlier work by \citet{Artymowicz1993} and \citet{subr&karas1999}. (There are also related studies on planets aligning with protoplanetary discs \citep{rein2012}.) We have focused on the alignment process itself, leaving the subsequent evolution of objects embedded in the disc to future work.

Our main results are summarized as follows: 
\begin{enumerate}
    \item We find analytic expressions for the alignment timescale of objects with an AGN disc that are accurate to within a factor of a few, accounting for both gas dynamical friction and geometric gas drag.
    \item Orbits evolve to become aligned and prograde with respect to the disc and circularize as they align. Low inclination orbits remain close to their initial semimajor axis, while high inclination orbits inspiral significantly. The final semimajor axis can be estimated analytically. This estimate is exact for circular orbits (see Appendix~\ref{app:coupleSol}, \citealt{rauch1995}, and \citealt{subr&karas1999}). 
    \item We find that up to $\sim 50\%$ of the main sequence stars and black holes  and all of the giants in the central 0.1 pc of our Galaxy would align with an AGN disc (not considering any feedback and heating of the disc). 
    As discussed in \S~\ref{sec:res}, we find uncertainties in the analytic prescriptions can affect this fraction by a factor of $\lsim1.2$.
    \item Our prescriptions for the alignment timescale do not account for mass accretion.
    In general, accretion will occur on a similar timescale, and can affect the alignment timescale by a factor of order unity. With accretion, the stellar mass formally diverges, as a star comes into alignment. However, we expect feedback effects not taken into account to halt accretion of material (see also Section 5 of \citealt{Artymowicz1993} and \citealt{leigh+2013}). While they are embedded in the disc, aligned stars can become ``immortal,'' experiencing no chemical/nuclear evolution. Such immortal stars can dramatically affect the chemistry of an AGN disc \citep{jermyn+2022}. Alignment of stars with AGN disc, and associated mass growth is a possible explanation for the observed young stellar disc with a top-heavy mass function in the Galactic Center (see also \citealt{panamarev+2018, davies&lin2020}.
\end{enumerate}

\section*{Acknowledgements}
We thank the anonymous referee for helpful comments. We thank 
Barry Ginat and Ya-Ping Li for pointing out useful references. We thank Bence Kocsis for pointing out a simplification to equation 7.
AG is supported at the Technion by a Zuckerman Fellowship.

\section*{Data Availability}
Data and models used in this study will be available upon reasonable request.



\bibliographystyle{mnras}
\bibliography{main} 

\begin{thebibliography}{}
\makeatletter
\relax
\def\mn@urlcharsother{\let\do\@makeother \do\$\do\&\do\#\do\^\do\_\do\%\do\~}
\def\mn@doi{\begingroup\mn@urlcharsother \@ifnextchar [ {\mn@doi@}
  {\mn@doi@[]}}
\def\mn@doi@[#1]#2{\def\@tempa{#1}\ifx\@tempa\@empty \href
  {http://dx.doi.org/#2} {doi:#2}\else \href {http://dx.doi.org/#2} {#1}\fi
  \endgroup}
\def\mn@eprint#1#2{\mn@eprint@#1:#2::\@nil}
\def\mn@eprint@arXiv#1{\href {http://arxiv.org/abs/#1} {{\tt arXiv:#1}}}
\def\mn@eprint@dblp#1{\href {http://dblp.uni-trier.de/rec/bibtex/#1.xml}
  {dblp:#1}}
\def\mn@eprint@#1:#2:#3:#4\@nil{\def\@tempa {#1}\def\@tempb {#2}\def\@tempc
  {#3}\ifx \@tempc \@empty \let \@tempc \@tempb \let \@tempb \@tempa \fi \ifx
  \@tempb \@empty \def\@tempb {arXiv}\fi \@ifundefined
  {mn@eprint@\@tempb}{\@tempb:\@tempc}{\expandafter \expandafter \csname
  mn@eprint@\@tempb\endcsname \expandafter{\@tempc}}}

\bibitem[\protect\citeauthoryear{{Aharon} \& {Perets}}{{Aharon} \&
  {Perets}}{2015}]{aharon&perets2015}
{Aharon} D.,  {Perets} H.~B.,  2015, \mn@doi [\apj]
  {10.1088/0004-637X/799/2/185}, \href
  {http://adsabs.harvard.edu/abs/2015ApJ...799..185A} {799, 185}

\bibitem[\protect\citeauthoryear{{Alexander}}{{Alexander}}{2017}]{alexander_tal2017}
{Alexander} T.,  2017, \mn@doi [\araa] {10.1146/annurev-astro-091916-055306},
  \href {https://ui.adsabs.harvard.edu/abs/2017ARA&A..55...17A} {55, 17}

\bibitem[\protect\citeauthoryear{{Alexander} \& {Hopman}}{{Alexander} \&
  {Hopman}}{2009}]{alexander&hopman2009}
{Alexander} T.,  {Hopman} C.,  2009, \mn@doi [\apj]
  {10.1088/0004-637X/697/2/1861}, \href
  {http://adsabs.harvard.edu/abs/2009ApJ...697.1861A} {697, 1861}

\bibitem[\protect\citeauthoryear{{Antonini} \& {Rasio}}{{Antonini} \&
  {Rasio}}{2016}]{antonini&rasio2016}
{Antonini} F.,  {Rasio} F.~A.,  2016, \mn@doi [\apj]
  {10.3847/0004-637X/831/2/187}, \href
  {https://ui.adsabs.harvard.edu/abs/2016ApJ...831..187A} {831, 187}

\bibitem[\protect\citeauthoryear{{Armitage}}{{Armitage}}{2010}]{Armitage2010}
{Armitage} P.~J.,  2010, {Astrophysics of Planet Formation}.
Cambridge University Press

\bibitem[\protect\citeauthoryear{{Artymowicz}, {Lin}  \&
  {Wampler}}{{Artymowicz} et~al.}{1993}]{Artymowicz1993}
{Artymowicz} P.,  {Lin} D.~N.~C.,   {Wampler} E.~J.,  1993, \mn@doi [\apj]
  {10.1086/172690}, \href
  {https://ui.adsabs.harvard.edu/abs/1993ApJ...409..592A} {409, 592}

\bibitem[\protect\citeauthoryear{{Bahcall} \& {Wolf}}{{Bahcall} \&
  {Wolf}}{1977}]{bahcall&wolf1977}
{Bahcall} J.~N.,  {Wolf} R.~A.,  1977, \mn@doi [\apj] {10.1086/155534}, \href
  {https://ui.adsabs.harvard.edu/abs/1977ApJ...216..883B} {216, 883}

\bibitem[\protect\citeauthoryear{{Bartko} et~al.}{{Bartko}
  et~al.}{2010}]{bartko+2010}
{Bartko} H.,  et~al., 2010, \mn@doi [\apj] {10.1088/0004-637X/708/1/834}, \href
  {http://adsabs.harvard.edu/abs/2010ApJ...708..834B} {708, 834}

\bibitem[\protect\citeauthoryear{{Bartos}, {Kocsis}, {Haiman}  \&
  {M{\'a}rka}}{{Bartos} et~al.}{2017}]{bartos+2017}
{Bartos} I.,  {Kocsis} B.,  {Haiman} Z.,   {M{\'a}rka} S.,  2017, \mn@doi
  [\apj] {10.3847/1538-4357/835/2/165}, \href
  {https://ui.adsabs.harvard.edu/abs/2017ApJ...835..165B} {835, 165}

\bibitem[\protect\citeauthoryear{{Baruteau}, {Cuadra}  \& {Lin}}{{Baruteau}
  et~al.}{2011}]{BaruteauCuadraLin2011}
{Baruteau} C.,  {Cuadra} J.,   {Lin} D.~N.~C.,  2011, \mn@doi [\apj]
  {10.1088/0004-637X/726/1/28}, \href
  {https://ui.adsabs.harvard.edu/abs/2011ApJ...726...28B} {726, 28}

\bibitem[\protect\citeauthoryear{{Chan}, {Piran}, {Krolik}  \& {Saban}}{{Chan}
  et~al.}{2019}]{chankrolik+2019}
{Chan} C.-H.,  {Piran} T.,  {Krolik} J.~H.,   {Saban} D.,  2019, \mn@doi [\apj]
  {10.3847/1538-4357/ab2b40}, \href
  {https://ui.adsabs.harvard.edu/abs/2019ApJ...881..113C} {881, 113}

\bibitem[\protect\citeauthoryear{{Chan}, {Piran}  \& {Krolik}}{{Chan}
  et~al.}{2021}]{chankrolik+2021}
{Chan} C.-H.,  {Piran} T.,   {Krolik} J.~H.,  2021, \mn@doi [\apj]
  {10.3847/1538-4357/abf0a7}, \href
  {https://ui.adsabs.harvard.edu/abs/2021ApJ...914..107C} {914, 107}

\bibitem[\protect\citeauthoryear{{Choi}, {Dotter}, {Conroy}, {Cantiello},
  {Paxton}  \& {Johnson}}{{Choi} et~al.}{2016}]{choi+2016}
{Choi} J.,  {Dotter} A.,  {Conroy} C.,  {Cantiello} M.,  {Paxton} B.,
  {Johnson} B.~D.,  2016, \mn@doi [\apj] {10.3847/0004-637X/823/2/102}, \href
  {https://ui.adsabs.harvard.edu/abs/2016ApJ...823..102C} {823, 102}

\bibitem[\protect\citeauthoryear{{Davies} \& {Lin}}{{Davies} \&
  {Lin}}{2020}]{davies&lin2020}
{Davies} M.~B.,  {Lin} D. N.~C.,  2020, \mn@doi [\mnras]
  {10.1093/mnras/staa2590}, \href
  {https://ui.adsabs.harvard.edu/abs/2020MNRAS.498.3452D} {498, 3452}

\bibitem[\protect\citeauthoryear{{DeLaurentiis}, {Epstein-Martin}  \&
  {Haiman}}{{DeLaurentiis} et~al.}{2022}]{deLaurentiis+2022}
{DeLaurentiis} S.,  {Epstein-Martin} M.,   {Haiman} Z.,  2022, arXiv e-prints,
  \href {https://ui.adsabs.harvard.edu/abs/2022arXiv221202650D} {p.
  arXiv:2212.02650}

\bibitem[\protect\citeauthoryear{{Dotter}}{{Dotter}}{2016}]{dotter+2016}
{Dotter} A.,  2016, \mn@doi [\apjs] {10.3847/0067-0049/222/1/8}, \href
  {https://ui.adsabs.harvard.edu/abs/2016ApJS..222....8D} {222, 8}

\bibitem[\protect\citeauthoryear{{Fabj}, {Nasim}, {Caban}, {Ford}, {McKernan}
  \& {Bellovary}}{{Fabj} et~al.}{2020}]{fabj+2020}
{Fabj} G.,  {Nasim} S.~S.,  {Caban} F.,  {Ford} K.~E.~S.,  {McKernan} B.,
  {Bellovary} J.~M.,  2020, \mn@doi [\mnras] {10.1093/mnras/staa3004}, \href
  {https://ui.adsabs.harvard.edu/abs/2020MNRAS.499.2608F} {499, 2608}

\bibitem[\protect\citeauthoryear{{Fogg} \& {Nelson}}{{Fogg} \&
  {Nelson}}{2007}]{fogg&nelson2007}
{Fogg} M.~J.,  {Nelson} R.~P.,  2007, \mn@doi [\aap]
  {10.1051/0004-6361:20077950}, \href
  {https://ui.adsabs.harvard.edu/abs/2007A&A...472.1003F} {472, 1003}

\bibitem[\protect\citeauthoryear{{Freitag}, {Amaro-Seoane}  \&
  {Kalogera}}{{Freitag} et~al.}{2006}]{freitag+2006}
{Freitag} M.,  {Amaro-Seoane} P.,   {Kalogera} V.,  2006, \mn@doi [\apj]
  {10.1086/506193}, \href {http://adsabs.harvard.edu/abs/2006ApJ...649...91F}
  {649, 91}

\bibitem[\protect\citeauthoryear{{Ginat}, {Panamarev}, {Kocsis}  \&
  {Perets}}{{Ginat} et~al.}{2022}]{ginat+2022}
{Ginat} Y.~B.,  {Panamarev} T.,  {Kocsis} B.,   {Perets} H.~B.,  2022, \mn@doi
  [arXiv e-prints] {10.48550/arXiv.2211.14784}, \href
  {https://ui.adsabs.harvard.edu/abs/2022arXiv221114784G} {p. arXiv:2211.14784}

\bibitem[\protect\citeauthoryear{{Hopman} \& {Alexander}}{{Hopman} \&
  {Alexander}}{2006}]{hopman&alexander2006}
{Hopman} C.,  {Alexander} T.,  2006, \mn@doi [\apjl] {10.1086/506273}, \href
  {https://ui.adsabs.harvard.edu/abs/2006ApJ...645L.133H} {645, L133}

\bibitem[\protect\citeauthoryear{{Jermyn}, {Dittmann}, {McKernan}, {Ford}  \&
  {Cantiello}}{{Jermyn} et~al.}{2022}]{jermyn+2022}
{Jermyn} A.~S.,  {Dittmann} A.~J.,  {McKernan} B.,  {Ford} K.~E.~S.,
  {Cantiello} M.,  2022, \mn@doi [\apj] {10.3847/1538-4357/ac5d40}, \href
  {https://ui.adsabs.harvard.edu/abs/2022ApJ...929..133J} {929, 133}

\bibitem[\protect\citeauthoryear{{Just}, {Yurin}, {Makukov}, {Berczik},
  {Omarov}, {Spurzem}  \& {Vilkoviskij}}{{Just} et~al.}{2012}]{just+2012}
{Just} A.,  {Yurin} D.,  {Makukov} M.,  {Berczik} P.,  {Omarov} C.,  {Spurzem}
  R.,   {Vilkoviskij} E.~Y.,  2012, \mn@doi [\apj]
  {10.1088/0004-637X/758/1/51}, \href
  {https://ui.adsabs.harvard.edu/abs/2012ApJ...758...51J} {758, 51}

\bibitem[\protect\citeauthoryear{{Karas} \& {{\v{S}}ubr}}{{Karas} \&
  {{\v{S}}ubr}}{2001}]{karas&subr2001}
{Karas} V.,  {{\v{S}}ubr} L.,  2001, \mn@doi [\aap]
  {10.1051/0004-6361:20011009}, \href
  {https://ui.adsabs.harvard.edu/abs/2001A&A...376..686K} {376, 686}

\bibitem[\protect\citeauthoryear{{Karas} \& {{\v{S}}ubr}}{{Karas} \&
  {{\v{S}}ubr}}{2007}]{karas&subr2007}
{Karas} V.,  {{\v{S}}ubr} L.,  2007, \mn@doi [\aap]
  {10.1051/0004-6361:20066068}, \href
  {https://ui.adsabs.harvard.edu/abs/2007A&A...470...11K} {470, 11}

\bibitem[\protect\citeauthoryear{{Kathirgamaraju}, {Barniol Duran}  \&
  {Giannios}}{{Kathirgamaraju} et~al.}{2017}]{kathirgamara+2017}
{Kathirgamaraju} A.,  {Barniol Duran} R.,   {Giannios} D.,  2017, \mn@doi
  [\mnras] {10.1093/mnras/stx846}, \href
  {https://ui.adsabs.harvard.edu/abs/2017MNRAS.469..314K} {469, 314}

\bibitem[\protect\citeauthoryear{{Kennedy}, {Meiron}, {Shukirgaliyev},
  {Panamarev}, {Berczik}, {Just}  \& {Spurzem}}{{Kennedy}
  et~al.}{2016}]{kennedy+2016}
{Kennedy} G.~F.,  {Meiron} Y.,  {Shukirgaliyev} B.,  {Panamarev} T.,  {Berczik}
  P.,  {Just} A.,   {Spurzem} R.,  2016, \mn@doi [\mnras]
  {10.1093/mnras/stw908}, \href
  {https://ui.adsabs.harvard.edu/abs/2016MNRAS.460..240K} {460, 240}

\bibitem[\protect\citeauthoryear{{Kocsis}, {Yunes}  \& {Loeb}}{{Kocsis}
  et~al.}{2011}]{kocsis+2011_emri}
{Kocsis} B.,  {Yunes} N.,   {Loeb} A.,  2011, \mn@doi [\prd]
  {10.1103/PhysRevD.84.024032}, \href
  {https://ui.adsabs.harvard.edu/abs/2011PhRvD..84b4032K} {84, 024032}

\bibitem[\protect\citeauthoryear{{Kozai}}{{Kozai}}{1962}]{kozai1962}
{Kozai} Y.,  1962, \mn@doi [\aj] {10.1086/108790}, \href
  {https://ui.adsabs.harvard.edu/abs/1962AJ.....67..591K} {67, 591}

\bibitem[\protect\citeauthoryear{{Leigh}, {B{\"o}ker}, {Maccarone}  \&
  {Perets}}{{Leigh} et~al.}{2013}]{leigh+2013}
{Leigh} N. W.~C.,  {B{\"o}ker} T.,  {Maccarone} T.~J.,   {Perets} H.~B.,  2013,
  \mn@doi [\mnras] {10.1093/mnras/sts554}, \href
  {https://ui.adsabs.harvard.edu/abs/2013MNRAS.429.2997L} {429, 2997}

\bibitem[\protect\citeauthoryear{{Levin} \& {Beloborodov}}{{Levin} \&
  {Beloborodov}}{2003}]{levin&beloborodov2003}
{Levin} Y.,  {Beloborodov} A.~M.,  2003, \mn@doi [\apjl] {10.1086/376675},
  \href {http://adsabs.harvard.edu/abs/2003ApJ...590L..33L} {590, L33}

\bibitem[\protect\citeauthoryear{{Li}, {Dempsey}, {Li}, {Li}  \& {Li}}{{Li}
  et~al.}{2021}]{li+2021}
{Li} Y.-P.,  {Dempsey} A.~M.,  {Li} S.,  {Li} H.,   {Li} J.,  2021, \mn@doi
  [\apj] {10.3847/1538-4357/abed48}, \href
  {https://ui.adsabs.harvard.edu/abs/2021ApJ...911..124L} {911, 124}

\bibitem[\protect\citeauthoryear{{Li}, {Dempsey}, {Li}, {Li}  \& {Li}}{{Li}
  et~al.}{2022}]{li+2022}
{Li} Y.-P.,  {Dempsey} A.~M.,  {Li} H.,  {Li} S.,   {Li} J.,  2022, \mn@doi
  [\apjl] {10.3847/2041-8213/ac60fd}, \href
  {https://ui.adsabs.harvard.edu/abs/2022ApJ...928L..19L} {928, L19}

\bibitem[\protect\citeauthoryear{{Lidov}}{{Lidov}}{1962}]{lidov1962}
{Lidov} M.~L.,  1962, \mn@doi [\planss] {10.1016/0032-0633(62)90129-0}, \href
  {https://ui.adsabs.harvard.edu/abs/1962P&SS....9..719L} {9, 719}

\bibitem[\protect\citeauthoryear{{Lu}, {Do}, {Ghez}, {Morris}, {Yelda}  \&
  {Matthews}}{{Lu} et~al.}{2013}]{lu+2013}
{Lu} J.~R.,  {Do} T.,  {Ghez} A.~M.,  {Morris} M.~R.,  {Yelda} S.,   {Matthews}
  K.,  2013, \mn@doi [\apj] {10.1088/0004-637X/764/2/155}, \href
  {http://adsabs.harvard.edu/abs/2013ApJ...764..155L} {764, 155}

\bibitem[\protect\citeauthoryear{{Marconi}, {Risaliti}, {Gilli}, {Hunt},
  {Maiolino}  \& {Salvati}}{{Marconi} et~al.}{2004}]{marconi+2004}
{Marconi} A.,  {Risaliti} G.,  {Gilli} R.,  {Hunt} L.~K.,  {Maiolino} R.,
  {Salvati} M.,  2004, \mn@doi [\mnras] {10.1111/j.1365-2966.2004.07765.x},
  \href {https://ui.adsabs.harvard.edu/abs/2004MNRAS.351..169M} {351, 169}

\bibitem[\protect\citeauthoryear{{Martini} \& {Weinberg}}{{Martini} \&
  {Weinberg}}{2001}]{martini&weinberg2001}
{Martini} P.,  {Weinberg} D.~H.,  2001, \mn@doi [\apj] {10.1086/318331}, \href
  {https://ui.adsabs.harvard.edu/abs/2001ApJ...547...12M} {547, 12}

\bibitem[\protect\citeauthoryear{{McKernan}, {Ford}, {Lyra}  \&
  {Perets}}{{McKernan} et~al.}{2012}]{McKernan2012}
{McKernan} B.,  {Ford} K.~E.~S.,  {Lyra} W.,   {Perets} H.~B.,  2012, \mn@doi
  [\mnras] {10.1111/j.1365-2966.2012.21486.x}, \href
  {https://ui.adsabs.harvard.edu/abs/2012MNRAS.425..460M} {425, 460}

\bibitem[\protect\citeauthoryear{{McKernan} et~al.,}{{McKernan}
  et~al.}{2018}]{McKernan2018}
{McKernan} B.,  et~al., 2018, \mn@doi [\apj] {10.3847/1538-4357/aadae5}, \href
  {https://ui.adsabs.harvard.edu/abs/2018ApJ...866...66M} {866, 66}

\bibitem[\protect\citeauthoryear{{McKernan}, {Ford}, {Cantiello}, {Graham},
  {Jermyn}, {Leigh}, {Ryu}  \& {Stern}}{{McKernan}
  et~al.}{2022}]{mckernan+2021_starfall}
{McKernan} B.,  {Ford} K.~E.~S.,  {Cantiello} M.,  {Graham} M.,  {Jermyn}
  A.~S.,  {Leigh} N.~W.~C.,  {Ryu} T.,   {Stern} D.,  2022, \mn@doi [\mnras]
  {10.1093/mnras/stac1310}, \href
  {https://ui.adsabs.harvard.edu/abs/2022MNRAS.514.4102M} {514, 4102}

\bibitem[\protect\citeauthoryear{{Miralda-Escud{\'e}} \&
  {Gould}}{{Miralda-Escud{\'e}} \& {Gould}}{2000}]{miralda-escude&gould2000}
{Miralda-Escud{\'e}} J.,  {Gould} A.,  2000, \mn@doi [\apj] {10.1086/317837},
  \href {https://ui.adsabs.harvard.edu/abs/2000ApJ...545..847M} {545, 847}

\bibitem[\protect\citeauthoryear{{Miralda-Escud{\'e}} \&
  {Kollmeier}}{{Miralda-Escud{\'e}} \&
  {Kollmeier}}{2005}]{miralde-escude&kollmeier2005}
{Miralda-Escud{\'e}} J.,  {Kollmeier} J.~A.,  2005, \mn@doi [\apj]
  {10.1086/426467}, \href
  {https://ui.adsabs.harvard.edu/abs/2005ApJ...619...30M} {619, 30}

\bibitem[\protect\citeauthoryear{{Nasim} et~al.,}{{Nasim}
  et~al.}{2022}]{nasim+2022}
{Nasim} S.~S.,  et~al., 2022, arXiv e-prints, \href
  {https://ui.adsabs.harvard.edu/abs/2022arXiv220709540N} {p. arXiv:2207.09540}

\bibitem[\protect\citeauthoryear{{Ostriker}}{{Ostriker}}{1999}]{ostriker1999}
{Ostriker} E.~C.,  1999, \mn@doi [\apj] {10.1086/306858}, \href
  {https://ui.adsabs.harvard.edu/abs/1999ApJ...513..252O} {513, 252}

\bibitem[\protect\citeauthoryear{{Pan} \& {Yang}}{{Pan} \&
  {Yang}}{2021}]{pan&yang2021}
{Pan} Z.,  {Yang} H.,  2021, \mn@doi [\prd] {10.1103/PhysRevD.103.103018},
  \href {https://ui.adsabs.harvard.edu/abs/2021PhRvD.103j3018P} {103, 103018}

\bibitem[\protect\citeauthoryear{{Panamarev}, {Shukirgaliyev}, {Meiron},
  {Berczik}, {Just}, {Spurzem}, {Omarov}  \& {Vilkoviskij}}{{Panamarev}
  et~al.}{2018}]{panamarev+2018}
{Panamarev} T.,  {Shukirgaliyev} B.,  {Meiron} Y.,  {Berczik} P.,  {Just} A.,
  {Spurzem} R.,  {Omarov} C.,   {Vilkoviskij} E.,  2018, \mn@doi [\mnras]
  {10.1093/mnras/sty459}, \href
  {https://ui.adsabs.harvard.edu/abs/2018MNRAS.476.4224P} {476, 4224}

\bibitem[\protect\citeauthoryear{{Paumard} et~al.,}{{Paumard}
  et~al.}{2006}]{paumard+2006}
{Paumard} T.,  et~al., 2006, \mn@doi [\apj] {10.1086/503273}, \href
  {http://adsabs.harvard.edu/abs/2006ApJ...643.1011P} {643, 1011}

\bibitem[\protect\citeauthoryear{{Paxton}, {Bildsten}, {Dotter}, {Herwig},
  {Lesaffre}  \& {Timmes}}{{Paxton} et~al.}{2011}]{paxton+2011}
{Paxton} B.,  {Bildsten} L.,  {Dotter} A.,  {Herwig} F.,  {Lesaffre} P.,
  {Timmes} F.,  2011, \mn@doi [\apjs] {10.1088/0067-0049/192/1/3}, \href
  {https://ui.adsabs.harvard.edu/abs/2011ApJS..192....3P} {192, 3}

\bibitem[\protect\citeauthoryear{{Paxton} et~al.,}{{Paxton}
  et~al.}{2013}]{paxton+2013}
{Paxton} B.,  et~al., 2013, \mn@doi [\apjs] {10.1088/0067-0049/208/1/4}, \href
  {https://ui.adsabs.harvard.edu/abs/2013ApJS..208....4P} {208, 4}

\bibitem[\protect\citeauthoryear{{Paxton} et~al.,}{{Paxton}
  et~al.}{2015}]{paxton+2015}
{Paxton} B.,  et~al., 2015, \mn@doi [\apjs] {10.1088/0067-0049/220/1/15}, \href
  {https://ui.adsabs.harvard.edu/abs/2015ApJS..220...15P} {220, 15}

\bibitem[\protect\citeauthoryear{{Paxton} et~al.,}{{Paxton}
  et~al.}{2018}]{paxton+2018}
{Paxton} B.,  et~al., 2018, \mn@doi [\apjs] {10.3847/1538-4365/aaa5a8}, \href
  {https://ui.adsabs.harvard.edu/abs/2018ApJS..234...34P} {234, 34}

\bibitem[\protect\citeauthoryear{{Preto} \& {Amaro-Seoane}}{{Preto} \&
  {Amaro-Seoane}}{2010}]{preto&amaro-seoane2010}
{Preto} M.,  {Amaro-Seoane} P.,  2010, \mn@doi [\apjl]
  {10.1088/2041-8205/708/1/L42}, \href
  {https://ui.adsabs.harvard.edu/abs/2010ApJ...708L..42P} {708, L42}

\bibitem[\protect\citeauthoryear{{Rauch}}{{Rauch}}{1995}]{rauch1995}
{Rauch} K.~P.,  1995, \mn@doi [\mnras] {10.1093/mnras/275.3.628}, \href
  {https://ui.adsabs.harvard.edu/abs/1995MNRAS.275..628R} {275, 628}

\bibitem[\protect\citeauthoryear{{Rein}}{{Rein}}{2012}]{rein2012}
{Rein} H.,  2012, \mn@doi [\mnras] {10.1111/j.1365-2966.2012.20869.x}, \href
  {https://ui.adsabs.harvard.edu/abs/2012MNRAS.422.3611R} {422, 3611}

\bibitem[\protect\citeauthoryear{{Rein} \& {Liu}}{{Rein} \&
  {Liu}}{2012}]{rein.liu2012}
{Rein} H.,  {Liu} S.~F.,  2012, \mn@doi [\aap] {10.1051/0004-6361/201118085},
  \href {https://ui.adsabs.harvard.edu/abs/2012A&A...537A.128R} {537, A128}

\bibitem[\protect\citeauthoryear{{Rowan}, {Boekholt}, {Kocsis}  \&
  {Haiman}}{{Rowan} et~al.}{2022}]{rowan+2022}
{Rowan} C.,  {Boekholt} T.,  {Kocsis} B.,   {Haiman} Z.,  2022, arXiv e-prints,
  \href {https://ui.adsabs.harvard.edu/abs/2022arXiv221206133R} {p.
  arXiv:2212.06133}

\bibitem[\protect\citeauthoryear{{Rozner} \& {Perets}}{{Rozner} \&
  {Perets}}{2022}]{rozner&perets2022}
{Rozner} M.,  {Perets} H.~B.,  2022, \mn@doi [\apj] {10.3847/1538-4357/ac6d55},
  \href {https://ui.adsabs.harvard.edu/abs/2022ApJ...931..149R} {931, 149}

\bibitem[\protect\citeauthoryear{{Rozner}, {Generozov}  \& {Perets}}{{Rozner}
  et~al.}{2023}]{rozner+2022}
{Rozner} M.,  {Generozov} A.,   {Perets} H.~B.,  2023, \mn@doi [\mnras]
  {10.1093/mnras/stad603}, \href
  {https://ui.adsabs.harvard.edu/abs/2023MNRAS.521..866R} {521, 866}

\bibitem[\protect\citeauthoryear{{Secunda}, {Hernandez}, {Goodman}, {Leigh},
  {McKernan}, {Ford}  \& {Adorno}}{{Secunda} et~al.}{2021}]{secunda+2021}
{Secunda} A.,  {Hernandez} B.,  {Goodman} J.,  {Leigh} N. W.~C.,  {McKernan}
  B.,  {Ford} K.~E.~S.,   {Adorno} J.~I.,  2021, \mn@doi [\apjl]
  {10.3847/2041-8213/abe11d}, \href
  {https://ui.adsabs.harvard.edu/abs/2021ApJ...908L..27S} {908, L27}

\bibitem[\protect\citeauthoryear{{Stone}, {Metzger}  \& {Haiman}}{{Stone}
  et~al.}{2017}]{Stone2017}
{Stone} N.~C.,  {Metzger} B.~D.,   {Haiman} Z.,  2017, \mn@doi [\mnras]
  {10.1093/mnras/stw2260}, \href
  {https://ui.adsabs.harvard.edu/abs/2017MNRAS.464..946S} {464, 946}

\bibitem[\protect\citeauthoryear{{Syer}, {Clarke}  \& {Rees}}{{Syer}
  et~al.}{1991}]{syer+1991}
{Syer} D.,  {Clarke} C.~J.,   {Rees} M.~J.,  1991, \mn@doi [\mnras]
  {10.1093/mnras/250.3.505}, \href
  {https://ui.adsabs.harvard.edu/abs/1991MNRAS.250..505S} {250, 505}

\bibitem[\protect\citeauthoryear{{Tagawa}, {Haiman}  \& {Kocsis}}{{Tagawa}
  et~al.}{2020}]{Tagawa2020}
{Tagawa} H.,  {Haiman} Z.,   {Kocsis} B.,  2020, \mn@doi [\apj]
  {10.3847/1538-4357/ab9b8c}, \href
  {https://ui.adsabs.harvard.edu/abs/2020ApJ...898...25T} {898, 25}

\bibitem[\protect\citeauthoryear{{Tamayo}, {Rein}, {Shi}  \& {Hernand
  ez}}{{Tamayo} et~al.}{2020}]{tamayo+2019}
{Tamayo} D.,  {Rein} H.,  {Shi} P.,   {Hernand ez} D.~M.,  2020, \mn@doi
  [\mnras] {10.1093/mnras/stz2870}, \href
  {https://ui.adsabs.harvard.edu/abs/2020MNRAS.491.2885T} {491, 2885}

\bibitem[\protect\citeauthoryear{{Thompson}, {Quataert}  \&
  {Murray}}{{Thompson} et~al.}{2005}]{thompson+2005}
{Thompson} T.~A.,  {Quataert} E.,   {Murray} N.,  2005, \mn@doi [\apj]
  {10.1086/431923}, \href
  {https://ui.adsabs.harvard.edu/abs/2005ApJ...630..167T} {630, 167}

\bibitem[\protect\citeauthoryear{{Vasiliev}}{{Vasiliev}}{2017}]{vasiliev2017}
{Vasiliev} E.,  2017, \mn@doi [\apj] {10.3847/1538-4357/aa8cc8}, \href
  {http://adsabs.harvard.edu/abs/2017ApJ...848...10V} {848, 10}

\bibitem[\protect\citeauthoryear{{Vokrouhlicky} \& {Karas}}{{Vokrouhlicky} \&
  {Karas}}{1998}]{vokrouhlicky&karas1998}
{Vokrouhlicky} D.,  {Karas} V.,  1998, \mn@doi [\mnras]
  {10.1046/j.1365-8711.1998.01564.x}, \href
  {https://ui.adsabs.harvard.edu/abs/1998MNRAS.298...53V} {298, 53}

\bibitem[\protect\citeauthoryear{{{\v{S}}ubr} \& {Karas}}{{{\v{S}}ubr} \&
  {Karas}}{1999}]{subr&karas1999}
{{\v{S}}ubr} L.,  {Karas} V.,  1999, \aap, \href
  {https://ui.adsabs.harvard.edu/abs/1999A&A...352..452S} {352, 452}

\bibitem[\protect\citeauthoryear{{von Fellenberg} et~al.,}{{von Fellenberg}
  et~al.}{2022}]{vonFellenberg+2022}
{von Fellenberg} S.~D.,  et~al., 2022, \mn@doi [\apjl]
  {10.3847/2041-8213/ac68ef}, \href
  {https://ui.adsabs.harvard.edu/abs/2022ApJ...932L...6V} {932, L6}

\bibitem[\protect\citeauthoryear{{von Zeipel}}{{von
  Zeipel}}{1910}]{von-Zeipel1910}
{von Zeipel} H.,  1910, \mn@doi [Astronomische Nachrichten]
  {10.1002/asna.19091832202}, \href
  {https://ui.adsabs.harvard.edu/abs/1910AN....183..345V} {183, 345}

\makeatother
\end{thebibliography}




\appendix

\section{Orbital nodes}
\label{app:node}
Here we compile useful expressions for the stellar velocity at the nodes of the orbit. The velocity of a star at the node crossing is 

\begin{align}
    &v_{\rm node}=\sqrt{\frac{G M}{a}} \sqrt{\frac{1+e^2\pm 2 e \cos(\omega)}{1-e^2}},\nonumber\\
\end{align}
where the positive (negative) sign corresponds to the ascending (descending) node. The distance between the node and the central mass is 
\begin{equation}
    r_{\rm node}=\frac{a (1-e^2)}{1+e \cos(\omega)}.
\end{equation}
The velocity of the gas at the orbital nodes is 
\begin{equation}
    v_d\approx \sqrt{\frac{G M}{a}} \sqrt{\frac{1 \pm e \cos(\omega)}{1-e^2}} \left[1-\left(\frac{h}{r} \right)^2 \right],
\end{equation}
where the term in brackets corrects for the pressure within the disc. This correction will be unimportant, except for orbits that are aligned or nearly aligned with the disc. Neglecting this, the relative velocity between the star and the gas at the orbital node is
\begin{equation}
    \resizebox{0.96\columnwidth}{!}{
    $v_{\rm rel}=\sqrt{\frac{G M}{a}}\sqrt{\frac{e^2 \pm e \cos (\omega ) \left(2 \cos (i) \sqrt{1\pm e \cos (\omega )}-3\right)-2 \cos (i) \sqrt{1\pm e \cos (\omega )}+2}{1-e^2}}$
    }
\end{equation}

\section{Approximate analytic solutions for inclination and semimajor axis evolution.}
\label{app:coupleSol}
In this appendix we derive a relationship between the angular momentum and inclination of a nearly Keplerian orbit, under the influence of a drag force of form 
\begin{align}
    \mathbf{f_{\rm drag}} = -g(v_{\rm rel}, m_\star, \rho_g,...)  \mathbf{v_{\rm rel}},
\end{align}
while it crosses through a thin, gas disc. Above, $v_{\rm rel}$ is the relative velocity between and the gas, $g$ is an arbitrary function of $v_{\rm rel}$, the gas density, the object's mass, and other parameters. Both geometric drag and GDF have the above form. Thus, our results will be applicable to both.

The time derivative of the binding energy is
\begin{align}
    \frac{d E}{d t}=g \mathbf{v_{\rm rel} \cdot v_{\rm node}},
    \label{eq:dEdtGen}
\end{align}
The time derivative of the angular momentum is
\begin{align}
    \frac{d J}{d t}=-g \mathbf{\left(r_{\rm node} \times v_{\rm rel}\right)\cdot \jhat}
    \label{eq:dJdtGen}
\end{align}
The time derivative of the z-component of the angular momentum is 
\begin{align}
    \frac{d J_z}{d t}=-g \mathbf{\left(r_{\rm node} \times v_{\rm rel}\right)_z}.
    \label{eq:dJzdtGen}
\end{align}
Note $J_z=J \cos(i)$, where $i$ is the inclination of the stellar orbit. From equations~\eqref{eq:dJzdtGen} and~\eqref{eq:dJdtGen}, and Appendix~\ref{app:node},
\begin{align}
    \frac{d J_z}{d J}=\frac{1-\cos(i) \sqrt{1\pm e \cos(\omega)}}{\cos(i) -\sqrt{1\pm e \cos(\omega)}}
\end{align}
For $\omega =\pi/2$\footnote{Geometrically this corresponds to an orbit rotated over its minor axis.} or $e=0$,
\begin{align}
    \frac{d J_z}{d J}=-1.
\end{align}
In this case,
\begin{align}
    J-J_o=J_{z,o}-J_{z},
\end{align}
where the subscript `o' denotes the initial condition. In general, $J_z$ increases while $J$ decreases.\footnote{For retrograde orbits, while $J_z$ is increasing its absolute value is decreasing.} Thus,
\begin{align}
    \left(\frac{J}{J_o}\right)^{-1}=\frac{1+\cos(i)}{1+\cos(i_o)}
     \label{eq:jfGen}
\end{align}
 We expect aligned orbits to circularize as discussed in \S~\ref{sec:ecc}. Thus, equation~\eqref{eq:jfGen} completely determines the final semimajor axis of the orbit.  
 
 Equation~\eqref{eq:jfGen} will in only an approximation. Empirically, the final angular momentum in the GDF regime can be estimated more accurately using
 \begin{align}
    \left(\frac{J}{J_o}\right)^{-1}=\frac{1+\cos(i)}{1+\cos(i_o)} \left[(1+e_o \cos(\omega_o))(1-e_o \cos(\omega_o))\right]^{p(\gamma)},
     \label{eq:jfGenCorr}
\end{align}
where $p(\gamma) \approx -0.2 \gamma + 0.8$.

\section{Eccentricity evolution}
\label{app:ecc}
At each orbital node the energy and angular momentum are dissipated. 
In the GDF regime, the change in the binding energy after a node crossing is
\begin{equation}
    \delta E = c_1 \frac{\rho(r_1) \mathbf{v_{\rm rel} \cdot v_{\rm node}}}{v_{\rm rel}^3} \frac{t_{\rm cross}}{P_{\rm orb}},
\end{equation}
where $c_1$ is a constant; $r_1$ is the radius of the node; $v_{\rm node}$ is the orbital velocity at the node; and $v_{\rm rel}$ is the relative velocity between the gas and the star at the node. Note the binding energy is always positive and is increasing as energy is dissipated.

The change in the (specific) angular momentum is 
\begin{align}
    \delta \mathbf{J}= c_1 \frac{\rho(r_1) \mathbf{r_{1} \times v_{\rm rel}}}{v_{\rm rel}^3} \frac{t_{\rm cross}}{P_{\rm orb}}.
\end{align}
The change in eccentricity is 
\begin{align}
    \delta e=\left(\frac{\partial e}{\partial E}\right)_{J} \delta E+\left(\frac{\partial e}{\partial J}\right)_{E} \delta J.
    \label{eq:deltaE}
\end{align}
 Equation~\eqref{eq:deltaE} can be used to evaluate the direction of eccentricity evolution

 \section{Additional numerical validation}
 \label{app:numValid2}
Here we provide additional numerical validation of our analytic alignment time (equation~\ref{eq:talignGen}).

First, we construct a grid of orbits in eccentricity-inclination space for two different density profiles ($r^{-3}$ and $r^{-1.5}$) and two different arguments of pericenter (0 and 90$^\circ$). The semimajor axis is 1 pc. Then we simulate each orbit until it aligns with the disc. To better isolate the individual timescales in equation~\ref{eq:talignGen}, we perform two sets of simulations: (i) one with only GDF and (ii) one with only geometric drag. To speed up the numerical calculations we assume an unrealistically high gas density 
of $1.9\times 10^9 M_{\odot}$ pc$^{-3}$ at 1 pc for the geometric regime and for inclinations above $40^{\circ}$ in the GDF regime. For inclinations below 40$^{\circ}$ in the GDF regime, we reduce the density by two orders of magnitude to ensure the alignment time remains longer than orbital time.

\begin{figure*}
    \includegraphics[width=\columnwidth]{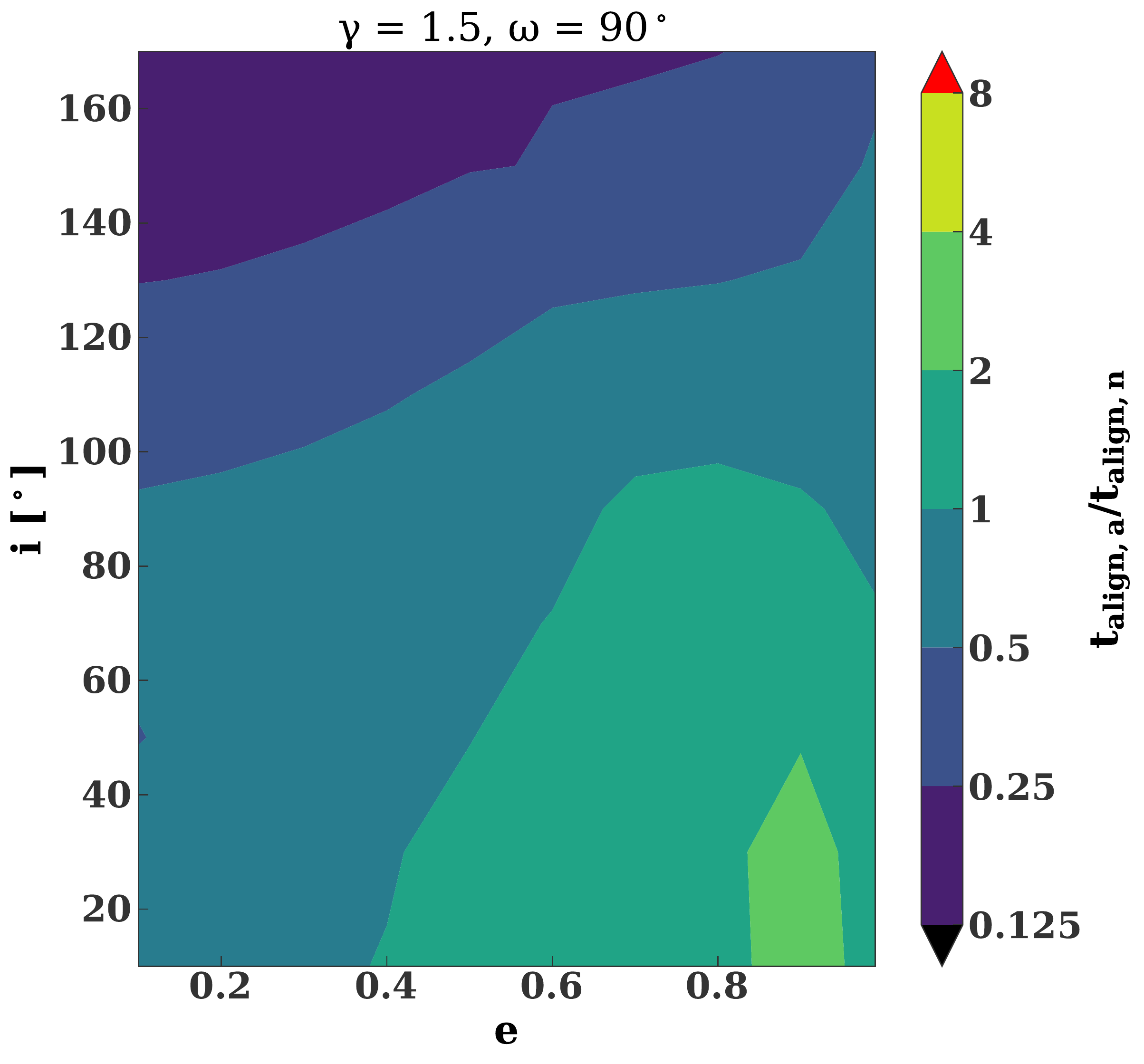}
    \includegraphics[width=\columnwidth]{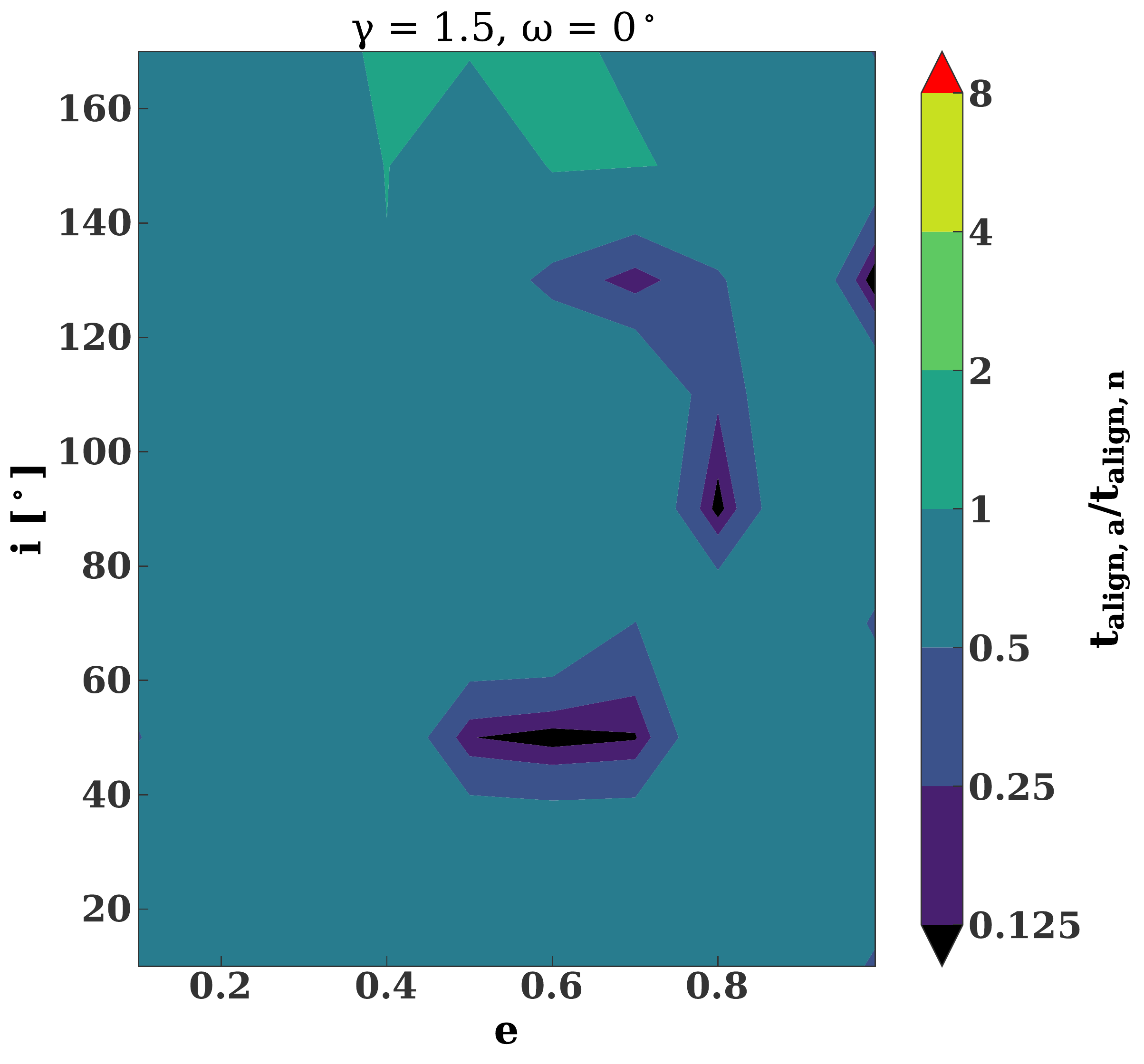}
    \includegraphics[width=\columnwidth]{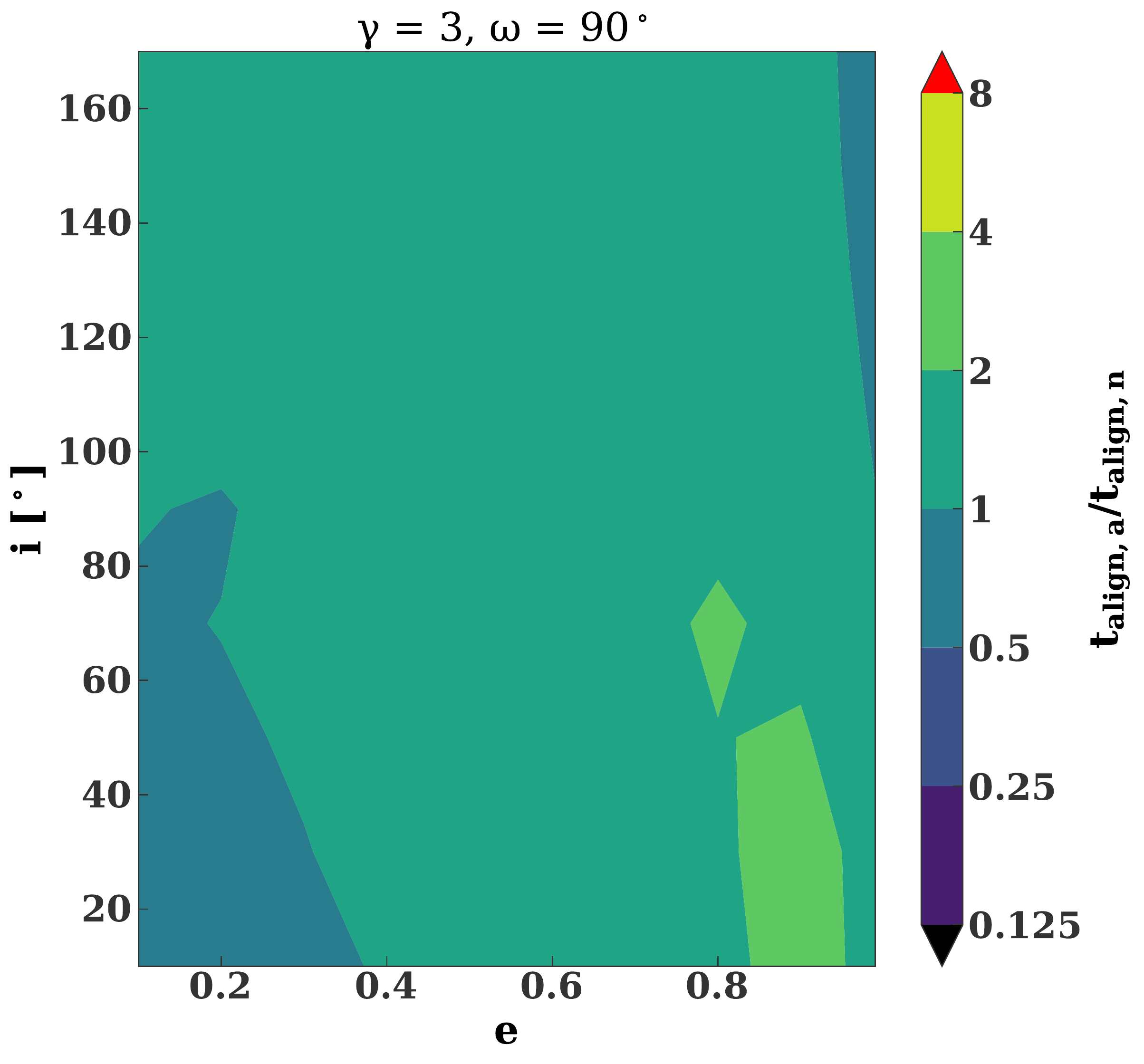}
    \includegraphics[width=\columnwidth]{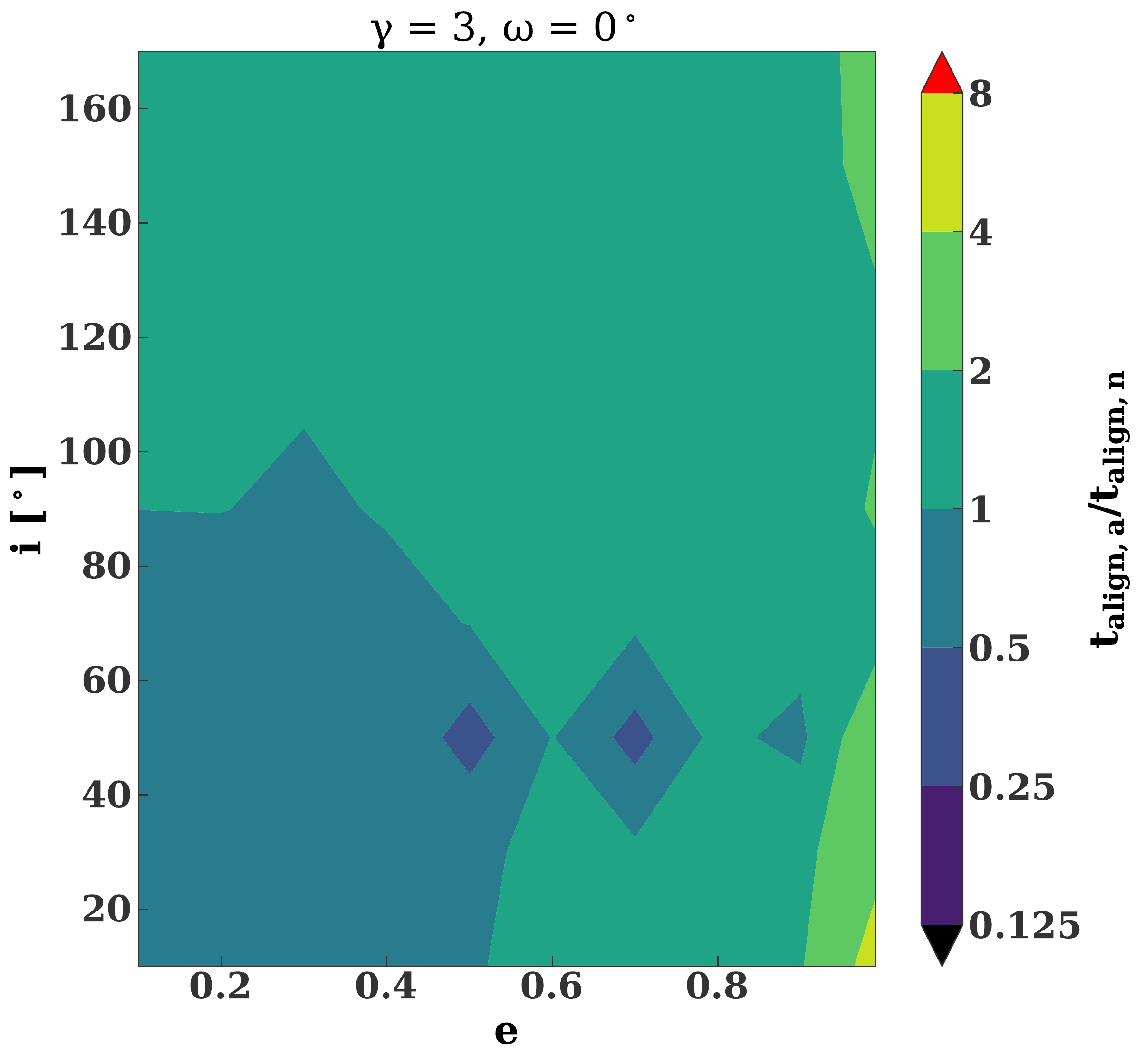}
    \caption{Ratio of analytic alignment (eq~\ref{eq:talignGen}) to the numerical alignment time in the GDF regime as a function of eccentricity and inclination. The different panels correspond to different density profiles and arguments of pericenter.}
    \label{fig:numValid2a}
\end{figure*}

\begin{figure*}
    \includegraphics[width=\columnwidth]{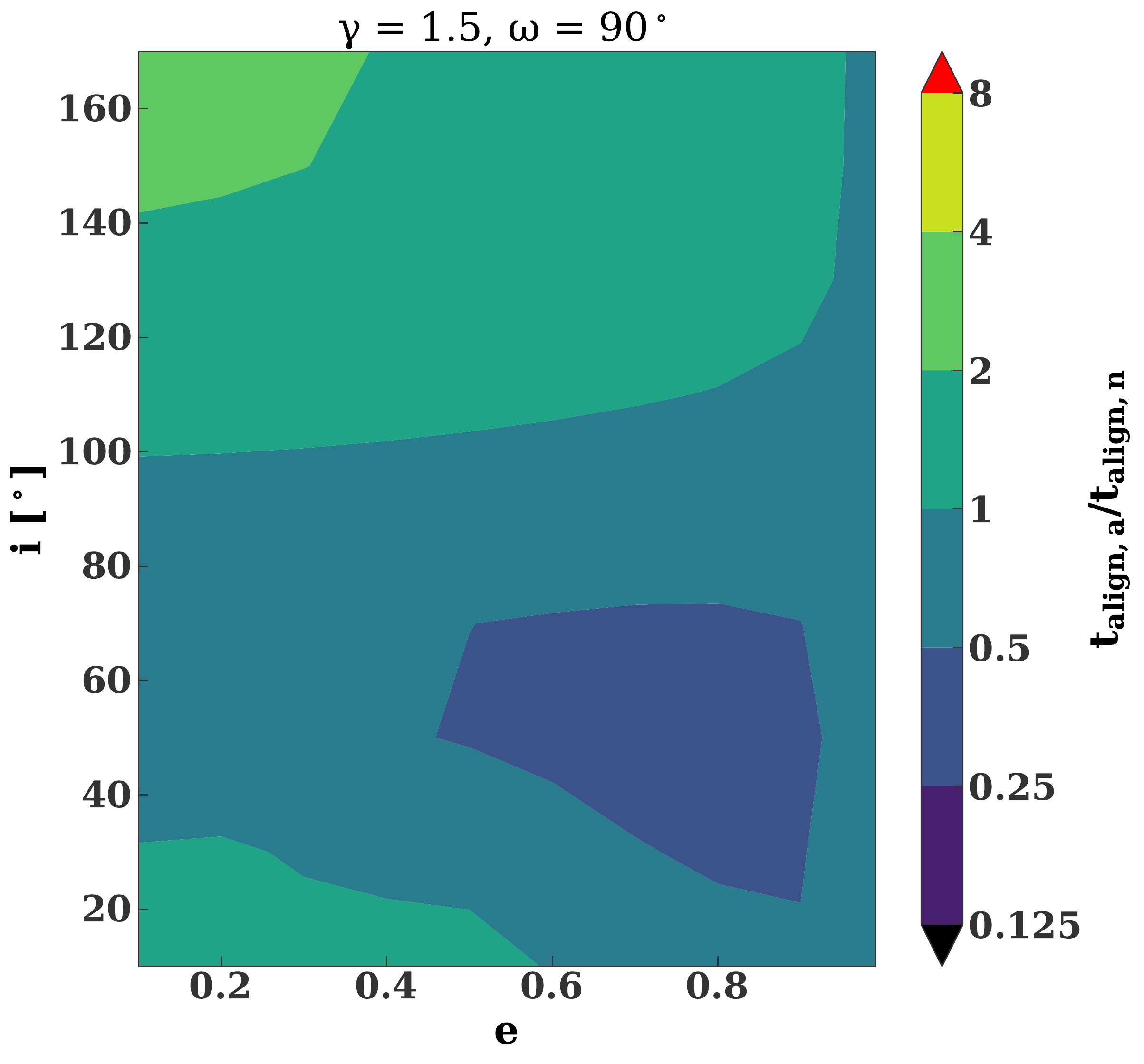}
    \includegraphics[width=\columnwidth]{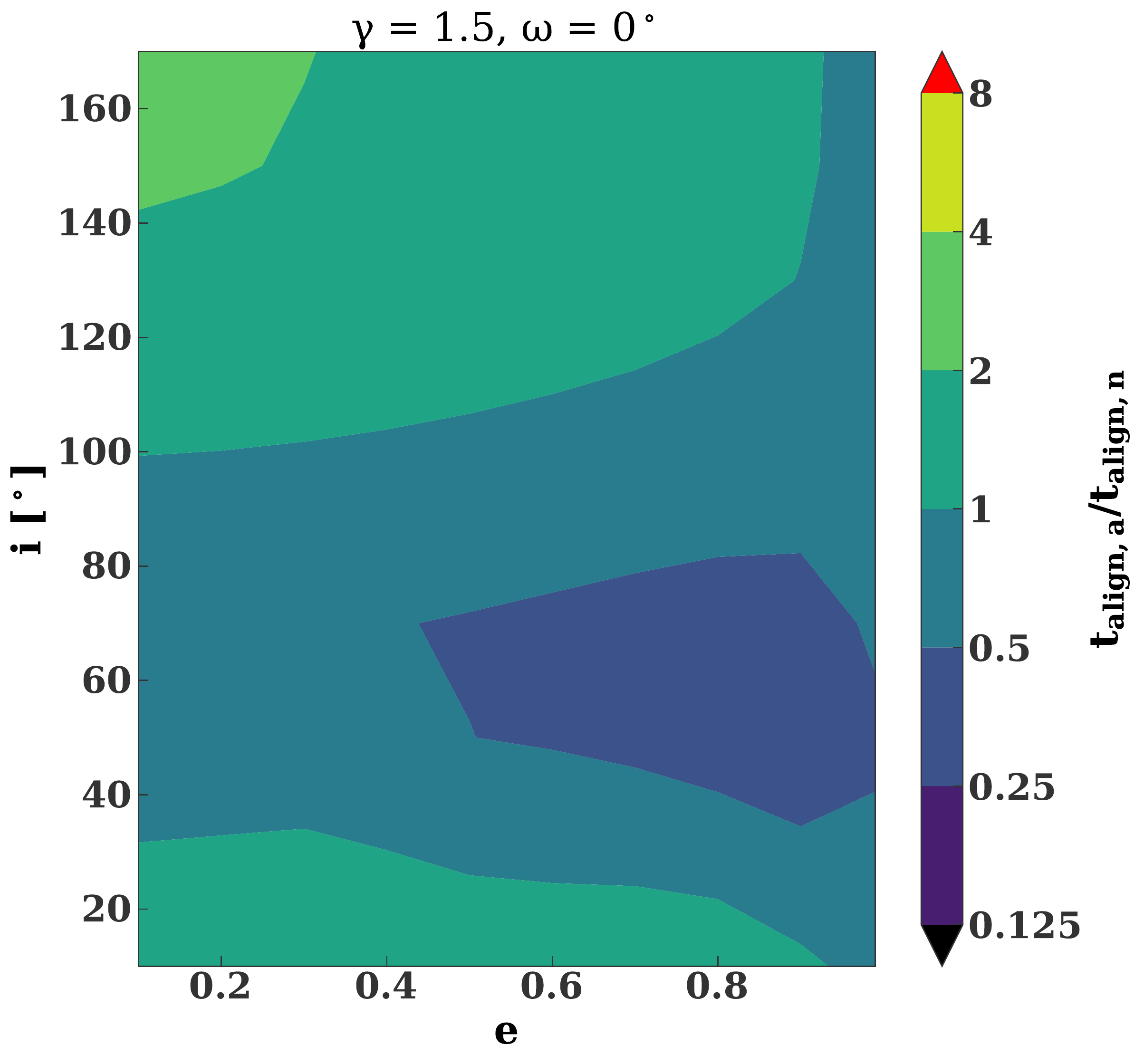}
    \includegraphics[width=\columnwidth]{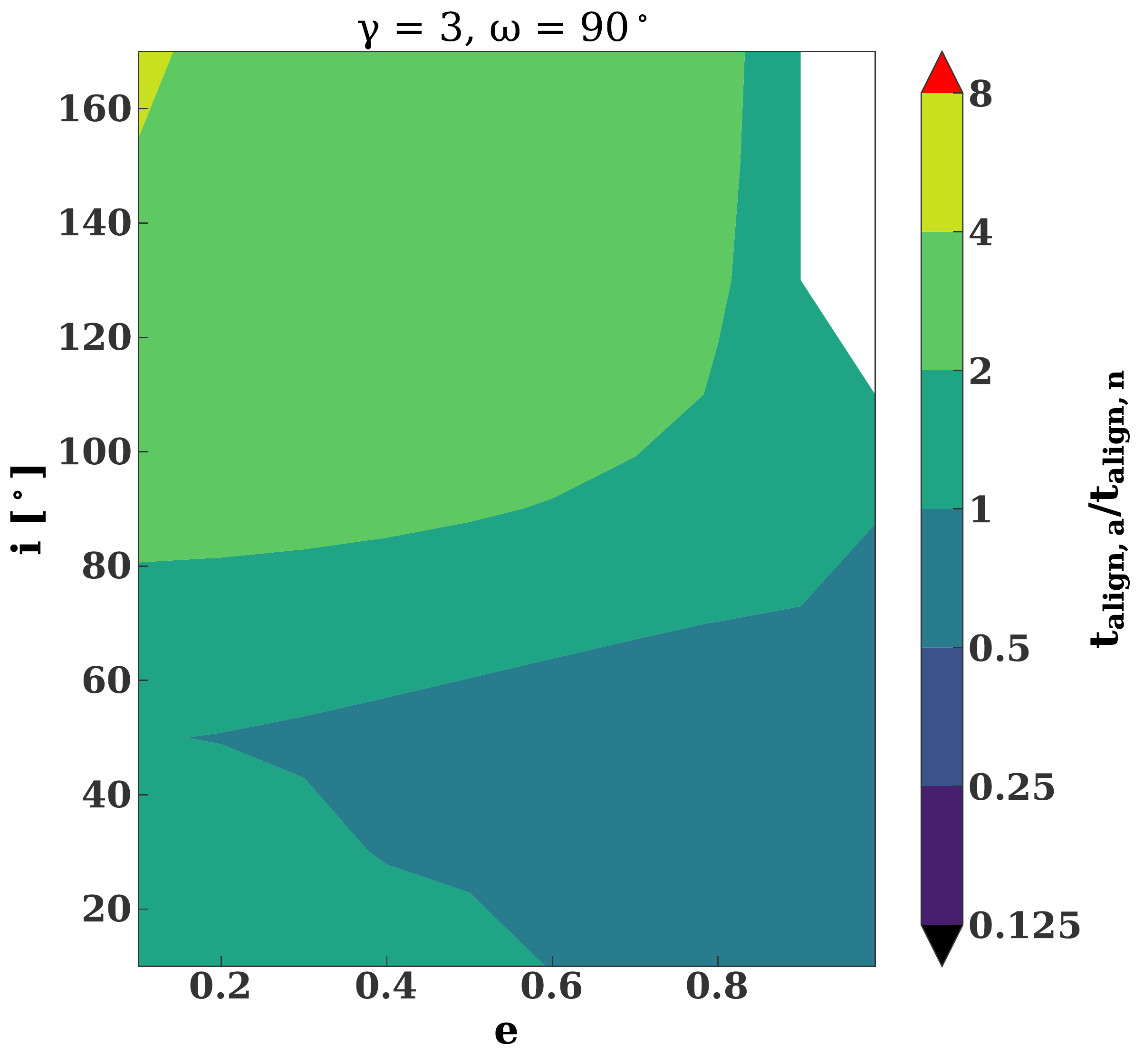}
    \includegraphics[width=\columnwidth]{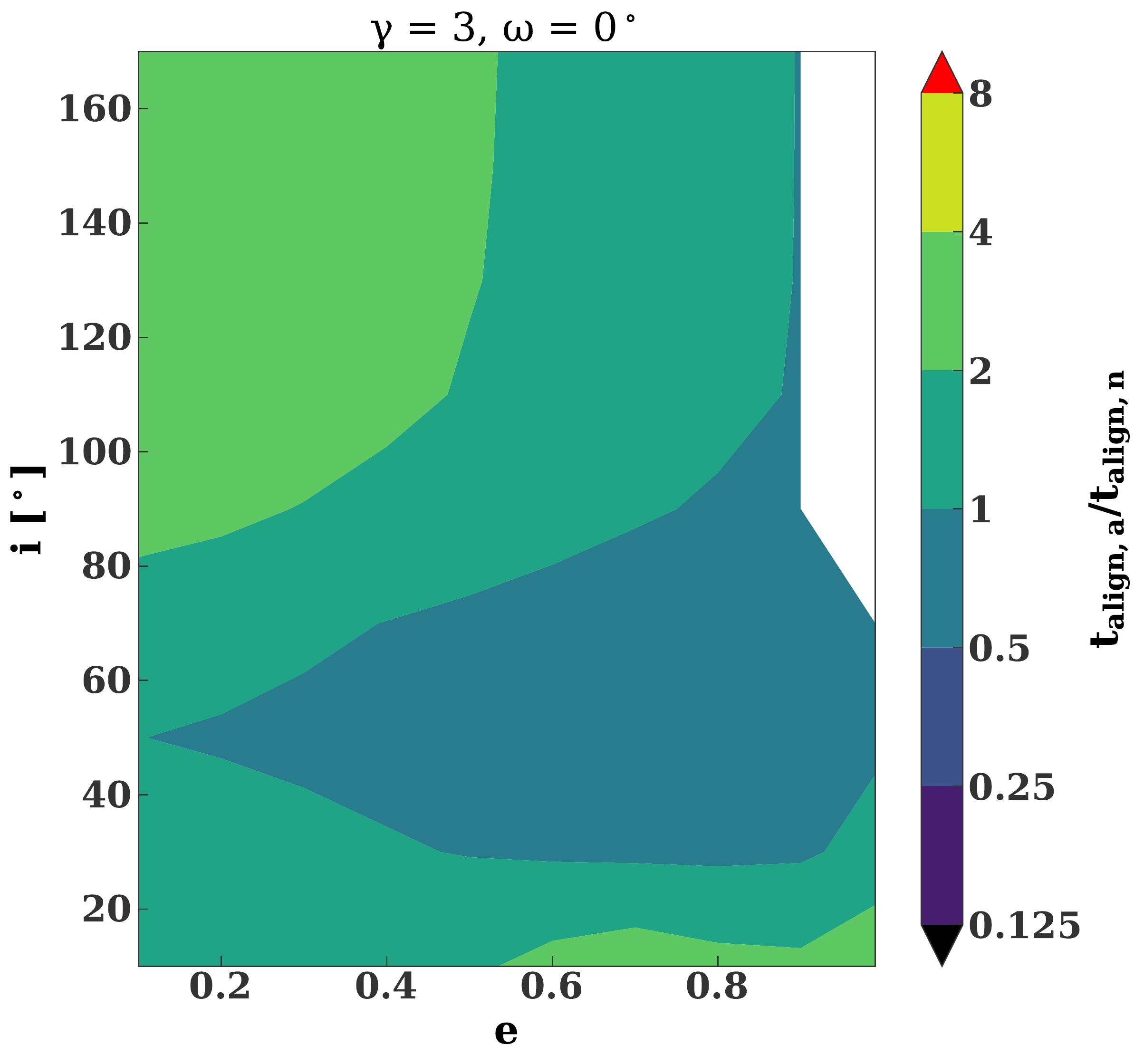}
    \caption{Same as Figure~\ref{fig:numValid2a} except for the geometric regime. The white regions are excluded, because the alignment time would be less than the orbital time.}
    \label{fig:numValid2b}
\end{figure*}

Figures~\ref{fig:numValid2a} and~\ref{fig:numValid2b} show the ratio between the analytic alignment time (equation~\ref{eq:talignGen}) and the alignment time in simulations.  Generally, the numerical results fall within a factor of a few of our analytic prescription. For
retrograde orbits with $\omega \approx$ 90$^\circ$, our analytic alignment time differs by up to a factor of $\sim$6 from the numerical results. Further experiments suggest this large discrepancy is confined to a narrow interval of $\omega$. For an eccentricity of 0.1 and inclination of 150$^{\circ}$, the timescales agree within a factor of 3, except if $\omega \in (80^{\circ}, 100^{\circ})$ or $\omega \in (260^{\circ}, 280^{\circ})$.

\begin{figure*}
    \includegraphics[width=\columnwidth]{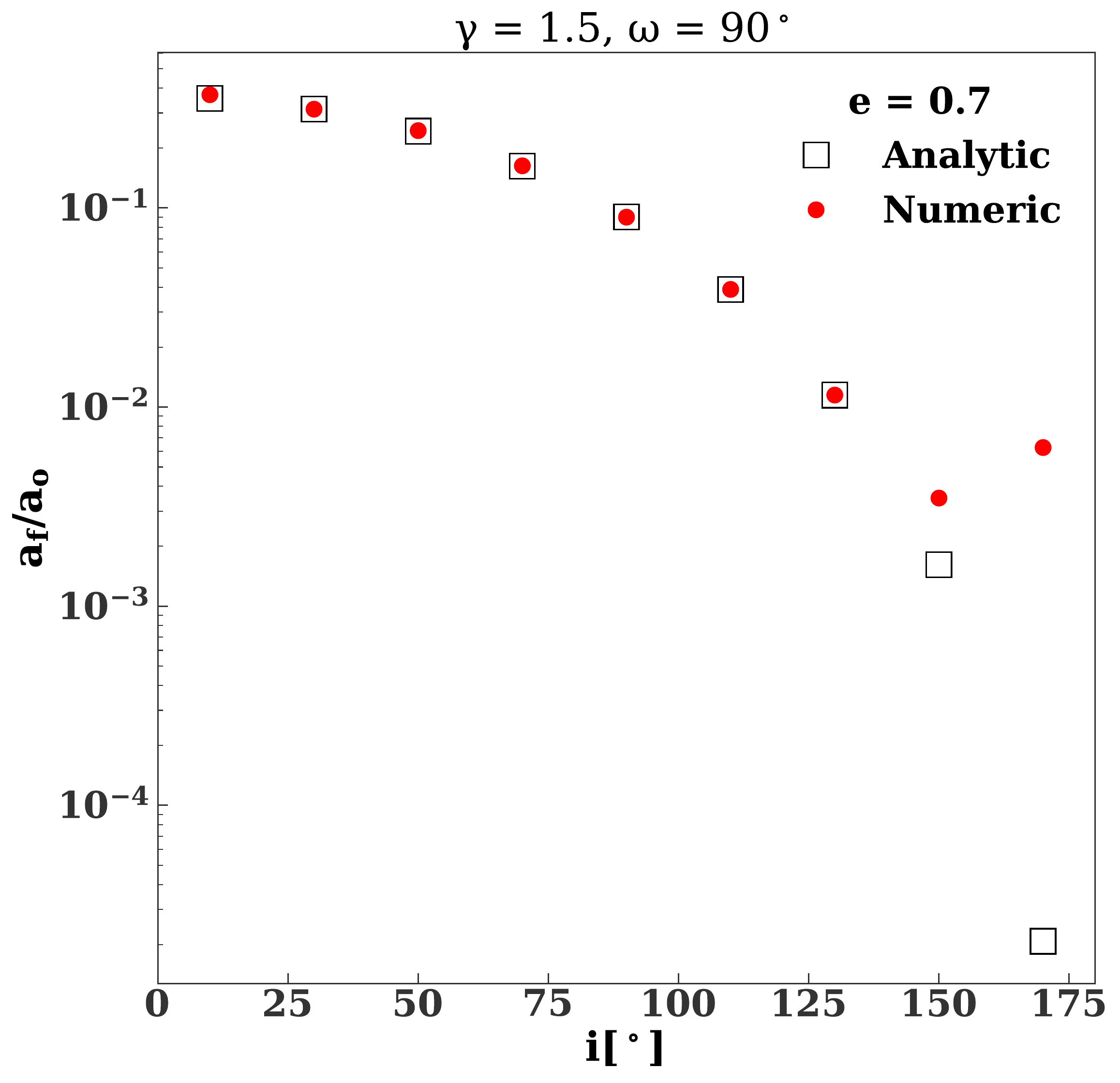}
    \includegraphics[width=\columnwidth]{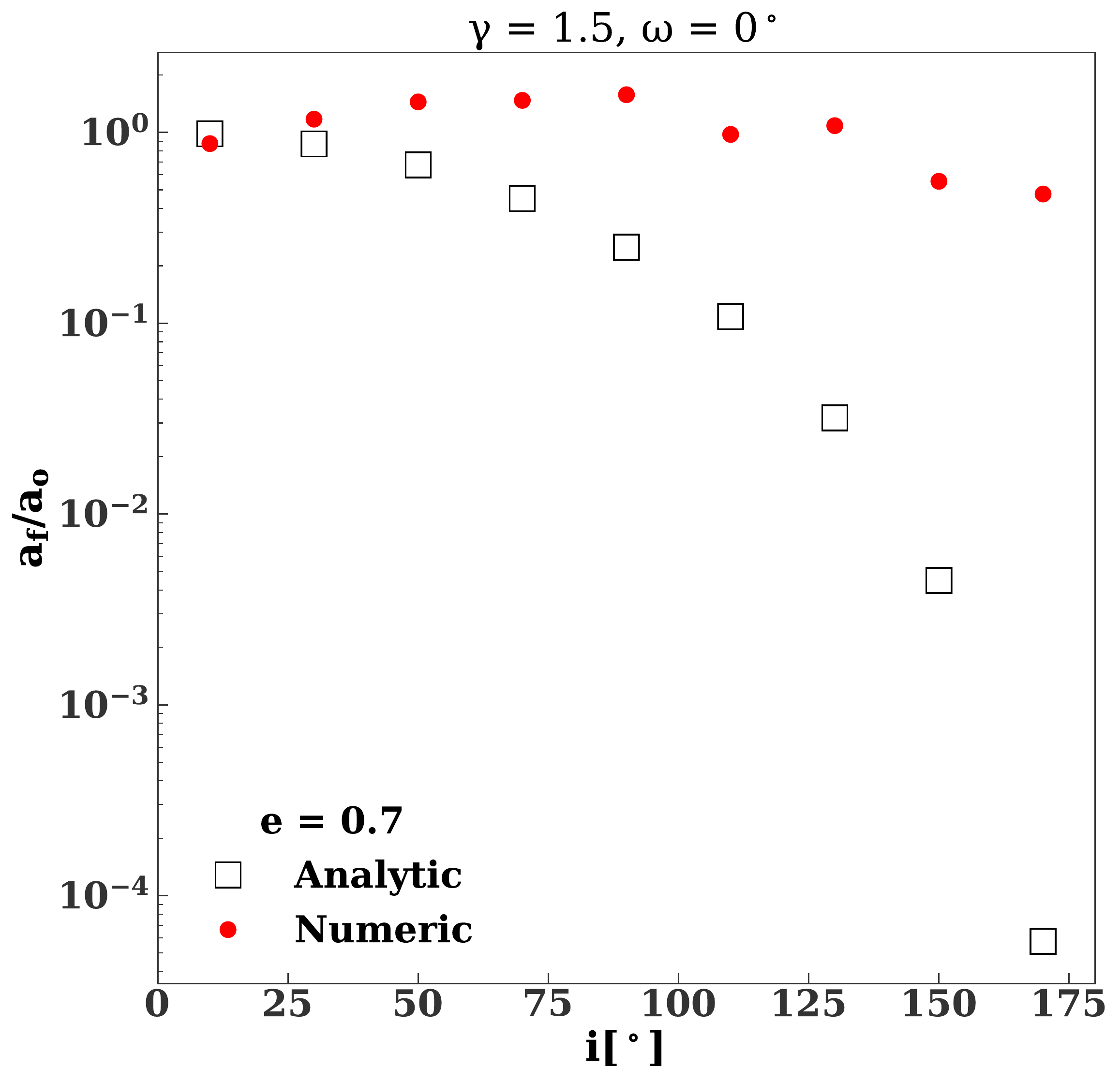}
    \includegraphics[width=\columnwidth]{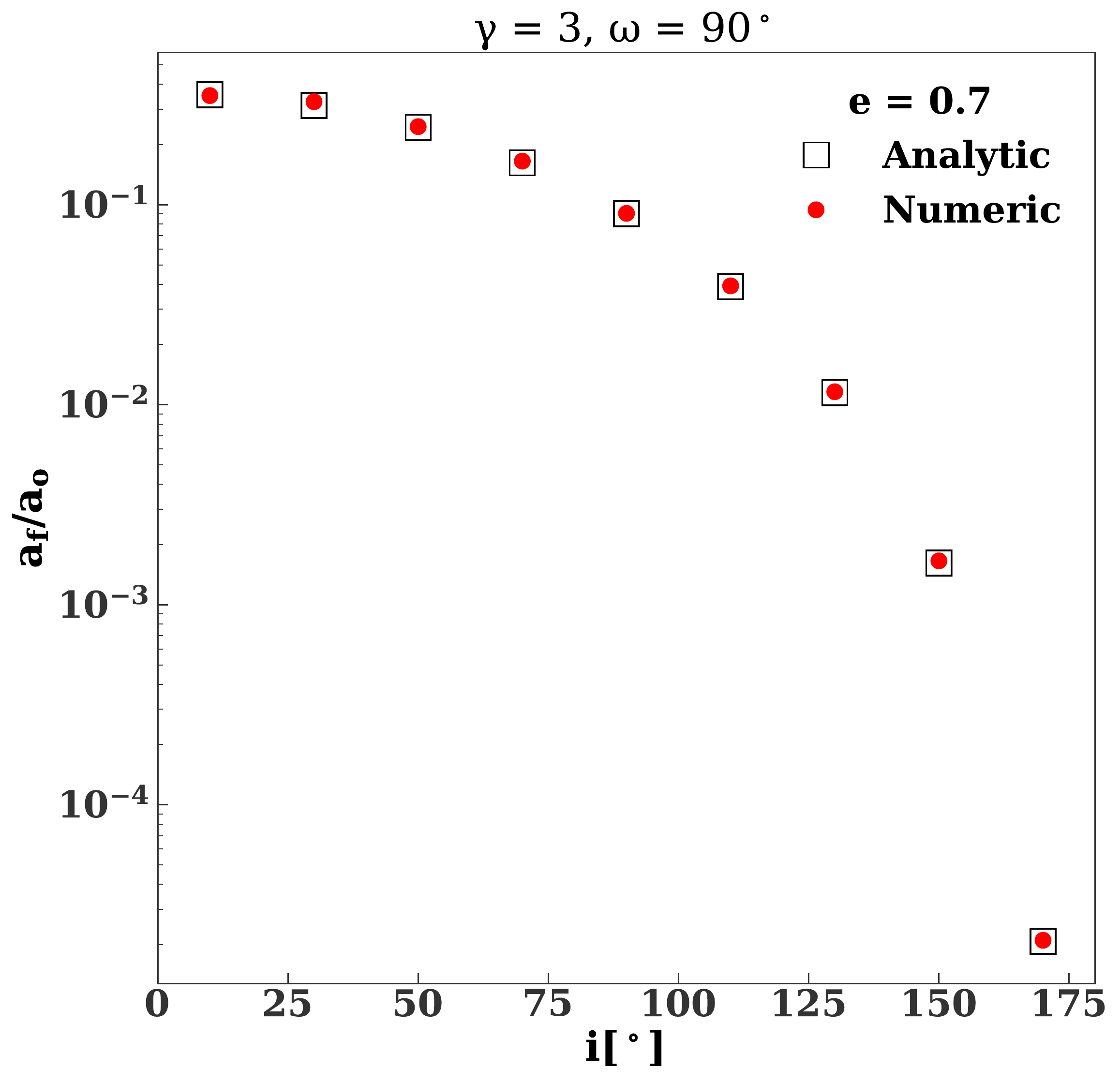}
    \includegraphics[width=\columnwidth]{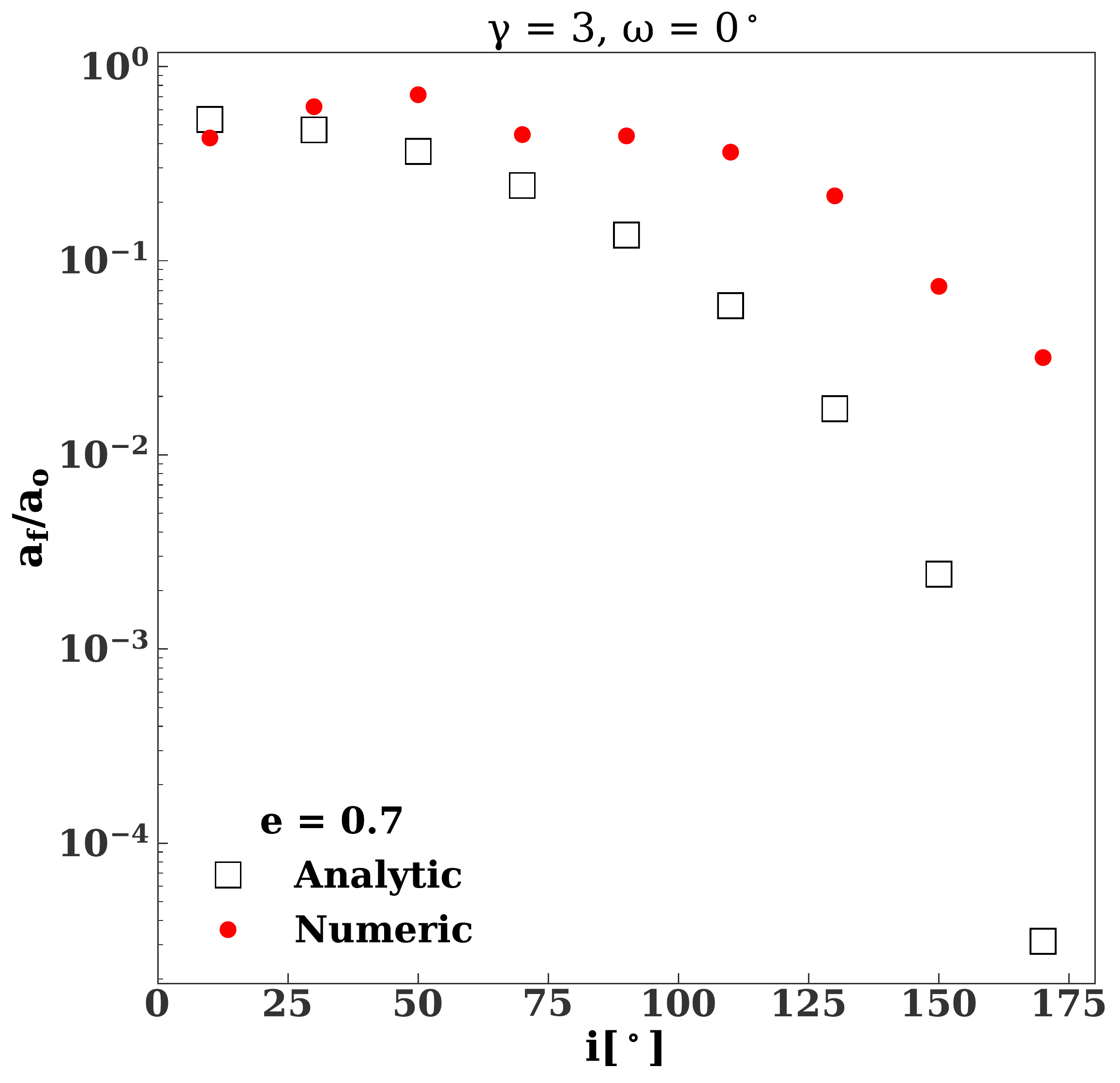}
    \caption{Ratio of final to initial semimajor axis as a function of inclination in the GDF regime. The red circles (black squares) correspond to analytic (numerical results). The different panels correspond to different density profiles and arguments of pericenter. The eccentricity is 0.7.}
    \label{fig:numValidSma1}
\end{figure*}

Also, our prediction for the final semimajor axis of retrograde orbits can be inaccurate, particularly in the GDF regime. Equation~\eqref{eq:jfGenCorr} predicts the final semimajor axis decreases monotonically with inclination and goes to 0 as the orbital inclination approaches $180^{\circ}$. Numerically, this is not always the case as shown in Figure~\ref{fig:numValidSma1}.
This discrepancy could be relevant for predictions of the final semimajor axis distribution of stellar mass black holes, considering a non-negligible fraction of retrograde black holes can align via GDF.

\begin{figure*}
    \includegraphics[width=\columnwidth]{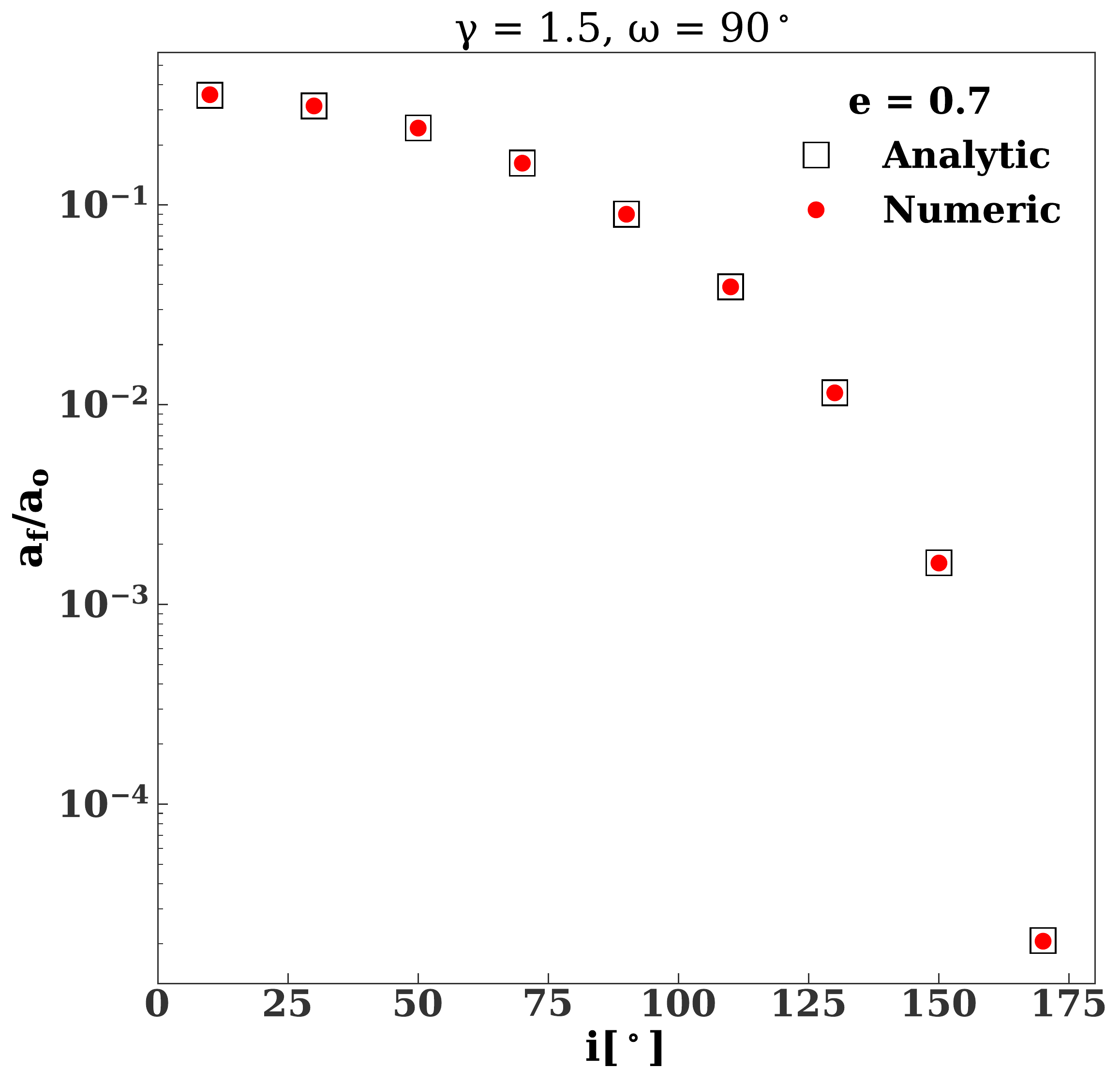}
    \includegraphics[width=\columnwidth]{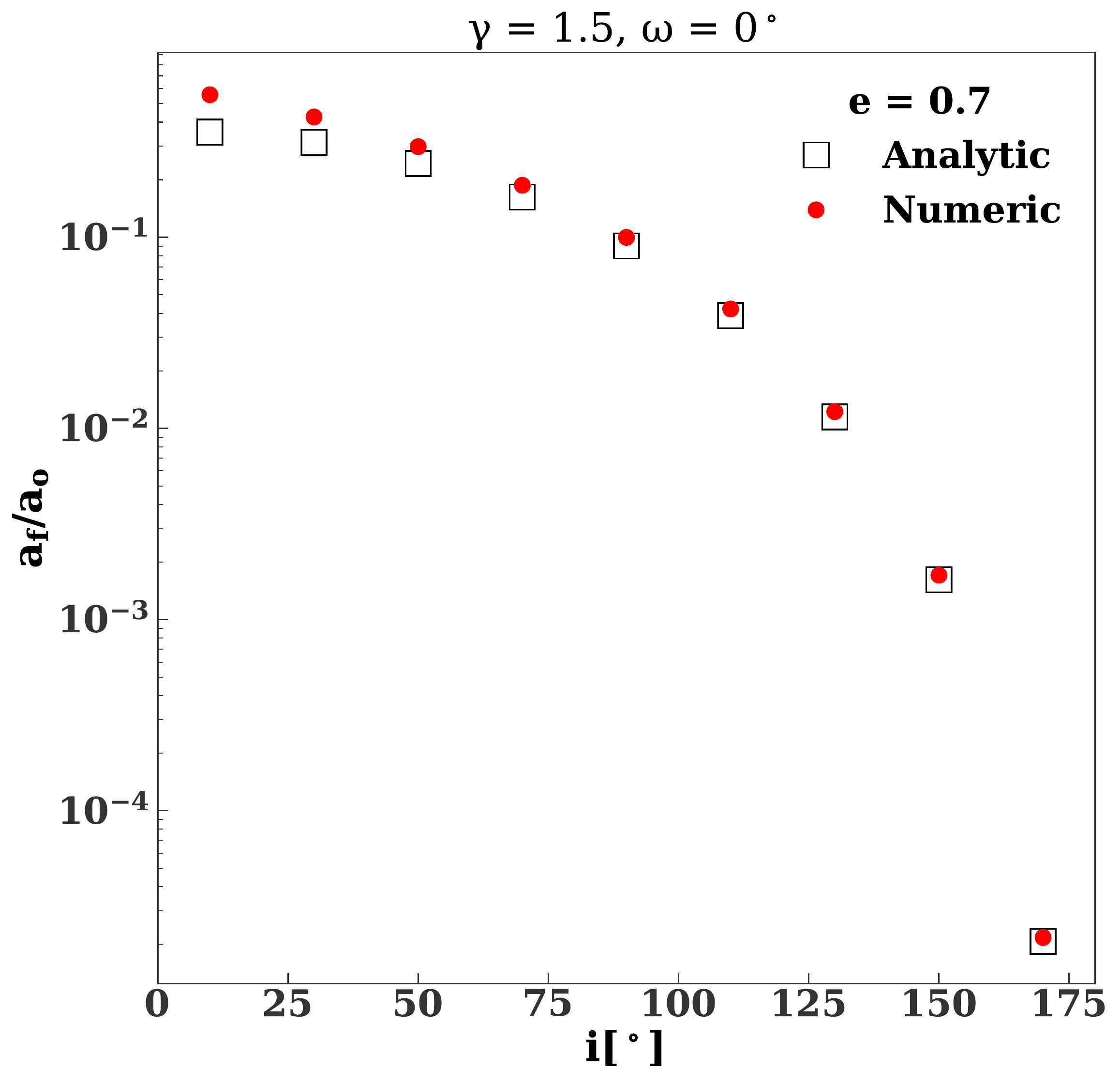}
    \includegraphics[width=\columnwidth]{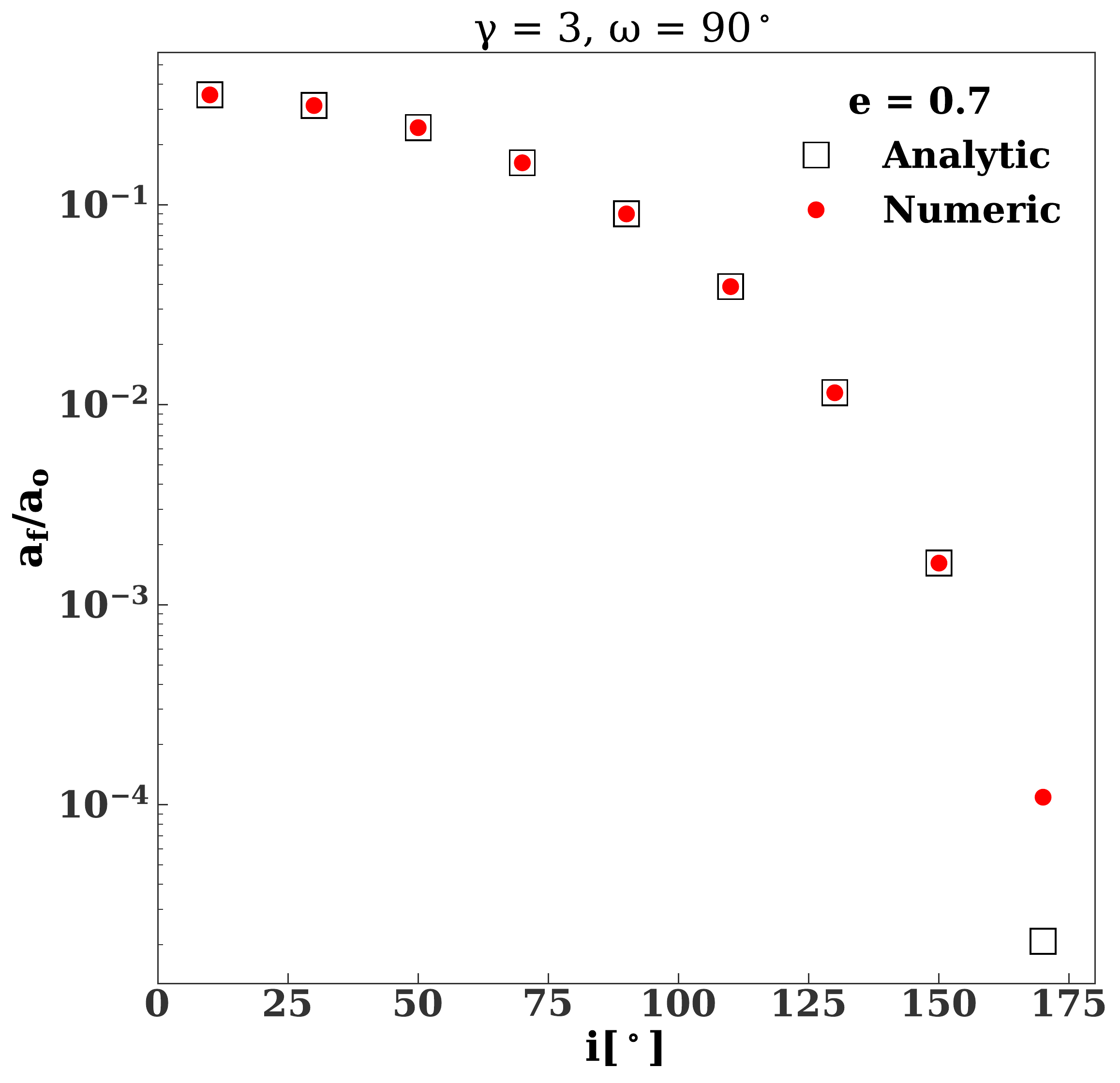}
    \includegraphics[width=\columnwidth]{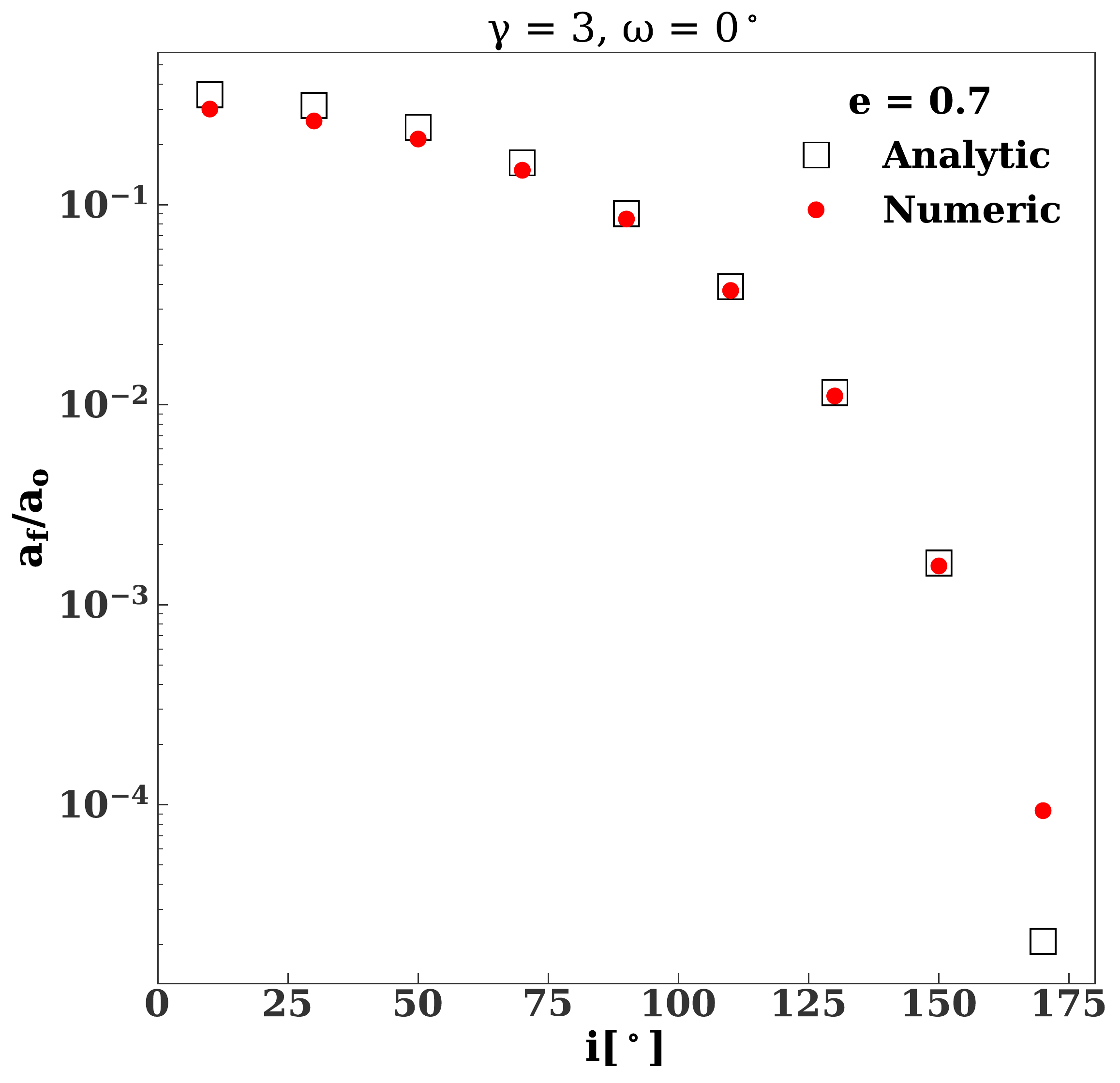}
    \caption{Same as Figure~\ref{fig:numValidSma1} except in the geometric regime.}
    \label{fig:numValidSma2}
\end{figure*}


\bsp	
\label{lastpage}
\end{document}